\documentclass[longauth]{aa}  
\usepackage{graphicx}
\usepackage{hyperref}
\hypersetup{
    colorlinks=true,
    linkcolor=blue,
    citecolor=blue,
    filecolor=magenta,      
    urlcolor=cyan,
    }
\usepackage{txfonts}
\usepackage{pdflscape}
\usepackage{rotating}
\usepackage{multirow}
\usepackage{longtable}
\usepackage{xspace}

\newcommand{\Te}{\ensuremath{T_\text{e}}\xspace}

\newcommand{\mstar}{M$_{\star}$}
\newcommand{\msun}{M$_{\sun}$}

\newcommand{\NII}{{[N\,{\sc ii}]}}

\newcommand{\NIIl}{{[N\,{\sc ii}]\,$\lambda$}}
\newcommand{\NeIII}{{[Ne\,{\sc iii}]}}

\newcommand{\NeIIIl}{{[Ne\,{\sc iii}]\,$\lambda$}}
\newcommand{\SII}{{[S\,{\sc ii}]}}

\newcommand{\SIIll}{{[S\,{\sc ii}]\,$\lambda\lambda$}}

\newcommand{\OIII}{{[O\,{\sc iii}]}}

\newcommand{\OIIIl}{{[O\,{\sc iii}]\,$\lambda$}}
\newcommand{\OII}{{[O\,{\sc ii}]}}

\newcommand{\OIIl}{{[O\,{\sc ii}]\,$\lambda$}}
\newcommand{\OIIll}{{[O\,{\sc ii}]\,$\lambda\lambda$}}

\newcommand{\Ha}{H$\alpha$}

\newcommand{\Hb}{H$\beta$}

\newcommand{\Hy}{H$\gamma$}

\newcommand{\ppxf}{{\sc ppxf}\xspace} 

\begin{document}

\title{JADES: Insights on the low-mass end of the mass - metallicity - star-formation rate relation at 3 < z < 10 from deep JWST/NIRSpec spectroscopy}

   \titlerunning{Low-mass metallicity scaling relations at z>3}

   \author{Mirko Curti
          \inst{1}\fnmsep\inst{2}\fnmsep\inst{3}
          \thanks{E-mail: mirko.curti@eso.org}          
\and
Roberto Maiolino\inst{2}\fnmsep\inst{3}
\and
Emma Curtis-Lake\inst{4}
\and
Jacopo Chevallard\inst{5}
\and 
Stefano Carniani\inst{6}
\and
Francesco D'Eugenio\inst{2}\fnmsep\inst{3}
\and
Tobias J. Looser\inst{2}\fnmsep\inst{3}
\and
Jan Scholtz\inst{2}\fnmsep\inst{3}
\and
Stephane Charlot\inst{7}
\and
Alex Cameron\inst{5}
\and
Hannah \"Ubler\inst{2}\fnmsep\inst{3}
\and
Joris Witstok\inst{2}\fnmsep\inst{3}
\and
Kristian Boyett\inst{8}\fnmsep\inst{9}
\and
Isaac Laseter\inst{10}
\and
Lester Sandles\inst{2}\fnmsep\inst{3}
\and
Santiago Arribas\inst{11}
\and
Andrew Bunker\inst{5}
\and
Giovanna Giardino\inst{12}
\and
Michael V. Maseda\inst{10}
\and
Tim Rawle\inst{13}
\and
Bruno Rodríguez Del Pino\inst{11}
\and
Renske Smit\inst{14}
\and
Chris J.\ Willott\inst{15}
\and
Daniel J.\ Eisenstein\inst{16}
\and
Ryan Hausen\inst{17}
\and
Benjamin Johnson\inst{16}
\and
Marcia Rieke\inst{18}
\and
Brant Robertson\inst{19}
\and
Sandro Tacchella\inst{2}\fnmsep\inst{3}
\and
Christina C. Williams\inst{20}
\and
Christopher Willmer\inst{18}
\and
William M. Baker\inst{2}\fnmsep\inst{3}
\and
Rachana Bhatawdekar\inst{13}\fnmsep\inst{21}
\and
Eiichi Egami\inst{18}
\and
Jakob M. Helton\inst{18} 
\and
Zhiyuan Ji\inst{18}
\and
Nimisha Kumari\inst{22}
\and
Michele Perna\inst{11}
\and
Irene Shivaei\inst{18}
\and
Fengwu Sun\inst{18}
          }

   \institute{
  European Southern Observatory, Karl-Schwarzschild-Strasse 2, 85748 Garching, Germany\\
\and 
Kavli Institute for Cosmology, University of Cambridge, Madingley Road, Cambridge, CB3 0HA, UK\\
\and
Cavendish Laboratory - Astrophysics Group, University of Cambridge, 19 JJ Thomson Avenue, Cambridge, CB3 0HE, UK\\
\and
Centre for Astrophysics Research, Department of Physics, Astronomy and Mathematics, University of Hertfordshire, Hatfield AL10 9AB, UK\\
\and
Department of Physics, University of Oxford, Denys Wilkinson Building, Keble Road, Oxford OX1 3RH, UK\\
\and
Scuola Normale Superiore, Piazza dei Cavalieri 7, I-56126 Pisa, Italy\\
\and
Sorbonne Universit\'e, CNRS, UMR 7095, Institut d'Astrophysique de Paris, 98 bis bd Arago, 75014 Paris, France\\
\and
School of Physics, University of Melbourne, Parkville 3010, VIC, Australia\\
\and
ARC Centre of Excellence for All Sky Astrophysics in 3 Dimensions (ASTRO 3D), Australia\\
\and
Department of Astronomy, University of Wisconsin-Madison, 475 N. Charter St., Madison, WI 53706 USA\\
\and
Centro de Astrobiolog\'ia (CAB), CSIC–INTA, Cra. de Ajalvir Km.~4, 28850- Torrej\'on de Ardoz, Madrid, Spain\\
\and
ATG Europe for the European Space Agency, ESTEC, Noordwijk, The Netherlands\\
\and
European Space Agency (ESA), European Space Astronomy Centre (ESAC), Camino Bajo del Castillo s/n, 28692 Villanueva de la Cañada, Madrid, Spain \\
\and
Astrophysics Research Institute, Liverpool John Moores University, 146 Brownlow Hill, Liverpool L3 5RF, UK\\
\and
NRC Herzberg, 5071 West Saanich Rd, Victoria, BC V9E 2E7, Canada\\
\and
Center for Astrophysics $|$ Harvard \& Smithsonian, 60 Garden St., Cambridge MA 02138 USA\\
\and
Department of Physics and Astronomy, The Johns Hopkins University, 3400 N. Charles St. Baltimore, MD 21218\\
\and
Steward Observatory University of Arizona 933 N. Cherry Avenue Tucson AZ 85721, USA\\
\and
Department of Astronomy and Astrophysics, University of California, Santa Cruz, 1156 High Street, Santa Cruz, CA 95064, USA\\
\and
NSF’s National Optical-Infrared Astronomy Research Laboratory, 950 North Cherry Avenue, Tucson, AZ 85719, USA\\
\and
European Space Agency, ESA/ESTEC, Keplerlaan 1, 2201 AZ Noordwijk, NL\\
\and
AURA for European Space Agency, Space Telescope Science Institute, 3700 San Martin Drive. Baltimore, MD, 21210\\
             }
   \authorrunning{M. Curti et al.}
   \date{} 
 
  \abstract
{We analyse the gas-phase metallicity properties of a sample of low stellar mass (log\ \mstar/\msun$\lesssim 9$) galaxies at $3<z<10$, observed with JWST/NIRSpec as part of the JADES programme in its \emph{deep} GOODS-S tier.
By combining this sample with more massive
galaxies at similar redshifts from other programmes, we study the scaling relations between stellar mass (\mstar), oxygen abundance (O/H), and star-formation rate (SFR) for $146$ galaxies, spanning across three orders of magnitude in stellar mass and out to the epoch of early galaxy assembly.
We find evidence for a shallower slope at the low-mass-end of the mass-metallicity relation (MZR), with 12 + log(O/H) = (7.72$\pm$0.02) + (0.17$\pm$0.03) log(\mstar/10$^8$\msun),
in good agreement with the MZR probed by local analogues of high-redshift systems like `Green Pea' and `Blueberry' galaxies.
The inferred slope is well matched by models including `momentum-driven' SNe winds, suggesting that feedback mechanisms in dwarf galaxies (and at high-z) might be different from those in place at higher masses.  
The evolution in the normalisation is observed to be relatively mild compared to previous determinations of the MZR at $z\sim3$ ($\sim0.1-0.2$~dex across the explored mass regime).
We observe a deviation from the local fundamental metallicity relation (FMR) for our sample at high redshift, especially at $z>6$, with galaxies significantly less enriched (with a median offset in log(O/H) of $\sim 0.5$~dex, significant at $\sim 5\sigma$) than predicted given their \mstar\ and SFR.
These observations are consistent with an enhanced stochasticity in the star-formation history, and/or with an increased efficiency in metal removals by outflows, prompting us to reconsider the nature of the relationship between \mstar, O/H, and SFR in the early Universe. 
}
   \maketitle
%


\section{Introduction}

The \emph{James Webb Space Telescope} (\emph{JWST}) has already begun to revolutionise our view of the high-redshift Universe, opening a new window onto early galaxy formation. 
In particular, the high-sensitivity Near Infrared Spectrograph \citep[NIRSpec,][]{jakobsen_nirspec_2022,ferruit_nirspec_2022, boeker_nirspec_2023}
now allows the spectroscopic characterisation of the properties of the interstellar medium (ISM) and the nature of ionizing spectra of primordial galaxies. 
In this context, the study of the gas-phase metallicity\footnote{as traced by the oxygen abundance, and usually quoted as 12+log(O/H)} via rest-frame optical spectroscopy, which has been limited for decades to $z \leq 3.5$ due to the intrinsic limitations of ground-based observatories, is now possible out to z$\sim$10.
Such observations are providing precious
constraints on cosmological simulations and chemical evolution models, enabling scientists to chart the processes that shaped the formation of galaxy structures in the early Universe \citep{somerville_dave_2015, maiolino_re_2019}.
For instance, characterising the scaling relation between stellar mass and gas-phase metallicity in galaxies (the so-called mass-metallicity relation, MZR, \citealt{lequeux_chemical_1979, tremonti_origin_2004, lee_extending_2006, yates_present-day_2019, baker_mzr_2023}) at high redshift is critical to constrain the processes regulating the growth of early galaxies, as this relation 
is shaped by the interplay between gas accretion, star formation, metal enrichment, and outflows driving the baryon cycle.
Furthermore, as a consequence of the long-lasting interplay between these different processes, the metallicity of galaxies has been observed to correlate also with different galactic properties. 
In particular, an anti-correlation between metallicity and star-formation rate, at fixed stellar mass, has been observed and characterised in detail on both global and local scales, in what is usually referred to as the `Fundamental Metallicity Relation'  
 \citep[FMR, ][]{ellison_clues_2008, mannucci_fundamental_2010, yates_relation_2012, salim_mass-metallicity-star_2015, telford_exploring_2016, curti_massmetallicity_2020, baker_resolved_fmr_2023}.
 
Several studies leveraged the new spectroscopic capabilities of JWST/NIRSpec to start characterising the properties of strong line emitters at $z>3$, exploiting the data collected in the framework of Early Release Observations \citep[ERO]{pontoppidan_ERO_2022}, Early Release Science (ERS) and Cycle 1 General Observer (GO) and Guaranteed Time Observations (GTO) programmes such as GLASS \citep[Proposal ID: 1324; ][]{treu_glass_survey_2022}, Cosmic Evolution Early Release Science \citep[CEERS, Proposal ID: 1345; ][Finkelstein et al., in preparation]{finkelstein_ceers_paper1_2022}, and \emph{JWST} Advanced Deep Extragalactic Survey \citep[JADES, Proposal ID: 1210; ][Eisenstein et al., in preparation]{robertson_jades_2022, curtis-lake_2023}. 
Recent works have indeed investigated the ionisation properties of galaxies beyond $z=3$, showing that these sources exhibit emission line ratios consistent with hard ionizing spectra and low metallicities \citep{mascia_reionisation_2023, matthee_eiger_2023, sanders_ceers_2023, cameron_jades_bpt_2023}. 
These kind of analyses have been pushed to some of the highest redshift galaxies discovered, with \cite{williams_z9_2022} reporting a $z=9.5$ galaxy with a very high \OIII/\OII\ ratio and relatively low metallicity, while \cite{bunker_gnz11_2023} recently reporting the detection of multiple emission lines (including several high-ionisation metal lines) in the NIRSpec spectrum of the luminous $z\sim$10.61 galaxy GN-z11 \citep{oesch_gnz11_2016}.
The wealth of emission lines detected in these high-z galaxy spectra can hence be exploited to assess the chemical enrichment of their ISM.

In the past, several studies have investigated the evolution in the metallicity scaling relations out to z$\sim$3, finding signatures of clear trend of decresing metallicity with redshift at fixed stellar mass \citep[e.g.,][]{shapley_chemical_2005, erb_mass-metallicity_2006,maiolino_amaze_2008,mannucci_lsd_2009,zahid_mass-metallicity_2011,henry_low_mass_mzr_2013, yabe_mzr_z1_2014, zahid_universal_2014, wuyts_consistent_2014, guo_stellar_2016, sanders_mosdef_mzr_2021, topping_mzr_2021}.
However, when reported to the framework of the FMR, no evidence of redshift evolution has been observed, suggesting that, on average, galaxies evolved through smooth secular processes over the past $\sim10$ Gyr as driven by the interplay of gas flows, star-formation, and metal enrichment, which is reflected on such scaling relation \citep{mannucci_fundamental_2010, belli_testing_2013, nakajima_ionization_2014, maier_fmr_z2_2014, salim_mass-metallicity-star_2015, hirschauer_metal_2018, sanders_mosdef_mzr_2021, hayden-pawson_NO_2022}.

More recent studies attempted to push the investigation of the chemical properties of galaxies up to $z\sim10$, either from the detection of  emission lines in individual sources, or from the analysis of composite spectra, finding initial evidence for a relatively mild evolution of the oxygen abundance with redshift at fixed stellar mass \citep[e.g.,][]{curti_smacs_2023, Schaerer_ero_2022, arellano-corodva_2023, taylor_2022, trump_2022, rhoads_2023, langeroodi_mzr_2022, matthee_eiger_2023, heintz_fmr_2022, nakajima_mzr_ceers_2023, shapley_mzr_2023}. 
Despite the efforts, the majority of previous studies focused on relatively high stellar mass systems, whereas probing the evolution of the metallicity scaling relations at low stellar masses, a regime strongly sensitive to feedback processes due to the shallower gravitational potential, remained challenging \citep{wuyts_low_mass_mzr_2012, henry_mzr_2013} until the arrival of \emph{JWST}
\citep{Li_mzr_dwarfs_z3_2022}.

Moreover, it is worth to recall that the vast majority of such metallicity determinations are based on strong emission line ratios that have been calibrated on local sample of galaxies \citep{pettini_oiiinii_2004, maiolino_amaze_2008, curti_new_2017} or on local analogues mimicking the physical conditions of high-z sources \citep{bian_ldquodirectrdquo_2018, nakajima_empress_2022}, and the applicability of these methods at high redshift still needs to be carefully assessed, as different ionization properties in the ISM of high-z galaxies could bias the metallicity measurement \citep{kewley_cosmic_2013, steidel_strong_2014, strom_nebular_2017,  sanders_mosdef_2018, sanders_ceers_2023, cameron_jades_bpt_2023}.
Nonetheless, the rate of auroral lines detections at high-z is now rapidly increasing thanks to the \emph{JWST}, and novel attempts to re-calibrate the classical strong line diagnostics for the high-redshift Universe are being provided \citep{sanders_calibrations_2023}.

In this paper, we aim to explore further the cosmic evolution of the metallicity scaling relations, leveraging deep spectroscopy with JWST/NIRSpec to probe the low stellar mass regime \mstar $\approx 10^{6.5}$-$10^{9.5}$ \msun\ from $z=3$ 
out to the highest redshift in which rest-frame optical nebular lines are accessible to NIRSpec ($z\sim 9.5$).
By combining observations from JADES, which enables us to explore the very low-mass regime down to \mstar $\sim 10^{6.5}$, with existing datasets, we assess the evolution of the MZR slope and normalisation, and its implications on the physical processes in place in early galaxies. Moreover, we test the FMR framework down to the epoch when the Universe was $\lesssim 500$ Myr old.  

The paper is organised as follows.
In Section~\ref{sec:data}, we describe observations, data reduction, and spectral fitting procedures.
Section~\ref{sec:quantities} we present the analysis performed to derive the physical quantities of interest, i.e., \mstar, SFR, and metallicity.
In Section~\ref{sec:scaling_relations_mzr}, we assess the mass-metallicity relation (MZR) at $z>3$ combining our JADES sample with galaxies at higher masses drawn from the CEERS programme, whereas in Section~\ref{sec:fmr_evol} we investigate any evolution in the framework of the fundamental metallicity relation, and discuss possible inferences about the interplay between gas flows and star-formation regulating early galaxy assembly.
In Section~\ref{sec:summary}, we summarise our conclusions.

Throughout this paper, we adopt a \cite{planck_2020} cosmology, a \cite{chabrier_galactic_2003} initial mass function, and a solar metallicity 12+log(O/H)=8.69 \citep{allende_prieto_forbidden_2001, asplund_solar_2009}.

\begin{table}
 \centering
   \caption{Definitions of line ratios adopted throughout the paper.}
 \begin{tabular}{lc}
  \hline
  Diagnostics & Line Ratio\\
  \hline
  R2 & [\ion{O}{II}]$\lambda\lambda$3727,29 / H$\beta$\\
  R3 & [\ion{O}{III}]$\lambda$5007 / H$\beta$\\
  N2 & [\ion{N}{II}]$\lambda$6584 / H$\alpha$\\
  S2 & [\ion{S}{II}]$\lambda\lambda$6717,31 / H$\alpha$\\
  R23 &( [\ion{O}{II}]$\lambda\lambda$3727,29 + [\ion{O}{III}]$\lambda\lambda$4959, 5007) / H$\beta$\\
  \^R & 0.47 $\times$ log$_{10}$(R2) + 0.88 $\times$ log$_{10}$(R3) \\ 
  O32 & [\ion{O}{III}]$\lambda$5007 / [\ion{O}{II}]$\lambda\lambda$3727,29\\
  Ne3O2 & [\ion{Ne}{III}]$\lambda$3869 / [\ion{O}{II}]$\lambda\lambda$3727,29\\
   
  \hline
 \end{tabular}
  \label{tab:line_ratios_def}
\end{table}

\begin{figure*}
    \centering
    \includegraphics[width=0.95\textwidth]{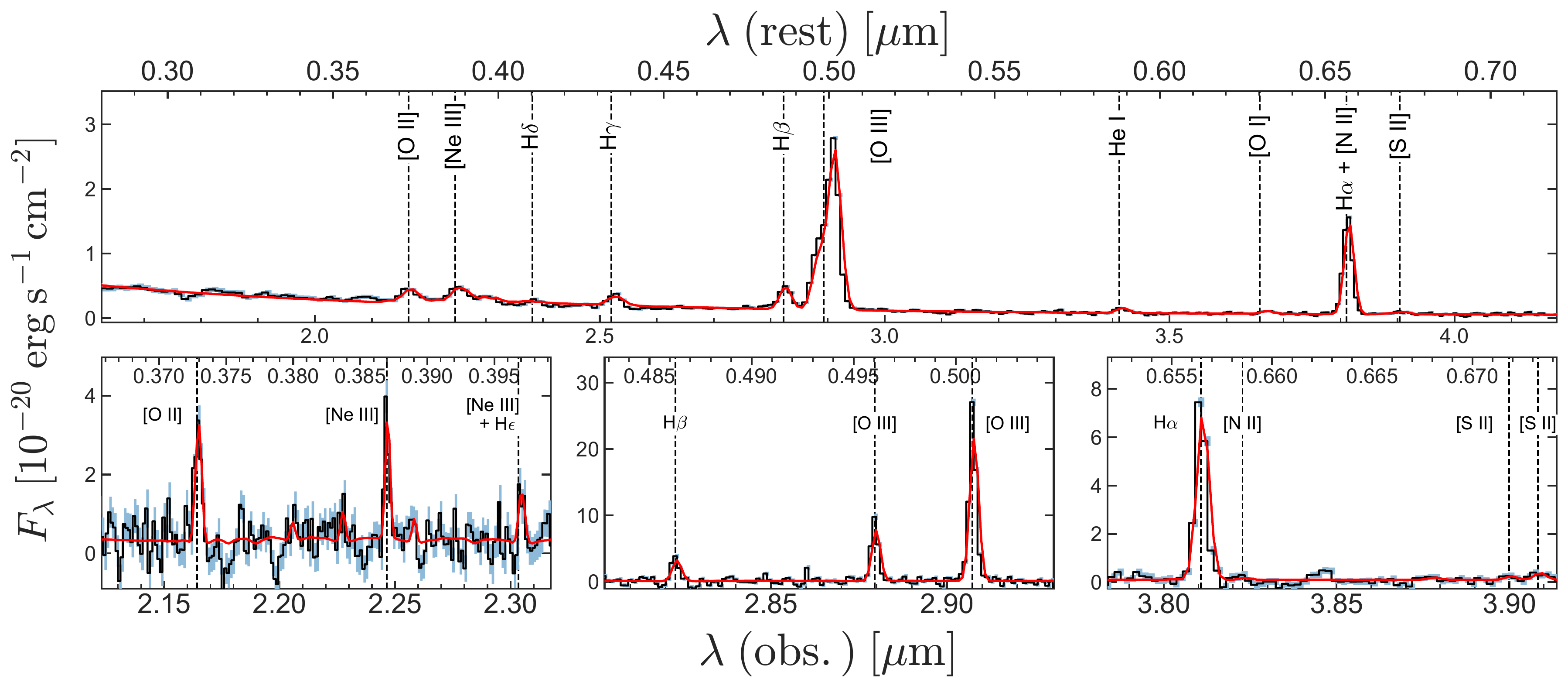}
    \caption{\textit{Top panel:} Example fit to the PRISM (R100) spectrum (in black) of the $z=4.805$ galaxy JADES-GS+53.16268-27.80237, zoomed-in on the region of the main rest-frame optical emission lines. The best-fit from \textsc{ppxf} is shown in red.
    \textit{Bottom panels:} Medium resolution grating (R1000) spectrum (and its best-fit) for the same galaxy. From left to right, the panels show a zoom-in on the regions of the \OII\ and \NeIII, \Hb\ and \OIII, \Ha\, \NII\ and \SII\ emission lines, respectively. In all panels, the error spectrum is marked by the cyan shaded region. 
    }
    \label{fig:fit_spectrum}
\end{figure*}

\section{Observations and data processing} 
\label{sec:data}

\subsection{Observations}
\label{sub:observations}

The data presented in this paper have been obtained via multi-object spectroscopy observations conducted with the micro-shutter assembly (MSA) of NIRSpec \citep{jakobsen_nirspec_2022, ferruit_nirspec_2022, boeker_nirspec_2023} on the \emph{JWST}.
Observations were performed in three visits carried out between the 21st-25th of October 2022 (Program ID: 1210; PI: N. Luetzgendorf) in the Great Observatories Origins Deep Survey South (GOODS-S) legacy field \citep{giavalisco_goodss_2004}, as part of one of the \emph{deep} tiers of the \emph{JWST}-GTO JADES programme \citep{Eisenstein_JADES_2023, bunker_hst_deep_DR_2023}.
Each visit consisted of 33,613 s of integration in the PRISM/CLEAR configuration (hereinafter just `PRISM'), and 8,403 s integration in each of G140M/F070LP, G235M/F170LP, G395M/F290LP (hereinafter `medium resolution gratings'). 
Across three visits, this totals 28 hours of integration in the PRISM, providing continuous spectral coverage from $0.6 - 5.3 \mu $m at $R\sim30-300$, and $\sim$7 hours in each of the medium resolution gratings, providing $R\sim1000$ across the full spectral range of NIRSpec\footnote{Further 7 hours were spent in the high-resolution G395H/F290LP grating, providing $R\sim2700$ between $\sim 2.8 - 5.1 \mu $m; however, we are not exploiting such observations in this specific work.}.

Observations within each visit were performed adopting a 3-shutter nodding pattern, with the central pointing of each visit dithered by $<1$ arcsec to sample different areas of the detector.
A total of 253 unique targets were observed with the PRISM within the three pointings; among these, 67 targets were observed in all three MSA configurations, whereas 62 targets featured in two pointings, and the remaining 124 were observed for only one-third of the total exposure time. 
Each pointing had a bespoke MSA configuration, and target allocation was performed via the eMPT\footnote{\url{https://github.com/esdc-esac-esa-int/eMPT_v1}} software \citep{bonaventura_empt_2023} to maximise the number of targets in common between all the three pointings, with special attention for rare objects included in the highest priority classes.

We note that in the medium and high resolution modes, individual spectra are dispersed over a large area on the detector. To minimise the possibility of overlapping emission, in our grating observations we have isolated our highest-priority targets by closing the shutters of low-priority targets on the same row (i.e., targets that could cause overlapping spectra).
For this reason, in the grating modes we observe only 198 unique targets. 

\subsection{Data Reduction}
Flux-calibrated 2D and 1D spectra have been produced using the pipeline developed by the ESA NIRSpec Science Operations Team (SOT) and the NIRSpec GTO Team. 
Most of the processing steps in the pipelines adopt the same algorithms included in the official STScI pipeline used to generate the MAST archive products \citep{ferruit_nirspec_2022}.
Initially, we processed the raw data (i.e, level 1a data from the MAST archive) with a ramp-to-slope pipeline that estimates the count rate per pixel by using all unsaturated groups in the ramp, and which is optimised to reject cosmic rays on the basis of the slope of the individual ramps \citep[for more details see ][]{birkmann_nirspec_2011,boeker_nirspec_2012, giardino_nirspec_CRs_2019, ferruit_nirspec_2022}.
All the pre-processed count-rate images were then processed using a data reduction pipeline which includes both ESA NIRSpec SOT codes and bespoke developed NIRSpec GTO algorithms.
We briefly outline here the main steps, while for a more detailed description we refer to \cite{bunker_hst_deep_DR_2023} and to a forthcoming paper of the NIRSpec/GTO collaboration (Carniani et al., in preparation). 
In brief, the pipeline consists of 11 main steps: i) identification of non-target galaxies intercepting the open shutters; ii) pixel-level background subtraction; iii) extraction of the spectral trace of each target and wavelength and spatial coordinate assignments to each pixel in the 2D maps; iv) pixel-to-pixel flat-field correction; v) spectrograph optics and disperser correction; vi) absolute flux calibration; vii) path-losses correction; viii) rectification of 2D spectra; ix) extraction of 1D spectra; x) combination of 1D spectra generated from each exposure, nod, and pointing; xi) combination of 2D spectra.
Therefore, the data processing workflow returns both a combined 1D and 2D spectrum for each target. 
We note that the combined 1D spectra are not extracted from the combined 2D maps, but are the result of a weighted average of 1D spectra from all integrations, which allowed us to implement bad pixels and outliers rejection algorithms more efficiently.
We adopt an irregular wavelength grid for the 1D and 2D spectra of the PRISM configuration, in order to avoid oversampling of the line spread function at short wavelengths ($\lambda\sim1~\mu$m). 
For the G140M/F070LP grating/filter configuration we extended the calibration of the spectrum up to $1.84\mu$m, taking into account the
transmission filter throughput beyond the nominal wavelength
range of this configuration ($0.70\mu$m– $1.27\mu$m).
Finally, path-loss corrections are applied by modelling galaxies as point-like sources, taking into account the relative intra-shutter position of each source by leveraging on a full forward-modelling of the telescope and instrument optical paths \citep{jakobsen_nirspec_2022, ferruit_nirspec_2022}.

\subsection{Spectral fitting}

Both continuum and line emission are modelled simultaneously adopting the penalised pixel fitting algorithm, \ppxf
\citep{cappellari_improving_2017, cappellari_ppxf_2022}, which models the continuum as a linear superposition of simple stellar-population (SSP) spectra, using
non-negative weights and matching the spectral resolution of the observed spectrum. We used the high-resolution (R=10,000) SSP library
combining MIST isochrones \citep{choi_MIST_2016} and the C3K theoretical atmospheres \citep{conroy_stellar_halo_2019}. The flux blueward of the Lyman break was manually set to zero.
These templates are complemented by a 5\textsuperscript{th}-degree multiplicative Legendre polynomial, to take into account systematic differences
between the SSPs and the data (e.g., dust extinction, mismatch between the SSP models and high-redshift stellar populations, as well as residual flux calibration issues). 
The emission lines are modelled as pixel-integrated Gaussians, matching the observed spectral resolution. To reduce the number of degrees
of freedom, we divide all emission lines in four kinematic groups, constrained to have the same redshift and \emph{intrinsic} broadening. These
are UV lines (blueward of 3000~\AA), the Balmer series of Hydrogen emission lines, non-Hydrogen optical lines (blueward of 9000~\AA), and near-infrared (NIR) lines. The stellar continuum has
the same kinematics as the Balmer lines.
Furthermore, we tie together doublets that have fixed flux ratios as determined by the atomic physics (e.g., the \OIIIl 5007/4959 ratio, which is set equal to 2.97), and constrain variable-ratio doublets to their physical ranges. 
We note that, at the resolution of the PRISM/CLEAR mode, the two components of the \OIIll 3726, 3729 doublet are completely unresolved and therefore are fit as a single Gaussian component centred at a rest-frame wavelength of $3728.42 \AA$, whereas for the gratings configurations the two emission lines are fitted separately, and their ratio is constrained to the physical range allowed by the density of the ISM ($\in [0.38-1.46]$); the same applies for the \SIIll 6718,6732 doublet ($\in [0.44-1.45]$) \citep{osterbrock_astrophysics_2006}.

We show an example fit for a $z=4.805$ galaxy in Figure~\ref{fig:fit_spectrum}.
The upper panel shows the fit (in red) to the full PRISM (R100) spectrum in the wavelength region covering the main rest-frame optical emission lines, whereas in the three bottom panels we show a zoom-in of the best-fit to the R1000 spectrum for the same galaxy on the region of \OII\ and \NeIII, \Hb\ and \OIII, and \Ha, \NII\ and \SII emission lines, respectively.

\section{Derived physical quantities} 
\label{sec:quantities}

\begin{figure}
    \includegraphics[width=0.95\columnwidth]{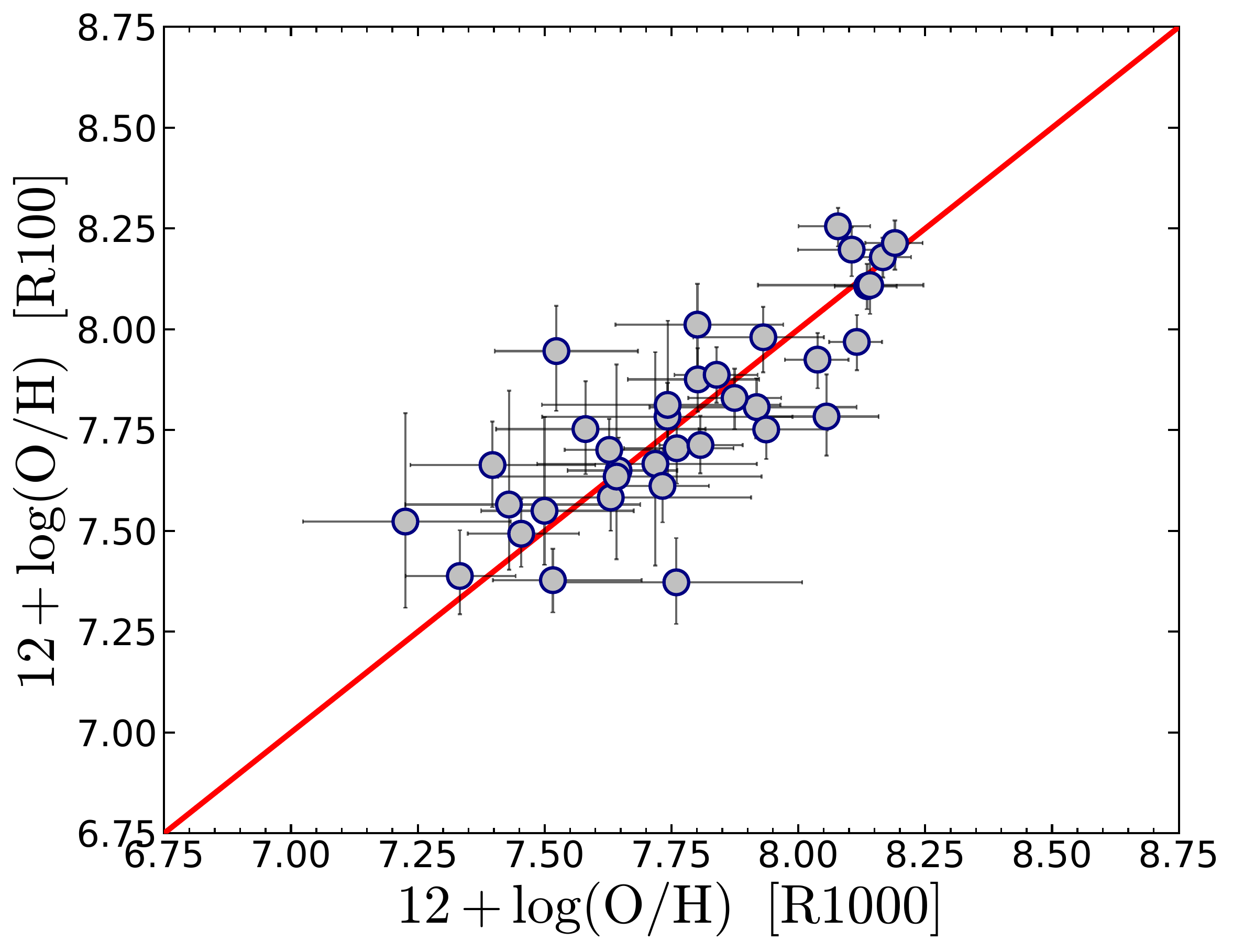}
    \caption{
    Comparison between the metallicity derived from medium resolution gratings and PRISM spectra, for objects in which both configurations satisfy the requirements described in Section~\ref{sec:metallicity}.
    The two distributions scatter across the equality line (in red), with a median offset of $0.01$~dex and a standard deviation of $0.15$~dex. In $75\%$ of the cases the two measurements are consistent within their $1\sigma$ uncertainties.
    }
    \label{fig:oh_compare_M_P}
\end{figure}

\begin{figure}
    \centering
    \includegraphics[width=0.9\columnwidth]{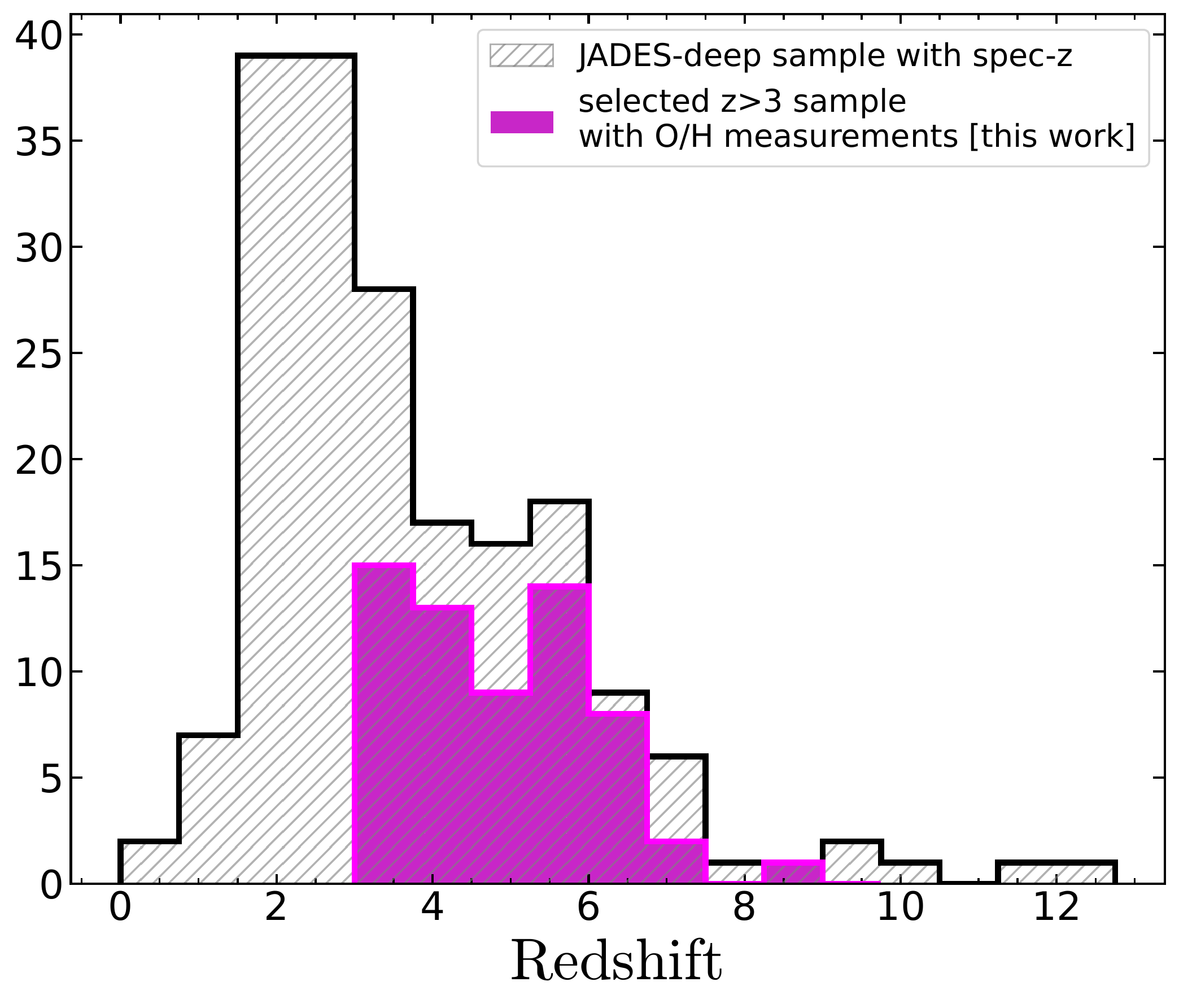}
    \caption{The redshift distribution of the selected sample of 62 JADES galaxies with metallicity measurements analysed in this paper, compared to the distribution of the total `JADES-Deep' sample presented in Section~\ref{sub:observations} and \citealt{bunker_hst_deep_DR_2023} with spectroscopic redshift determination.
    }
    \label{fig:z_histo}
\end{figure}

\subsection{Gas-phase metallicity}
\label{sec:metallicity}

For the purposes of deriving the gas-phase metallicity, we use the observations obtained with both PRISM (R$\sim$100) and medium resolution gratings (R$\sim$1000), treating however the two spectral configurations independently.
For each spectral configuration, emission line fluxes as measured from \ppxf\ are reddening corrected on the basis of the decrement measured from the available Balmer lines. 
More specifically, we exploit the \Ha/\Hb\ ratio at redshifts $z<6.75$, where \Ha\ is spectrally covered by NIRSpec, and \Hy/\Hb\ otherwise, adopting a \cite{gordon_LMC_attenuation_2003} attenuation law and assuming the theoretical ratios from Case B recombination at T$=1.5\times 10^4$K (i.e., \Ha/\Hb=2.86; \Hy/\Hb=0.47).
In case no Balmer decrement could be measured, we adopt the nebular attenuation inferred from the SED fitting performed on PRISM spectra with the \textsc{beagle} code \citep[][, Chevallard et al., in preparation]{chevallard_beagle_2016}, which incorporates a two-component dust attenuation model that accounts for the differential attenuation of nebular and stellar emission in a self-consistent way.

The gas-phase metallicity is derived exploiting a revisited version of the calibrations presented in \cite{curti_massmetallicity_2020} (which are anchored to the \Te abundance scale), refined using a sample of local, metal-poor galaxies to better sample the low-metallicity regime. 
Such calibration set is similar to that presented in \cite{nakajima_empress_2022}.
In addition, a novel diagnostics labeled \^R (see Table~\ref{tab:line_ratios_def}), and based on a linear combination of R2 and R3 which further reduces the scatter of the calibration sample at fixed metallicity and span a wider dynamical range compared to the `standard' R23 diagnostic, is implemented in place of the latter.
For a detailed description of the rationale behind the new diagnostic, and for a more thorough comparison of classical strong-line calibrations against individual auroral line detections within the hereby presented JADES dataset, we refer to \cite{laseter_auroral_jades_2023}.
A minimum signal-to-noise (S/N) of 5 is required on \Ha\ and \OIII, whereas of 3 on fainter emission lines, in order to consider a given emission line (and its associated diagnostics) in the metallicity derivation.

We visually inspected the spectra to remove sources affected by a poor fitting of the emission lines, as well as two galaxies identified as AGNs based on evidence for broad line emission, as detailed in \cite{maiolino_jades_agn_2023}. 
Moreover, additional 14 AGN candidates identified via significant emission detected in high-ionisation transitions like He II$\lambda4686$ and N IV$\lambda1485$ (Scholtz et al., in preparation) are also removed from the analysis, as AGN-powered ionisation would compromise the standard metallicity calibrations tuned for star-forming galaxies, as well as complicate the derivation of stellar masses and star-formation rates.
However, it is relevant to note that most of these galaxies still show line ratios consistent with the star-forming population according to the `BPT' diagram \citep{baldwin_classification_1981, kewley_theoretical_2001, kauffmann_dependence_2003} , whose applicability in discriminating among different ionising sources at high redshift has been in fact recently questioned, especially at low metallicity \citep{Ubler_AGN_z5_2023, maiolino_jades_agn_2023}.
Considering the challenges in quantitatively assessing the relative contribution of AGN ionisation with respect to the host galaxy in the observed spectra of these sources, we apply a conservative choice by removing the entire sample of narrow-line AGN candidates.
Nonetheless, all the conclusions of the present paper are robust against this choice, and in Appendix~\ref{sec:appendix_A} we report the results obtained by including also this sub-sample of galaxies.

In this work, all available strong-line diagnostics are included for each galaxy in the metallicity calculation, in order to reduce potential biases associated with the use of an individual, specific line ratio. 
The procedure explores the log(O/H) parameter space with a Markov Chain Monte Carlo (MCMC) algorithm exploiting the \textsc{emcee} package \citep{emcee_2013}, 
with the logarithmic likelihood defined as
\begin{equation}
    \text{log}(\text{L}) \propto \sum_{i} \frac{(\text{R}_{\text{obs},i} - R_{\text{cal},i} )^2}{(\sigma^2_{\text{obs},i} + \sigma^2_{\text{cal},i})} \ ,
\end{equation}
where the sum is performed over the set of available diagnostics used, R$_{\text{obs}}$ are
the observed line ratios,  R$_{\text{cal}}$ are the line ratios predicted by each calibration at a given metallicity, $\sigma_{\text{obs}}$ are the uncertainties on the observed line ratios, and $\sigma_{\text{cal}}$ are the dispersions of each calibrated line ratio at fixed metallicity.

For the majority of the sources under study however, we note that the \NII\ and \SII\ emission lines are either undetected, or falls outside of the NIRSpec wavelength coverage (see also \citealt{cameron_jades_bpt_2023}); moreover, \NIIl6584\ and \Ha\ are severely blended in PRISM spectra at the lowest redshifts probed by our sample.
Therefore, we ultimately and effectively involve only diagnostics based on `alpha' elements (i.e., R3, \^R, O32; when available, \NeIII/\OII\ is also included). 
In some cases, \OIIIl5007 and \Hb\ are the only emission lines significantly (above three-sigma) detected in the spectrum, which means that R3 is the only available diagnostic. 
In such cases, further information from upper limits on the \NIIl6584 (from R1000 spectra, where available) and on the \OIIl 3727,29 emission lines can be placed on the N2 and O32 diagnostics, helping to discriminate between the two solutions provided by the double-branched R3 calibration.
We here further note however that adopting instead a single, double-branched diagnostics (like R3 or R23) for the whole sample, although increasing the self-consistency of the metallicity derivation, might artificially introduce scatter in the log(O/H) distribution, as high-z galaxies typically tend to occupy the high-excitation region of the calibration, close or above the calibration plateau. This means that, even if the degeneracy between the two branches is somehow broken, the calibration itself carries less information about the metallicity in that region (being almost flat and hence not very sensitive to metallicity variations).
Therefore, forcing either the low- or high-solution introduces a `gap' in metallicity between populations of galaxies which share very similar line ratios \citep[see e.g., the discussion in][]{guo_stellar_2016}, as the offset between the low- and high-metallicity solution is large at fixed line ratio.  
This can have a non-negligible impact, for instance, on the inferred slope of the metallicity scaling relations.

The procedure outlined above is applied separately for each galaxy to both PRISM and gratings spectra (where the latter are available). 
As mentioned above, we note that the N2 diagnostic is not included in any of the metallicity calculations based on PRISM spectra, as the R100 spectral resolution is not enough to resolve \NIIl6584 from \Ha; similarly, S2-based diagnostics are not included for PRISM spectra, as they might be affected by \NII\ contamination of the \Ha \ line. 
As for the analysis carried out in the present paper, a fiducial metallicity value is then assumed, for each source, on the basis of the following scheme: 
\begin{itemize}
\item we take the O/H inferred from medium resolution (R1000) gratings if at least two diagnostics were available, else
\item we take the O/H inferred from the PRISM if at least two diagnostics were available in the PRISM spectrum, else
\item we take the O/H inferred from medium resolution gratings as based on one individual diagnostics (typically R3), exploiting the information from 3-sigma upper/lower limits on other line ratios (e.g., N2, O32) to break the degeneracy of double-valued calibrations, else
\item we take the O/H inferred from the PRISM as based on one diagnostics (typically R3) in combination with the information from the upper limits on \OIIll3727,29.
\end{itemize}
The hereby adopted scheme leverages the higher resolution grating spectra for galaxies with strong line emission, whereas the much deeper PRISM spectra for fainter targets.
We opt for not considering `hybrid' gratings-to-prism line ratios because of potential issues associated with small wavelength or flux inter-calibration uncertainties between the two configurations that are not yet fully understood.
However, we compare the metallicity derived from gratings and prism spectra for galaxies where both measurements are available in the right-hand panel of Figure~\ref{fig:oh_compare_M_P}, i.e., $35$ galaxies.
The two measurements are in agreement within the $1\sigma$ individual uncertainties in $\sim 75\%$ of the cases (and in $\sim 95\%$ within $2\sigma$), with negligible systematic offset (i.e.,  $0.01$~dex median offset towards higher grating metallicities) and a standard deviation of $0.15$~dex.

In this work, we focus specifically on galaxies at $z\geq 3$, in order to study the evolution of the metallicity scaling relations beyond the epochs probed by previous surveys from the ground.
Therefore, modulo the selections and caveats discussed above, our analysis includes a total of 62 galaxies from JADES with a metallicity measurement. 
The oxygen abundances measured in the selected JADES sample span between 12+log(O/H) = 7.21--8.54, with an average of 7.77 (i.e., $0.12\ \text{Z}_{\odot}$).
In Figure~\ref{fig:z_histo}, we report the redshift distribution of our final selected sample of galaxies, compared to the distribution of the sample from the JADES \emph{deep} GOODS-S tier observations with a spectroscopically confirmed redshift\footnote{i.e., with a redshift confirmed from either R1000 or R100 spectra}, which in turn constitutes $\sim 70$ per cent of the total sample introduced in Section~\ref{sub:observations}; for a more detailed description of the full JADES \emph{deep} galaxy sample, the selection function, and redshifts spectroscopic confirmation, we refer to  \cite{bunker_hst_deep_DR_2023}.
In particular, the final sub-sample selected for the current analysis represents the $\sim 57$ percent of the JADES \emph{deep} sample in GOODS-S at $z>3$ with a spectroscopic redshift, as remaining galaxies either do not have more than one emission line detected in their spectra, show evidence for contamination from AGN, or are lineless sources identified only through prominent spectral breaks \citep[e.g.,][]{curtis-lake_2023, looser_gs73_2023}.

\subsection{Stellar masses and star-formation rates}
\label{sec:mass_sfr}

\begin{figure*}
    \centering
    \includegraphics[width=0.9\textwidth]{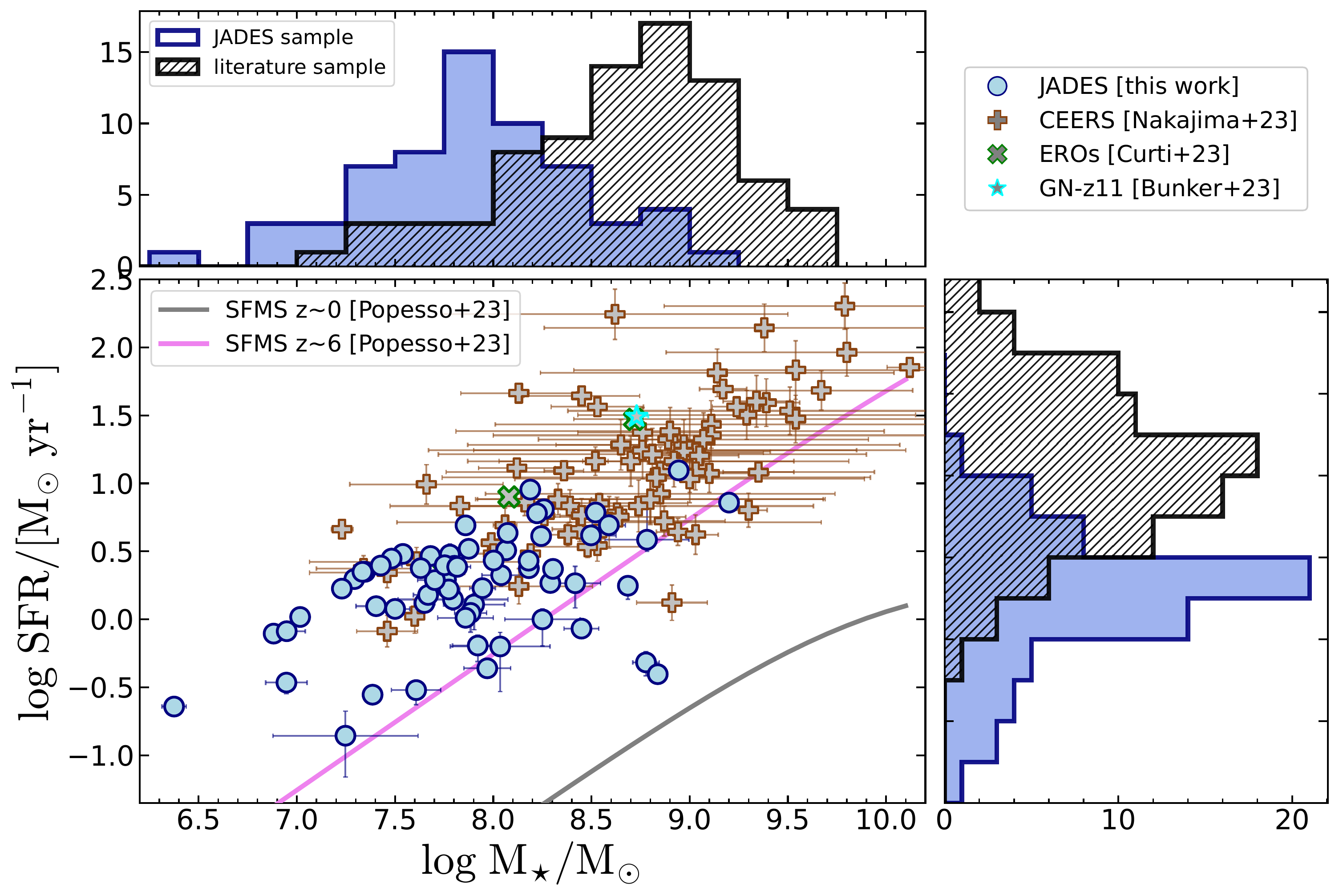}
    \caption{The distribution in the stellar mass versus star-formation rate plane for our combined \emph{JWST} sample, including JADES galaxies presented in this work and the literature sample compiled from \citealt{nakajima_mzr_ceers_2023} (CEERS), \citealt{curti_smacs_2023} (EROs), and \citealt{bunker_gnz11_2023} (GN-z11), respectively. 
    The top and right-hand inset panels show the histograms of the distribution in both parameters.
    The SFR for JADES galaxies is derived from \textsc{beagle} fitting to both PRISM spectra and NIRCAM photometry (where available, and to PRISM spectra only otherwise).
    For CEERS galaxies, the SFRs are compiled from \citealt{nakajima_mzr_ceers_2023} as inferred from the \Hb\ luminosity and the \cite{kennicutt_star_2012} calibration, but have been here scaled down by $0.23$~dex to account for the mean offset between the \Ha(\Hb)-based and \textsc{beagle}-based SFRs measured for the JADES sample (we refer to Section~\ref{sec:mass_sfr} for more details).
    For both ERO galaxies and GN-z11 the \mstar and SFR have been derived via \textsc{beagle} fitting, consistently with what done for JADES sources. 
    The parametrisation (and its extrapolation at low-mass)} of the main sequence of star-formation (SFMS) at z$\sim$0 and z$\sim$6 from \citealt{popesso_SFMS_2023} is also shown for reference.
    
    \label{fig:m-sfr_hist}
\end{figure*}



To measure the stellar masses for our selected sample of galaxies in JADES, 
we employ full spectral fitting to PRISM spectra performed with the \textsc{beagle} code \citep{chevallard_beagle_2016}. The PRISM spectra that have been employed only included slit-loss corrections based on the assumption of a point source, which is not appropriate for all objects in the sample.  To address this, where objects have NIRCam photometry from the JADES survey, the \textsc{beagle} fits are run with both PRISM spectroscopy and total, Kron-based photometry \citep{rieke_jades_DR_2023}.  In these cases, a low-order calibration polynomial is included in the fit to match the shape and normalisation of the spectrum to the photometry, whereby the total stellar mass and star-formation rate estimates are adjusted to the total fluxes measured for the objects. Where NIRCam photometry is not available we resort to the derived quantities from PRISM spectroscopy alone.   
We assume a delayed-exponential star-formation history (SFH), a \cite{chabrier_galactic_2003} IMF with an upper mass limit of 100\msun, and adopt the updated \cite{bruzual_stellar_2003} stellar population models described in \cite{vidal_garcia_2017}.
We assume the total mass currently locked into stars as our fiducial estimate for the stellar mass, which accounts for the fraction of mass returned to the ISM, instead of the integrated SFH. 
We refer to Chevallard et al. (in preparation) for further information of the \textsc{beagle} fitting procedure.

The star-formation rates of our galaxies are estimated in two different ways, i.e., from the output of the \textsc{beagle} SED fitting run described above, which provides an estimate of the star-formation rate averaged over the past ten Myr, and from the attenuation-corrected\footnote{as described in Section~\ref{sec:metallicity} we exploited the measured Balmer decrements with a \citet{gordon_LMC_attenuation_2003} law where available, whereas the output Av from \textsc{beagle} otherwise} \Ha\ luminosity, following the recipe\footnote{assuming a conversion factor of 10$^{-41.67}$ (M$_{\odot}$yr$^{-1}$)/(erg s$^{-1}$)} outlined in \cite{reddy_lyA_2022,shapley_balmer_2023}, better suited for low-metallicity galaxies characterised by an increased rate of ionising photons.
For galaxies in which \Ha\ is not observed, either because it is shifted out of the spectral coverage of NIRSpec (i.e., at z$\gtrsim$7), or because it falls in one of the detector gaps (in medium resolution gratings), we exploited the dust-corrected \Hb\ flux re-scaled by the theoretical case B recombination factor at T$\sim 1.5\times10^4$K (2.86). 
For consistency, for any given galaxy we use the \Ha(\Hb) flux as measured from either the PRISM or medium resolution gratings spectra according to the configuration exploited for the full metallicity calculation 
(as described in Section~\ref{sec:metallicity}). 

On average, the \Ha-based SFRs are $\sim$0.14~dex lower than those derived by \textsc{beagle}, with a dispersion in the deviation between the two measurements of $\sim$0.3~dex.
Such discrepancy is possibly driven by 
i) the different underlying stellar population synthesis models (i.e., BPASS for \citealt{reddy_lyA_2022}, revised \citealt{bruzual_stellar_2003} for \textsc{beagle}), ii) different prescriptions for the star-formation history, iii) residual contamination of the \Ha\ flux from \NII\ in PRISM spectra. 
We adopt the SED-based measurements as our fiducial value of the total SFR of a galaxy for the analysis presented in the current paper, more consistent with the stellar mass derivation.
Nonetheless, we verified that assuming either of the two SFR values does not alter any of the conclusions presented in this work, particularly when discussing the evolution in the fundamental metallicity relation (Section~\ref{sec:fmr_evol}), as shown in Appendix~\ref{sec:appendix_A}.

\subsection{High redshift galaxy samples from the literature}

The specific selection function of the \emph{deep} tier of the JADES spectroscopic campaign provides preferential coverage of the low-end of the stellar mass distribution (i.e., M$_{\star}$/M$_{\odot} \approx 10^{6.5}$-$10^{9}$), a regime poorly probed before, especially at such early cosmic epochs.
In order to extend the parameter space over which the metallicity scaling relations are analysed, here we complement our JADES sample with a sample of 80, $z=4-10$ galaxies from the CEERS programme as analysed and presented by \cite{nakajima_mzr_ceers_2023}, which are preferentially distributed across the log(\mstar/\msun) $\sim 8.5-10$ regime.

To preserve a good level of self-consistency in the analysis, we have re-computed the metallicity for these galaxies starting from the emission line ratios reported in Table B1 of \cite{nakajima_mzr_ceers_2023}, applying the procedure discussed in Section~\ref{sec:metallicity}.
Star-formation rates reported in \cite{nakajima_mzr_ceers_2023} are based on dust-corrected \Hb\ luminosity and the calibration reported by \cite{kennicutt_star_2012}. 
We scale down these values by the median offset (i.e., $0.23$~dex) between the \Ha-based SFR (this time derived assuming the \cite{kennicutt_star_2012} formula) and the \textsc{beagle}-SFR as measured for the JADES galaxies, in order to minimise systematic offsets between the SFRs compiled for CEERS galaxies and the fiducial values assumed for JADES objects.

Stellar masses in the CEERS sample are based on fits to NIRCAM photometry performed with the \textsc{prospector} code \citep{Johnson_prospector_2021}. 
A full assessment of the systematics involved in the stellar mass determination as based on different spectral fitting codes goes beyond the scope of this paper. 
Nonetheless, we note that the statistical uncertainties on \mstar quoted by \cite{nakajima_mzr_ceers_2023} are much larger than those associated with \textsc{beagle} measurements for JADES galaxies.
To reduce the impact of such differential measurement uncertainties, we do not apply inverse variance weighting based on \mstar uncertainties in any of our fitting analyses in the paper.

In addition, we include in the analysis four targets whose metallicity have been derived with the `direct' \Te-method, namely three galaxies observed in the framework of the Early Release Observations programme as described in \cite{curti_smacs_2023}, and the galaxy GN-z11 observed in the framework of JADES and presented in \cite{bunker_gnz11_2023,tacchella_gnz11_2023}, though the nature of its ionising source is debated and the presence of an AGN has been proposed \citep{maiolino_gnz11_2023}.
This brings the total number of galaxies analysed in this paper to 146.

For the three ERO obejcts, \mstar and SFR are derived via \textsc{beagle} fitting to NIRCAM photometry as described in \cite{curti_smacs_2023}, whereas the metallicities have been re-computed following the same data reduction implemented for the JADES galaxies analysed in this paper, as also described in \cite{laseter_auroral_jades_2023}.
For GN-z11, fiducial values for \mstar and SFR are taken from \cite{bunker_gnz11_2023} and are based on \textsc{beagle} fitting to PRISM spectrum assuming a \cite{chabrier_galactic_2003} IMF with an upper mass cut-off of $100$\msun, whereas the oxygen abundance is measured exploiting the detection of \OIIIl4363 in the PRISM spectrum and its ratio over \NeIIIl3869 and \Hy, as detailed in \cite{cameron_gnz11_2023}, which quote a fiducial value of 12+log(O/H) $=7.82$.
For comparison, we also obtain an indirect estimate of the \OIIIl5007 flux (which is not covered in the spectra of GN-z11) based on the observed \NeIII/\OII\ (from medium resolution gratings) and the \cite{witstok_lensed_z5_2021} \NeIII/\OII\ versus \OIII/\OII\ calibration, which we then use to derive the \^R diagnostic and use it conjunction with Ne3O2 and the methodology described in Section~\ref{sec:metallicity}: this approach provides 12+log(O/H) $=7.67\substack{+0.16 \\ -0.10}$.
At the same time, using the same indirect estimate of the \OIIIl5007 flux to apply the \Te method (and adopting \citealt{pilyugin_electron_2009} to derive the temperature of the low-ionisation zone) delivers a much higher 12+log(O/H)$=8.42$.
Given the large systematic uncertainties affecting both methods, we here assume the value reported by \cite{cameron_gnz11_2023} (7.82) as our fiducial value, with a conservative uncertainty of $0.35$~dex. 

In Figure~\ref{fig:m-sfr_hist} we show the distribution of both the JADES sample introduced in this work, and the compiled sample from the literature discussed above, in the stellar mass versus star-formation rate diagram; the marginalised histograms of the distribution of the two quantities are also shown in the upper and right-hand inset panels.
We note the two samples are quite complementary, and especially including galaxies from \cite{nakajima_mzr_ceers_2023} allows us to extend the mass regime probed up to \mstar $\sim 10^{9.5}$ \msun\ ($\langle$log(\mstar/\msun)$\rangle_{\text{CEERS}}$ = 8.72; $\langle$log(\mstar/\msun)$\rangle_{\text{JADES}}$ = 7.83).
Galaxies selected from CEERS are also distributed towards higher SFRs ($\langle \text{log(SFR)}\rangle_{\text{CEERS}}$ = 1.3; $\langle \text{log(SFR)}\rangle_{\text{JADES}}$ = 0.45), though the two galaxy samples probe similar specific star-formation rates (sSFR=SFR/\mstar).
For comparison, we plot the parametrisation of the `Main Sequence of Star Formation' \citep[SFMS,][]{noeske_star_2007, speagle_highly_2014,renzini_objective_2015} at $z\sim0$ and $z\sim6$ from \cite{popesso_SFMS_2023}, noting that the vast majority of our combined, low-mass \emph{JWST} galaxy sample is offset above the low-mass extrapolation of the relation at $z\sim6$, given our fiducial \mstar and SFR measurements.


\section{The evolution of the mass-metallicity relation beyond \lowercase{z}=3}
\label{sec:scaling_relations_mzr}

\begin{figure*}
    \centering
    \includegraphics[width=0.85\textwidth]{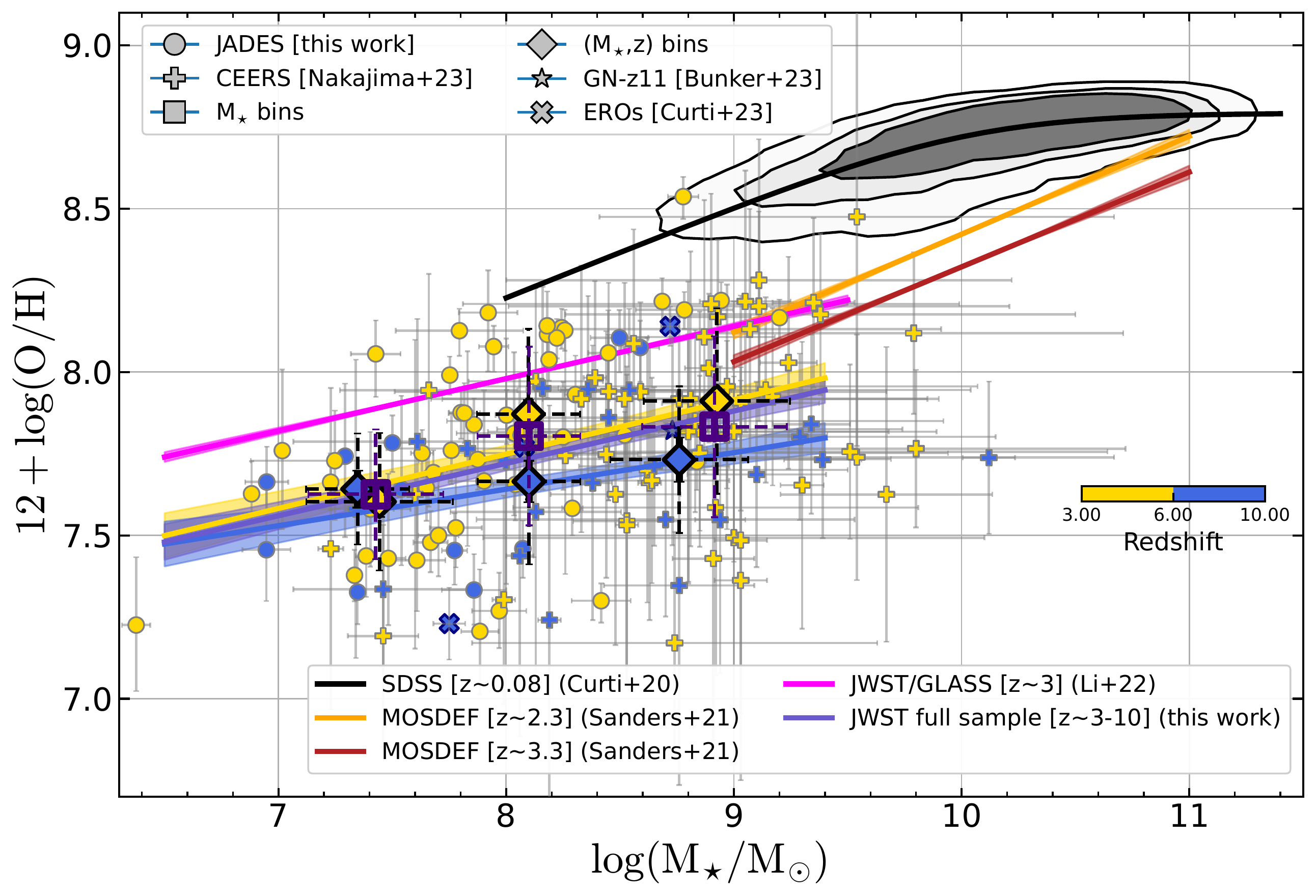}

  
    \caption{The mass-metallicity relation (MZR) for our full \emph{JWST} sample. Filled circles represent individual galaxies presented in this work from JADES, whereas filled crosses are galaxies observed in the framework of the CEERS programme and compiled from \citealt{nakajima_mzr_ceers_2023}, with metallicity derived as detailed in Section~\ref{sec:metallicity}. The star symbol reports the JADES/NIRSpec observations of GN-z11 (\citealt[at z=10.603]{bunker_gnz11_2023}), whereas 'X' symbols mark galaxies from the EROs as compiled from \citealt{curti_smacs_2023} and \citealt{laseter_auroral_jades_2023}.
    Large squared and diamond symbols mark the median values computed in bins of \mstar (full sample, in purple), and (\mstar\,z), as described in Table~\ref{tab:binned_values}. 
    An orthogonal linear regression fit to the median values in bins of \mstar\ for the different redshift sub-samples is shown by the purple (full sample), yellow ($z=3-6$) and blue ($z=6-10$) lines, respectively.
    We include a comparison with previous determinations of the MZR at lower redshifts from \citealt{curti_massmetallicity_2020} (SDSS at $z\sim 0.07$), and \citealt{sanders_mosdef_mzr_2021} (MOSDEF at $z\sim 2-3$), as well as the best-fit of the low-mass end of the MZR at $z\sim 3$ provided by \citealt{Li_mzr_dwarfs_z3_2022} and based on JWST/NIRISS slitless spectroscopy. The MZR curves at $z\sim2-3$ have been scaled down by $\sim0.1$~dex to account for the systematics differences between the metallicity calibrations used in this work and the \citealt{bian_ldquodirectrdquo_2018} calibrations adopted in the original papers.
    }
    \label{fig:mzr}
\end{figure*}

\begin{figure*}
     \includegraphics[width=0.48\textwidth]{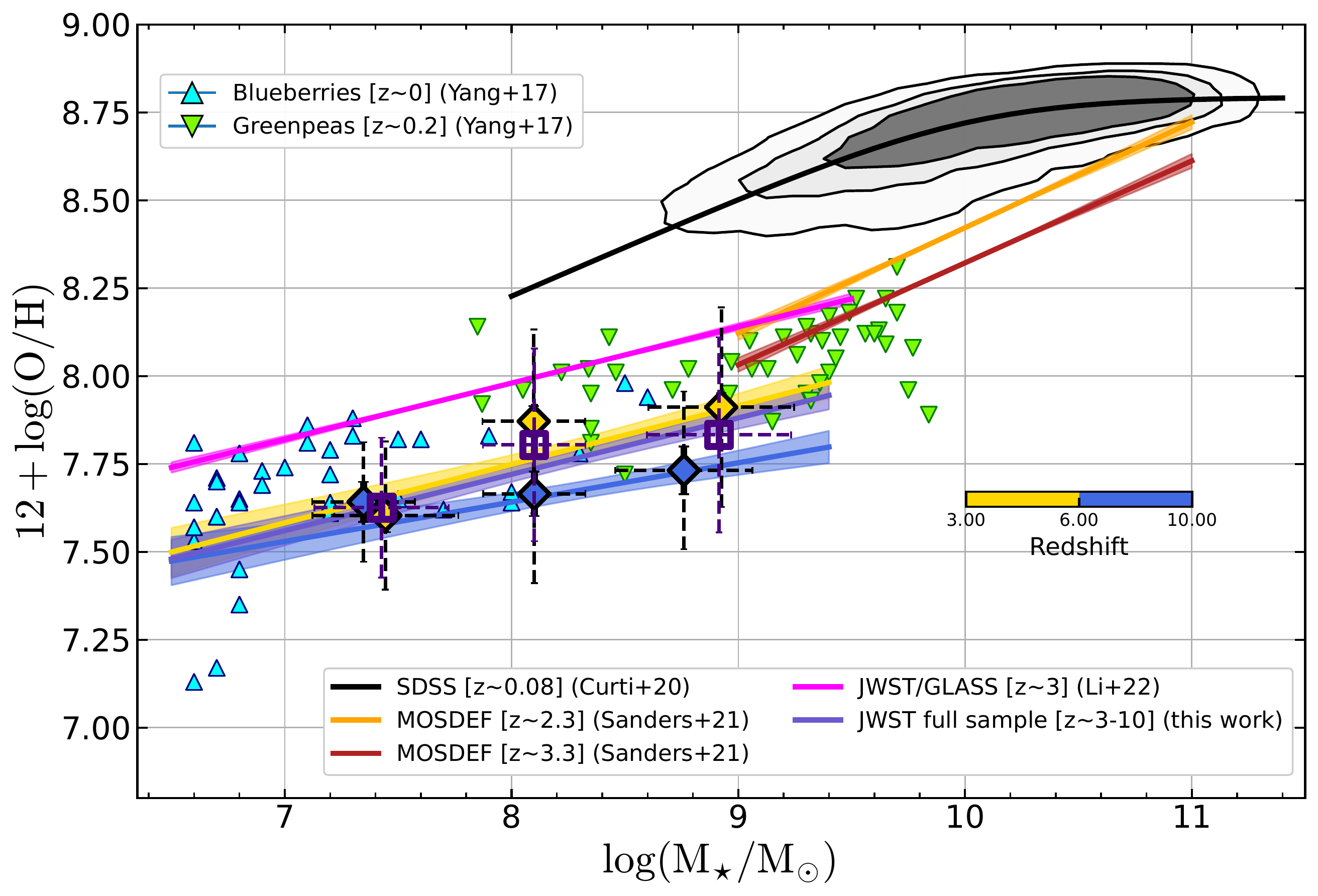}   
    \hspace{0.1cm}
    \includegraphics[width=0.48\textwidth]{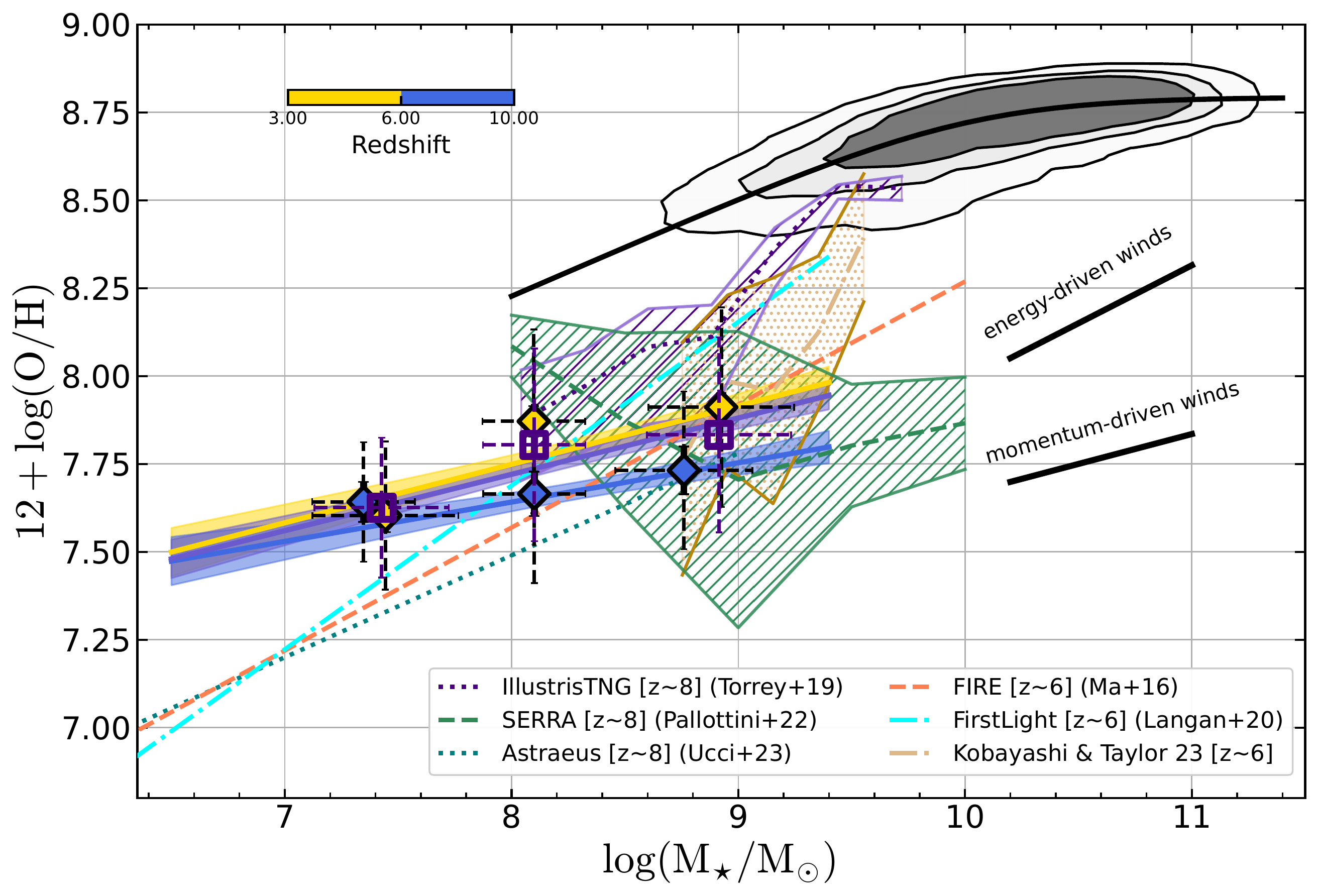}   

    \caption{\textit{Left-hand panel:} The best-fit MZR presented in this work (for both the full \emph{JWST} sample in purple, and in the $z=3-6$ and $z=6-10$ redshift binsin yellow and blue, respectively) is compared to the \Te-based mass-metallicity relation from local, metal-poor `Blueberry' and `Green Pea' galaxies (\citealt{yang_blueberries_2017,yang_greenpeas_2017}), which share similar excitation and emission line properties to our high-redshift galaxy sample (\citealt{cameron_jades_bpt_2023}). Their distribution agrees well in slope and normalisation (with $\sim 0.11$~dex offset in log(O/H)) with the average relation inferred in this work for our low-mass \emph{JWST} sample.
    \textit{Right-hand panel:} Our best-fit MZRs are compared with the predictions from different suites of cosmological simulations at z$\sim$6-8, namely FIRE (\citealt{ma_origin_2016}), IllustrisTNG (\citealt{torrey_mzr_TNG_2019}), FirstLight (\citealt{langan_theo_mzr_2020}), Astraeus (\citealt{ucci_astraeus_2023}), SERRA (\citealt{pallottini_serra_2022}), and with the chemodynamical simulations from \citealt{kobayashi_taylor_chemodyn_2023}. In addition, the typical MZR slopes predicted by (semi-) analytical chemical evolution models (e.g., \citealt{dave_mzr_model_2012}) implementing feedback via either `energy-driven' or `momentum-driven' winds are also shown (the normalisation is assumed arbitrary to aid visualisation). The latter agrees well with the slope observed at low stellar mass for galaxies at $z\geq 3$. }
    \label{fig:mzr_simul}

\end{figure*}

\begin{table*}
    \centering
    \caption{Median values, error on the median (and standard deviation) in stellar mass, SFR and metallicity for our sample of galaxies in each of the (\mstar-redshift) bins considered in this work. The number of galaxies in each bin and their median redshift is also reported.}

    \begin{tabular}{@{}lccccc@{}}
    \hline
    [log \mstar]$_{\text{bin}}$ & N$_{\text{gal}}$ & <z> & log \mstar/\msun & log SFR [\msun\ yr$^{-1}$] & 12+log(O/H)   \\
    \hline
    \multicolumn{6}{|c|}{Redshift z $\in$ [3,10]} \\
    $[6\ ;\ 7.75]$ & 29 & 5.80 & 7.43 $\pm$ 0.06 \ (0.32) & 0.22 $\pm$ 0.11 \ (0.46) & 7.63 $\pm$ 0.04 \ (0.20) \\
    $[7.75\ ;\ 8.5]$ & 52 & 5.18 & 8.10 $\pm$ 0.03 \ (0.23) & 0.48 $\pm$ 0.06 \ (0.39) & 7.80 $\pm$ 0.04 \ (0.27) \\
    $[8.5\ ;\ 10]$ & 64 & 5.00 & 8.92  $\pm$ 0.04 \ (0.32) & 1.16 $\pm$ 0.07 \ (0.53) & 7.83 $\pm$ 0.03 \ (0.28) \\
    \hline
    \multicolumn{6}{|c|}{Redshift z $\in$ [3,6]} \\
    $[6\ ;\ 7.75]$ & 20 & 5.25 & 7.44 $\pm$ 0.08 \ (0.36) & 0.15 $\pm$ 0.12 \ (0.51) & 7.60 $\pm$ 0.05 \ (0.36) \\
    $[7.75\ ;\ 8.5]$ & 36 & 4.41 & 8.10 $\pm$ 0.04 \ (0.24) & 0.41 $\pm$ 0.06 \ (0.35) & 7.87 $\pm$ 0.04 \ (0.24) \\
    $[8.5\ ;\ 10]$ & 52 & 4.78 & 8.93  $\pm$ 0.04 \ (0.33) & 1.16 $\pm$ 0.0 \ (0.56) & 7.91 $\pm$ 0.04 \ (0.33) \\
   \hline
    \multicolumn{6}{|c|}{Redshift z $\in$ [6,10]} \\
    $[6\ ;\ 7.75]$ & 10 & 6.71 & 7.35 $\pm$ 0.08 \ (0.24) & 0.34 $\pm$ 0.09 \ (0.30) & 7.64 $\pm$ 0.07 \ (0.24) \\
     $[7.75\ ;\ 8.5]$ & 15 & 6.54 & 8.10 $\pm$ 0.06 \ (0.22) & 0.63 $\pm$ 0.10 \ (0.40) & 7.67 $\pm$ 0.06 \ (0.22) \\
    $[8.5\ ;\ 10]$ & 11 & 6.93 & 8.76 $\pm$ 0.09 \ (0.30) & 1.28 $\pm$ 0.10 \ (0.32) & 7.73 $\pm$ 0.08 \ (0.30) \\
    \hline

    \end{tabular}
    \label{tab:binned_values}
\end{table*}

\begin{table}
    \centering
    \caption{Best-fit parameters of the mass-metallicity relation from equation~\ref{eq:mzr} for the full sample, and the two redshift bins at $z=3-6$ and $z=6-10$, for which we report both the slope $\beta_{\text{z}}$ and the normalisation at \mstar=10$^8$\msun, Z$_{\text{m8}}$.}
    \begin{tabular}{@{}lcc@{}}
    \hline
    Sample & $\beta_{\text{z}}$ & Z$_{\text{m8}}$   \\
    \hline
    z $\in$ [3,10] & 0.17\ $\pm$\ 0.03 & 7.72\ $\pm$\ 0.02 \\ 
    z $\in$ [3,6] & 0.18\ $\pm$\ 0.03 & 7.75\ $\pm$\ 0.03 \\ 
    z $\in$ [6,10] & 0.11\ $\pm$\ 0.05 & 7.65\ $\pm$\ 0.04 \\ 
    \hline
    \end{tabular}
    \label{tab:best-fit_mzr}
\end{table}


In Figure~\ref{fig:mzr}, we show our combined \emph{JWST} galaxy sample on the mass-metallicity plane.
Each point is marked with a different symbol according to its parent sample (as reported in the legend), and it is colour-coded on the basis of the following redshift binning scheme: yellow at $z\in [3-6]$, and blue at $z\in [6-10]$. 
The median redshift of the whole sample is $\langle z \rangle=5.10$, whereas the median redshift of the $z\in 3-6$ sub-sample is $\langle z \rangle_{3-6}=4.76$ and that of the $z\in 6-10$ sub-sample is $\langle z \rangle_{6-10}=6.73$.
In particular, filled circle points in the plot report galaxies from the JADES survey, while filled crosses symbols mark galaxies from CEERS \citep{nakajima_mzr_ceers_2023}, whose metallicity has been re-measured as described in Section~\ref{sec:metallicity}.
For both the total sample and each sub-sample in redshift we show the median (large diamond markers), error on the median (solid errorbar), and standard deviation (dashed errorbar) of the metallicity within three different bins in stellar mass.
The size of the stellar mass bins is not uniform to maintain a reasonable (i.e., at least 10) number of galaxies in each bin.
The average properties of the binned samples are reported in Table~\ref{tab:binned_values}.

Our galaxies present a median offset of $\sim -0.5$~dex ($-0.48$~dex and $-0.64$~dex at z=3-6 and z=6-10, respectively) compared to the low-mass end (and its extrapolation) of the MZR in the local Universe \citep[based on individual SDSS galaxies, ][]{curti_massmetallicity_2020}. 
We also compare our results with previous realisations of the MZR at $z\sim2-3$.
In particular, we first compare our observations with the mass-metallicity relation derived from stacked spectra of galaxies from the MOSDEF Survey as reported by \cite{sanders_mosdef_mzr_2021}.
To minimise the impact of systematics introduced by the use of different metallicity calibrations (the calibrations from \cite{bian_ldquodirectrdquo_2018} are adopted in \citealt{sanders_mosdef_mzr_2021}), we have re-computed the metallicity for the MOSDEF stacked spectra with the same methodology outlined in Section~\ref{sec:metallicity}, finding an average offset of $\sim0.088$~dex towards lower metallicities. Therefore, in Figure~\ref{fig:mzr} the mass-metallicity relations at $z\sim2.2$ and $z\sim3.3$ from \citealt{sanders_mosdef_mzr_2021} are lowered by $0.088$~dex compared to their original parametrisation.
Overall, the evolution in normalisation probed by our full galaxy sample at $z=3-10$ appears relatively mild if compared to the extrapolation at low stellar mass of the z$\sim$3.3 MZR from \cite{sanders_mosdef_mzr_2021}, with a mean offset for the full sample of $0.045$~dex ($0.13$~dex if we assume the fiducial MZR from \citealt{sanders_mosdef_mzr_2021} based on the \citealt{bian_ldquodirectrdquo_2018} calibrations).
However, such deviation is mass-dependent and is more prominent at higher \mstar, being almost zero for log(\mstar/\msun)$\lesssim 8$, while $\sim0.15$~dex at log(\mstar/\msun)$\sim 9$.
This behaviour suggests an evolution in the slope of the mass-metallicity relation in the mass and redshift regimes probed by this work.


In the attempt to characterise the evolution in the slope of the MZR at these redshifts, we perform an orthogonal linear regression fit to both the full \emph{JWST} sample of individual galaxies, and within the two redshift bins separately, in the functional form 
\begin{equation}
\text{12+log(O/H)}=\beta_\text{z} \  \text{log}(\text{M}_{\star}/10^8\text{M}_{\odot}) + \text{Z}_\text{m8} \, 
\label{eq:mzr}
\end{equation}
where $\beta_\text{z}$ is the slope at a given redshift interval, and $\text{Z}_\text{m8}$ is the normalisation at log(\mstar/\msun)=8.
The results are displayed in Figure~\ref{fig:mzr}, as shown by the solid purple (full sample), yellow ($z=3-6$), and blue ($z=6-10$) lines, whereas the shaded regions represent the 1-$\sigma$ confidence interval of each fit, derived from bootstrapping each sample (with replacement) and repeating the fitting procedure 300 times. 
The best-fit parameters are reported in Table~\ref{tab:best-fit_mzr}.
We note that the best-fit MZR (based on individual data points) agrees well with the median values computed in bins of \mstar for each redshift sub-sample.

We find indications of a flattening of the slope of the low-mass end of the MZR with redshift, with the best-fit slope for the full sample equal to $\beta_{\text{3-10}}=0.17\pm0.03$. 
If we focus more specifically on the two redshift bins described above, we find $\beta_{\text{3-6}}=0.18\pm0.03$, and $\beta_{\text{6-10}}=0.11\pm0.05$.
The low-mass end slope of the MZR as probed by our sample is flatter (at $3.6 \sigma$ significance) than both observed for the z$\sim$2-3 MZR at higher stellar masses ($\beta_{\text{2-3}}$=0.29), and also in the local Universe (0.28, as measured in \citealt{curti_massmetallicity_2020}). 
Such apparent invariance between $z\sim 0$ and $z\sim 3$ suggests that the same physical processes, and in particular how the metal removal efficiency of stellar winds scales with \mstar, governs the MZR slope over the last $\sim12$Gyr of cosmic time \citep{sanders_mosdef_mzr_2021}. 
We note however that most previous assessements of the MZR at high-z are based on near-infrared spectroscopic surveys from the ground, and were hence limited in the stellar mass range they could probe, struggling to access the regime below log(\mstar)$<10^9$\msun\ and thus to provide a sample matched in stellar mass to the \emph{JWST} sample discussed in the present work.
Further insights on the low-mass end slope of the MZR at $z\sim 2-3$ come from recent JWST/NIRISS observations within the GLASS-ERS programme, have been presented in \cite{Li_mzr_dwarfs_z3_2022}, and report similar evidence for a shallower slope ($\beta_{\text{2-3}}=0.17 \pm 0.03$, as shown in magenta in Figure~\ref{fig:mzr}) compared to the simple extrapolation of the $z\sim3.3$ relation inferred by \citep{sanders_mosdef_mzr_2021}.
In this work, we push the investigation of the MZR in the dwarf regime to much higher redshift, finding a slope at z$_{\text{3-6}}$ which is consistent with that reported by \cite{Li_mzr_dwarfs_z3_2022}.
Compared to that parametrisation of the MZR, we observe a more prominent evolution in the normalisation of $\sim0.25$~dex, after once again correcting for the systematics between the \cite{bian_ldquodirectrdquo_2018} calibrations and the set of calibrations adopted in this work.

Finally, it is interesting to note that the slope inferred at $z=6-10$ (in blue in Figure~\ref{fig:mzr}) is even flatter (and consistent with zero within $\sim 2\sigma$), suggesting that the mass-metallicity relation could be in an initial build-up phase. 
However, the high-redshift bins are only sparsely populated and probably subject to strong selection biases, hence it is difficult to draw any strong conclusion at this stage relative to any possible evolution in the MZR slope between $z=3-6$ and $z>6$.

\subsection{Comparison with local analogues of high redshift galaxies}

The best-fit relations (and median values in bins of \mstar\ and redshift) reported in this work show remarkable agreement with the MZR probed by samples of local star-forming galaxies with extreme line emission properties, e.g., the `Green Pea' and `Blueberry' galaxies compiled from \cite{yang_greenpeas_2017, yang_blueberries_2017}, whose metallicity has been derived with the ``direct'' \Te method; these objects are marked by green and cyan symbols, respectively, in the left panel of Fig.~\ref{fig:mzr_simul}.
In particular, the MZR slope inferred from the combined sample of `Green Pea' and `Blueberry' galaxies is $0.18 \pm 0.02$, fully consistent with the slope measured from our sample, whereas \emph{JWST} galaxies have, on average, $0.12$~dex lower metallicity.
These low-redshift sources are well matched in stellar mass with our \emph{JWST} sample (log(\mstar/\msun)$\approx$ 6.5-9.5), are very compact, characterised by very high equivalent width of the emission lines, low metallicities, and high ionisation parameters, and therefore have been long identified as potential analogues of very high-redshift sources. The good agreement between the emission line and metallicity properties in our $z\sim3-10$ \emph{JWST} sample and those observed in this sample of local, extreme line emitters corroborates such interpretation \citep[see also][]{cameron_jades_bpt_2023}.

\subsection{Comparison with models and simulations}
We here compare the observed MZR in our $z=3-10$ galaxy sample with the predictions of different cosmological simulations, as depicted in the right-hand panel of Figure~\ref{fig:mzr_simul}. 
In particular, we report the mass-metallicity relationship at $z \sim 6$ (close to the median redshift of our full sample, i.e, 5.86) from FIRE \citep{ma_origin_2016}, IllustrisTNG (TNG100) \citep{torrey_mzr_TNG_2019}, FirstLight \citep{langan_theo_mzr_2020}, and from the chemo-dynamical simulations by \cite{kobayashi_taylor_chemodyn_2023}, as well as predictions at higher redshift ($z \sim8$) from Astraeus \citep{ucci_astraeus_2023}, and Serra \citep{pallottini_serra_2022}.
For IllustrisTNG simulations 
we followed the approach described in \cite{torrey_mzr_TNG_2019} and already followed in \cite{curti_smacs_2023}, considering central galaxies and assuming that oxygen comprises 35 per cent of the SFR-weighted metal mass fraction within twice the stellar half-mass radius. 
Within FIRE simulations the gas-phase metallicity is defined as the mass-weighted metallicity of all gas particles that belong to the ISM, assuming solar abundance ratios \citep{asplund_solar_2009}.
We note that the FIRE results at $z=6$ are re-normalised to match the normalisation of the MZR in the local Universe.
Results from the FirstLight simulations at $z=6$ are compiled from
\cite[][, and Langan, private communication]{langan_theo_mzr_2020}, and based on the assumption that the unresolved nebular region around each star particle shares the same mass ratio of metals produced in Type II SN explosions as in the star particle. The galaxy metallicity is defined as the mass-weighted average nebular metallicity including all star particles younger than 100 Myr.
The predictions from chemo-dynamical simulations by \cite{kobayashi_taylor_chemodyn_2023} are instead based on the SFR-weighted gas-phase metallicity.
Finally, the Astraeus simulations \citep{ucci_astraeus_2023} accounts for the oxygen mass in the halo without any weighting,
whereas in SERRA gas and stellar metallicity are coupled and the metallicity is tracked as the sum of all heavy elements, \citep{pallottini_serra_2022}.

Overall, most simulations appear to reasonably match the normalisation of the observed MZR at \mstar$\approx 10^{8-9}$, though under-predicting the abundances observed at log(\mstar/\msun)<8, with the theoretical MZRs characterised by steeper slopes than inferred from our sample.  
Only the extrapolation of IllustrisTNG predictions to lower \mstar\ seems to match reasonably well (in normalisation) the average distribution of galaxies in the mass-metallicity plane below \mstar$=10^8$\msun, but diverges more dramatically at higher masses. 
Interestingly, the lack of a clear correlation (and the large scatter) at $z\sim8$ predicted by SERRA seems in line with our observations in  the highest redshift bin ($z>6$).

We also compare our inferred MZR slopes with those predicted by (semi-)analytical chemical evolution models under the assumption of different modes of stellar feedback.
In particular, in the framework of `equilibrium' models \citep[e.g.,][]{dave_mzr_model_2012, lilly_gas_2013}, an `energy-driven winds' scenario predict a slope for the MZR of $\beta_{\text{en}} \sim0.33$, whereas for `momentum-driven winds' the predicted slope is shallower, i.e. $\beta_{\text{mom}} \sim0.17$ \citep{guo_stellar_2016}; these two cases are shown in the right-hand panel of Figure~\ref{fig:mzr_simul} with thick black lines (here the normalisation is assumed arbitrary, and the two curves are artificially shifted along the x-axis to aid the visualisation and comparison of the various trends).
We observe the former to be in good agreement with the MZR slope probed by galaxies at intermediate and high masses (log(\mstar/\msun)$\geq 9$) and up to $z\sim3$, whereas the latter is more consistent with the observed MZR slope as probed by our \emph{JWST} sample at the low-mass end and at higher redshift.
Although preliminary, these results are consistent with a scenario in which the dominant feedback mechanism in galaxies changes in different mass regimes, producing a shallower MZR slope at low \mstar. 
At the same time, observing a similar MZR slope at \mstar$\sim 10^{7}-10^{9}$ between $z\sim3$ (from \citealt{Li_mzr_dwarfs_z3_2022}) and $z\sim6$ (from this work) suggests that the scaling of the mass-loading factor of outflows with stellar mass is invariant in this mass regime during this epoch.



Another critical parameter which contributes to setting the scaling relation between \mstar\ and metallicity is the gas fraction. 
The average gas fraction in galaxies is observed to evolve with redshift at fixed \mstar, as inferred from large existing datasets up to $z\sim2-3$ \citep[e.g.,][]{saintonge_2016, scoville_2017,tacconi_phibss_2018,tacconi_review_2020}.
If the evolution proceeds at a similar rate even at earlier epochs, the impact of metal dilution from larger gas reservoirs is expected to become more important as redshift increases.
However, this seems to be in tension with the relatively mild evolution in the MZR normalisation observed in our \emph{JWST} sample.

In the framework of analytical chemical evolution models that allow the gas fraction to vary \citep[see e.g.,][]{peeples_mzr_model_2011, lilly_gas_2013, sanders_mosdef_mzr_2021}, and under the assumption of (nearly-)pristine gas inflows, the metallicity of the ISM is the result of the balance between the nucleosynthetic stellar yield, the gas dilution, and the amount of metals lost due to outflows.
If we assume that the same scaling of the gas fraction $\mu_{\text{gas}}$ (here defined as $\mu_{\text{gas}} = \text{M}_{\text{gas}}/$\mstar) with redshift as inferred up to $z\sim3$ persists up to $z\sim6$ (e.g., log($\mu_{\text{gas}}$) $\propto 2.49\ $log(1+z) as based on the linear parametrisation of \citealt{tacconi_phibss_2018}), and fixing all the other parameters such as stellar yield, return fraction, and mass-loading factor of outflows, adopting the formalism of the \cite{peeples_mzr_model_2011} model (their equations (9) and (10)) one can predict the evolution in the metallicity at fixed \mstar\ in the low-mass regime (e.g., at \mstar $\sim$10$^8$\msun) as solely driven by the evolution in the gas fraction with redshift.

We perform these calculations for two different scenarios.
First, we assume the low-mass extrapolation of the MZR at $z\sim3.3$ from \cite{sanders_mosdef_mzr_2021} (with a slope of 0.30) to solve for the mass-loading factor of outflows $\zeta_{w}$ at \mstar $=10^{8}$\msun; then, assuming a constant $\zeta_{w}$ with redshift between $z=3$ and $z=6$, we derive the expected metallicity at \mstar $=10^{8}$\msun\ by evolving the gas fraction $\mu_{\text{gas}}$ with redshift according to the scaling relations of \cite{tacconi_phibss_2018}. From this, we predict an evolution in metallicity of $0.26$~dex, larger than the almost negligible evolution probed by the extrapolation of the \cite{sanders_mosdef_mzr_2021} MZR at $z\sim3.3$.
In the second case, we instead solve for $\zeta_{w}$ in order to match slope (0.17) and normalisation (12+log(O/H)=7.98 at \mstar $=10^{8}$\msun) of the MZR at $z=3$ in the dwarf regime from \cite{Li_mzr_dwarfs_z3_2022}; fixing $\zeta_{w}$ with redshift, we predict a decrease in log(O/H) as driven by the evolution in the gas fraction of $0.35$~dex, which is in slightly better agreement but still larger than observed in this mass regime (i.e, $0.25$~dex) between $z=3$ and $z=6$.
In both cases, we adopt the oxygen yield and return fraction provided by \cite{vincenzo_modern_2016} for a \cite{chabrier_galactic_2003} IMF with an upper cutoff of $100$\msun, i.e. $y=0.030$ and $R=0.30$, respectively.
Nonetheless, we note that in such analytical chemical evolution models the predicted normalisation of the MZR is highly sensitive to the large uncertainties on the assumed of oxygen stellar yield and the shape of the upper mass cutoff of the IMF \citep{vincenzo_modern_2016}, the relative contribution of atomic and molecular gas to the total gas mass at high redshift, which is generally estimated from relatively high-mass galaxy samples \citep{saintonge_xcoldgass_2017,catinella_xgass_2018}, the relative importance of star-formation efficiency \citep{baker_sfms_2023}, and the fraction of metals expelled from outflows, which may then be re-accreted onto galaxies via galactic fountains from the enriched CGM \citep{fraternali_accretion_2008,peroux_2020}, whereas on the other hand systematics in the adopted metallicity calibrations can shift the normalisation of the `observed' MZR by non-negligible amounts.


Finally, we note that the distribution of our data in the MZR plane is characterised by a large amount of scatter, at any given mass and redshift.
We note the sample size considered in the present work not being large enough (as well as being affected by selection biases) to perform a robust assessment of the scatter of the scaling relation, however we can tentatively measure the amount of intrinsic scatter in our galaxy sample as $\sigma_{\text{MZR}} = \sqrt{\sigma^{2}_{\text{obs}} - \sigma^{2}_{\text{meas}}}$, where $\sigma_{\text{obs}}$ is the standard deviation of the full \emph{JWST} sample around the best-fit relation and $\sigma_{\text{meas}}$ is the average measurement uncertainty associated with the metallicity determination\footnote{the dispersion of the adopted strong-line calibrations are included directly at the metallicity derivation step}.
We find $\sigma_{\text{MZR}} \sim0.073$~dex for our full sample, consistent with the scatter measured in the local MZR at higher masses (log(\mstar/\msun)>9.5) \citep[$\sigma_{\text{MZR},z=0}$ = 0.075][]{curti_massmetallicity_2020}. 
We note that the scatter at high masses in local galaxies is likely dominated by the flattening of the MZR at high masses (log(\mstar/\msun)$\geq$10), interpreted in classical gas-regulator models \citep{finlator_dave_2008, lilly_gas_2013} as a consequence of the metallicity approaching the stellar yield, whereas no clear `turnover' nor high-mass flattening is probed at higher redshifts.


\begin{figure*}
    \centering
    \includegraphics[width=0.85\linewidth]{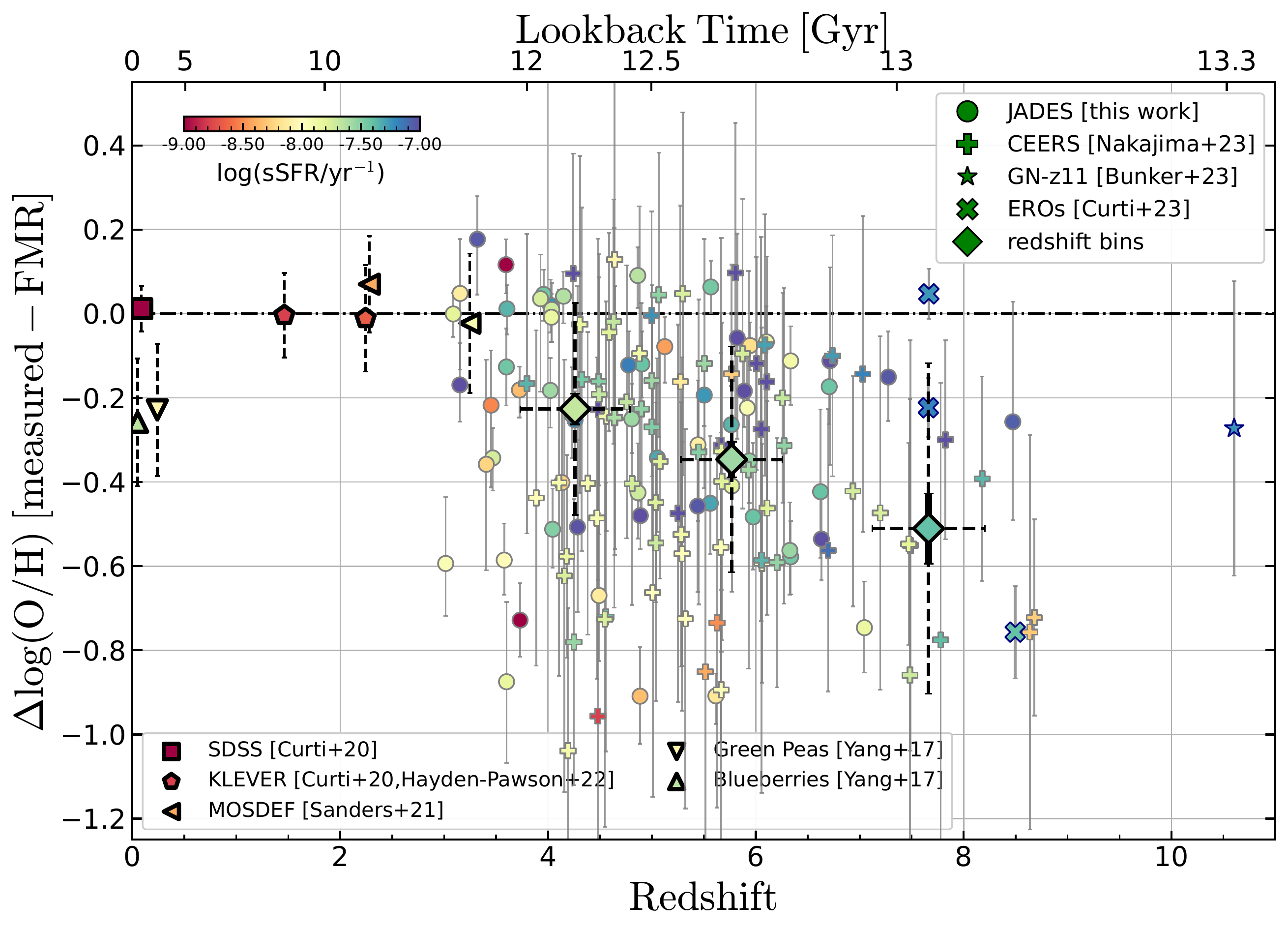}   

    \caption{Deviations of the \emph{JWST} sample from the predictions of the local Fundamental Metallicity Relation (FMR), here plotted as a function of redshift.
    Marker symbols are as in Figure~\ref{fig:mzr}, with each galaxy colour-coded by its specific star-formation rate (sSFR). The weighted average of the deviation in three redshift bins (colour-coded by the mean sSFR of the binned sample) for both the \emph{JWST} sample and for galaxy samples at lower redshift (SDSS at $z\sim 0$, KLEVER at $z\sim 1.5-2.5$, MOSDEF at $z\sim 3$) are marked by larger symbols. 
    Dashed error bars report the dispersion in $\Delta$log(O/H) of each binned sample, whereas thick, solid bars report the standard error on the mean (which is often smaller than the marker's size).
    Overall, $z>3$ galaxies observed by \emph{JWST} are scattered across the local FMR predictions, but show evidence of being preferentially offset towards lower metallicity values than expected with increasing redshift, with an average offset of $-0.4$~dex at $z>6$.
     }
 
    \label{fig:fmr}
\end{figure*}

\begin{figure*}
    \centering
   \includegraphics[width=0.48\textwidth]{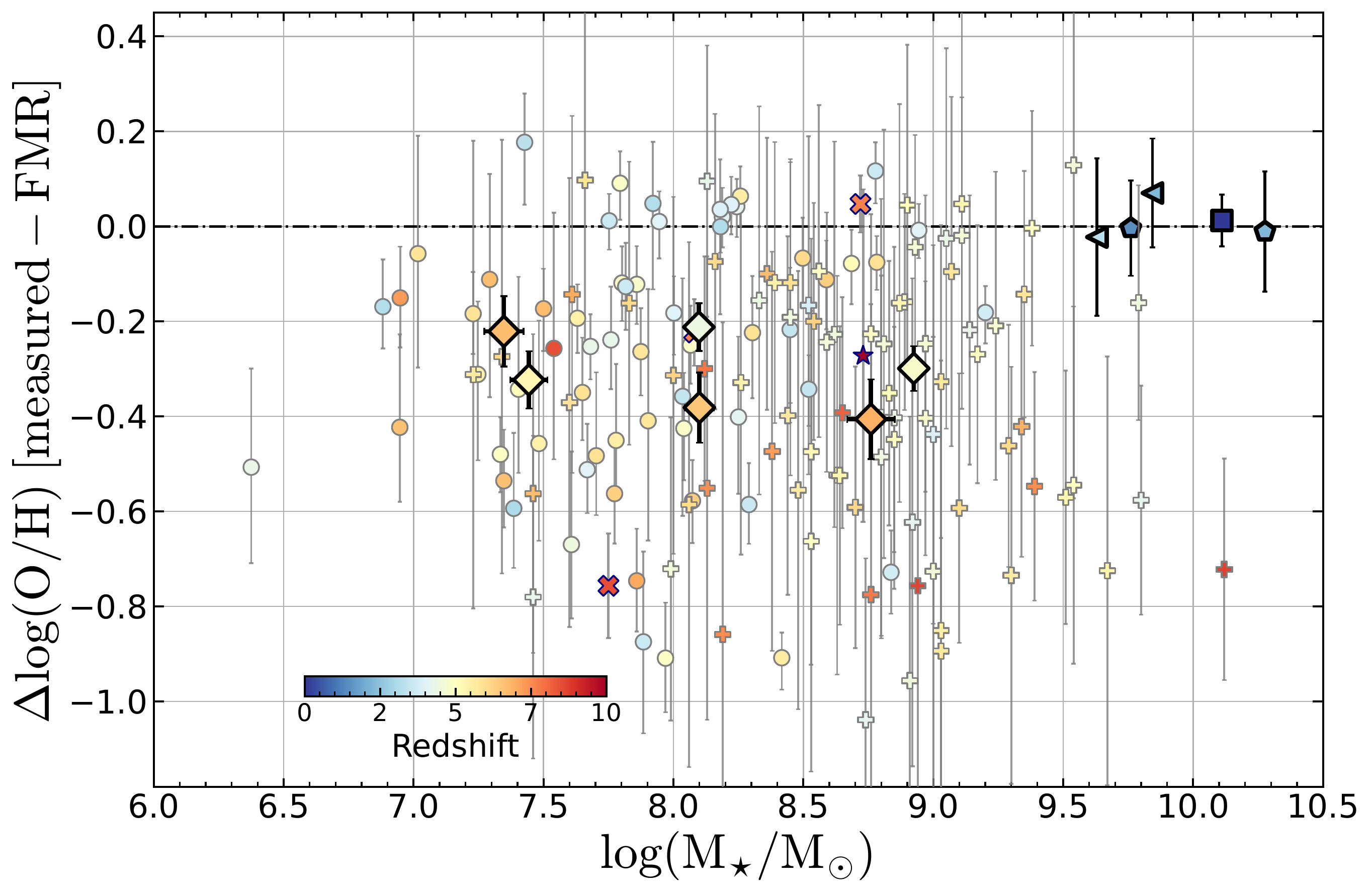}
    \hspace{0.05cm}
    \centering
    \includegraphics[width=0.48\textwidth]{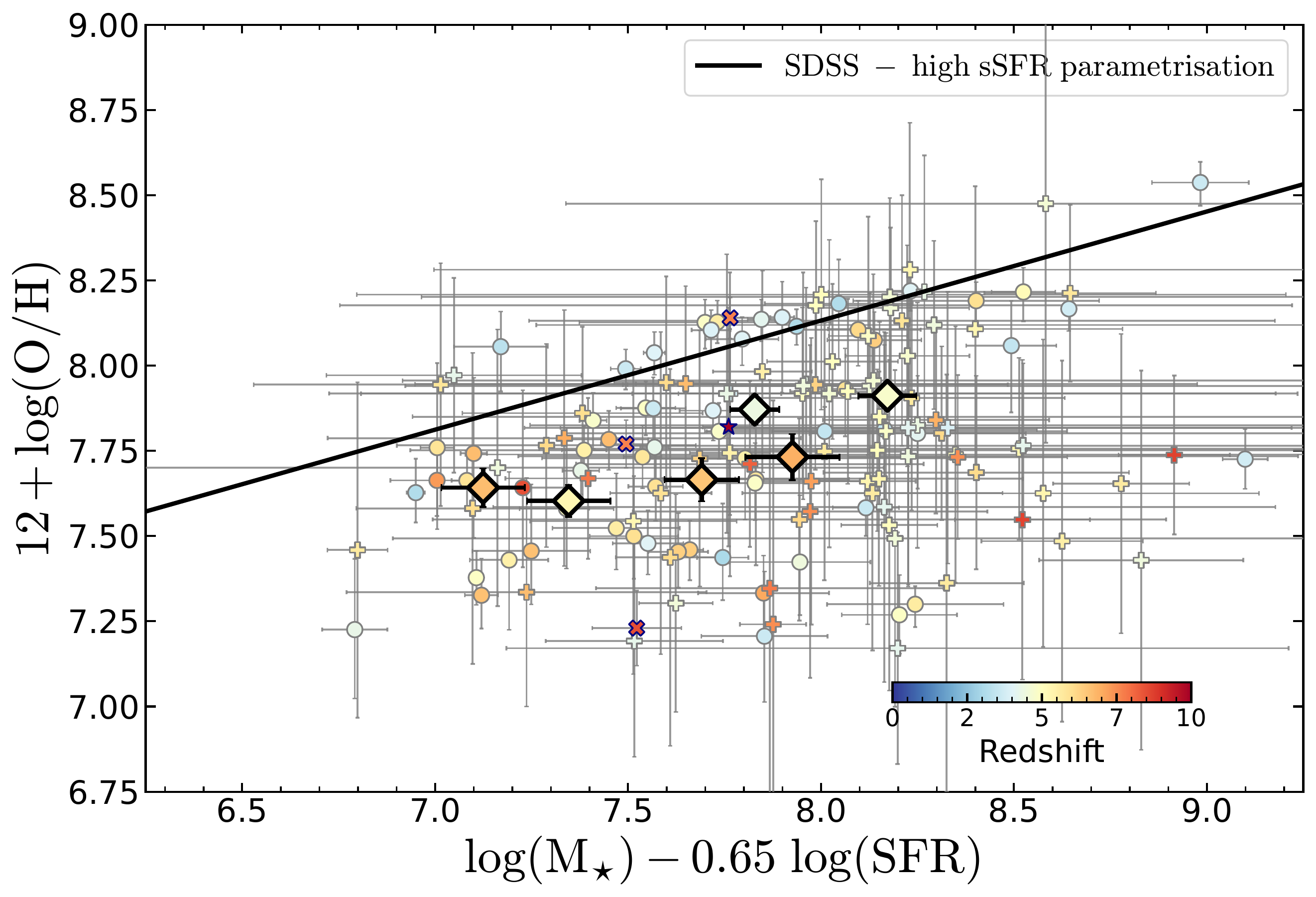}
    \caption{    
    \textit{Left-hand panel:} Deviations from the FMR plotted as a function of stellar mass. Symbols are as in Figure~\ref{fig:mzr} and \ref{fig:fmr}, with individual points (and \mstar-redshift bins) colour-coded by redshift.
    At fixed stellar mass, \emph{JWST} galaxies are preferentially offset towards lower O/H with incresing redshift, and a similar trend is also seen with increasing \mstar\ in the highest redshift bin.
    \textit{Right-hand panel:} Two-dimensional projection of the \mstar-SFR-O/H space in the O/H vs $\mu$=log(\mstar)-$\alpha$log(SFR) plane, assuming the parametrisation for high sSFR (log(sSFR/yr$^{-1}$)>$-9.5$) galaxies inferred from the analysis of the SDSS sample \citep[$\alpha=0.65$,][]{curti_massmetallicity_2020}. Most of the individual galaxies, as well as the median values in the \mstar-redshift bins, are offset below the relation, further suggesting that the local FMR parametrisation does not hold at the very high-z and low-\mstar\ probed in this work. 
}
    \label{fig:projections}
\end{figure*}

\section{Evidence for evolution in the fundamental metallicity relation ?}
\label{sec:fmr_evol}

It is known that the observed distribution of galaxies on the mass-metallicity plane reflects selection effects associated with the average star-formation rate probed by the samples under study \citep{yates_relation_2012, salim_mass-metallicity-star_2015, telford_exploring_2016, cresci_fundamental_2018}.
In this sense, a least biased picture of the chemical evolution stage of these early galaxies might be given by considering the full relationship between \mstar, O/H, and SFR, in what it is usually referred to as the Fundamental Metallicity Relation \citep[FMR, ][]{mannucci_fundamental_2010}, and which takes observationally into account the impact of the current star formation rate on the expected level of chemical enrichment of the ISM, as driven by the interplay between gas accretion, star formation, and outflows.

In Figure~\ref{fig:fmr} we report the deviation of the measured metallicity in our sample, as a function of redshift, from the predictions of the locally defined FMR as parametrised in \cite{curti_massmetallicity_2020}.
Individual points in Figure~\ref{fig:fmr} are colour-coded by their specific star-formation-rate (sSFR=SFR/\mstar).
The error bars on the y-axis includes not only the uncertainty on the individual metallicity measurements, but also an additional term (co-added in quadrature) associated with the FMR predictions as derived from the standard deviation of the distribution of log(O/H) obtained by varying one hundred times the input \mstar and SFR of each galaxy within their uncertainties.
The weighted median values in three bins of redshift ($z\in[3-5]$; $z\in[5-7]$; $z\in[7-10]$) for our combined \emph{JWST} sample are shown by larger diamond markers to highlight possible trends, with solid and dashed errorbars representing the error on the median (as derived from bootstrapping by varying each point 300 times within its uncertainty) and the dispersion of the sample in each bin, respectively.
The average offset from the FMR as measured by galaxy samples at lower redshifts ($z<3$) as compiled from the literature (SDSS at $z\sim0$, \citealt{curti_massmetallicity_2020}; `Blueberry' and `Green Pea' galaxies at $z\sim0.2$, \citealt{yang_blueberries_2017, yang_greenpeas_2017}; KLEVER at $z\sim 1.5-2.5$, \citealt{curti_klever_2020, hayden-pawson_NO_2022}; MOSDEF at $z\sim2-3$, \citealt{sanders_mosdef_mzr_2021}) are also shown with different symbols as outlined in the legend, and colour-coded by the average sSFR value of the relative sample.


Individual galaxies are mostly offset below the FMR at any redshift $z>3$, with $83$ over $146$ sources deviating from the FMR predictions at more than $1\sigma$ ($48$ percent of galaxies at $\langle z \rangle \sim 4$, $54$ percent at $\langle z \rangle \sim 6$, and $87$ percent at $\langle z \rangle \sim 8$), and $25$ at more than $3\sigma$ ($19$ percent of galaxies at $\langle z \rangle \sim 4$, $14$ percent at $\langle z \rangle \sim 6$, and $19$ percent at $\langle z \rangle \sim 8$).
We perform a two-sample Kolmogorov-Smirnov (KS) test on the two distributions, i.e. the JWST galaxy sample and the full SDSS sample used to parametrise the FMR from \cite{curti_massmetallicity_2020} (which is well approximated by a normal distribution with $mean=0$ and $\sigma=0.054$~dex), which are found to be significantly different (p-value $<<0.01$), both for the global sample and within all individual redshift bins.

Looking more closely at the trends in the redshift bins, the median $\Delta$log(O/H) at $\langle z \rangle \sim 4$ is $=-0.23\pm0.04$~dex (and $1-\sigma$ dispersion of $0.29$~dex), the median deviation at $\langle z \rangle \sim 4$ is $=-0.34\pm0.05$ (with a dispersion of $0.25$~dex), whereas the median deviation at $\langle z \rangle \sim 8$ is $=-0.50\pm0.10$ (with a dispersion of $0.38$~dex). 
A Z-test confirms that these are all significantly offset at more than $5\sigma$ from the FMR predictions. 
We note however that we are here neglecting any contribution from additional systematic uncertainties on the measured quantities (\mstar, SFR, O/H), as well as further uncertainties associated with the extrapolation of the FMR parametrisation outside the parameter space of its calibration sample. 
However, including a further $0.25$~dex systematic uncertainty on the individual metallicity predictions evaluated at the average \mstar and SFR of the \emph{JWST} sample (following Figure 11 in \citealt{curti_massmetallicity_2020}), the significance of the deviation of the median values from the predictions of the local FMR only slightly decreases to $\sim 4\sigma$ in each redshift bin.

Furthermore, we identify a tentative trend in which galaxies are seen to sit preferentially below the FMR predictions with increasing redshift, with in particular sources at $z>6$ that are significantly less enriched than their \mstar and SFR would predict were they local galaxies.
A Spearman correlation test between $\Delta$log(O/H) and redshift $z$ on the full \emph{JWST} sample provides evidence for a weak, though significant negative correlation, with $\rho = -0.18$, $\text{p-value}=0.03$. 
Moreover, high redshift galaxies are also characterised, on average, by higher specific star-formation rates (sSFR) than lower redshift systems, forming stars at a higher pace while rapidly building-up their mass.
A Spearman correlation test between $\Delta$log(O/H) and sSFR indeed reveals a moderate ($\rho=0.29$), significant correlation ($\text{p-value}<0.01$).

Finally, in the left-hand panel of Figure~\ref{fig:projections} we plot the same metallicity deviation from the FMR as a function of stellar mass, reporting both individual galaxies and median values in the same \mstar-z bins described in Table~\ref{tab:binned_values}, here colour-coded by their (median) redshift.
Although no clear trend can be identified from individual points, with galaxies showing different levels of offset from the FMR regardless of their stellar mass, from the median values we find weak evidence for higher redshift galaxies to be preferentially offset from the FMR at fixed stellar mass, especially in the two highest \mstar bins, where the offset between median points at different redshift is significant at $2.3\sigma$ and $1.2\sigma$, respectively.
No clear trend is instead identified with \mstar, at fixed redshift.


\subsection{Interpreting the apparent FMR evolution}
Although a more statistically representative sample of early galaxies assembled by \emph{JWST} is required to either robustly confirm or revisit such trends, these observations already suggest a few possible interpretations.
On the one hand, they might reveal prominent accretion of pristine gas at high-redshift, happening on timescales shorter than the gas depletion, star-formation, ISM enrichment and mixing timescales compared to lower redshift galaxies, increasing the gas reservoir and diluting the galaxy metal content at fixed \mstar\ and SFR \citep{dekel_cold_2009, lilly_gas_2013, somerville_dave_2015, dave_mufasa_2017}. 
This could also be reflected into an increased level of stochasticity in the star-formation history during such early phases of galaxy formation, especially at low stellar masses.
If the assembly of early galaxies is primarily driven by stochastic gas accretion from the cosmic web, as sometimes modelled with a scatter term in the baryonic accretion rate parameter \citep[e.g.,][]{forbes_fmr_models_2014}, the longer timescales over which the enrichment from supernovae can balance accretion and star-formation would make a system more likely to be observed out of the equilibrium which is at the basis of the scaling relations observed at lower redshift. \citep[][]{dave_galaxy_2011, zahid_metallicities_2012}.
On the other hand, the efficiency of metal removals could also be enhanced, such that the metal loading factor of outflows is higher at fixed SFR than the one required to reproduce the observed FMR at lower redshifts \citep{baker_resolved_fmr_2023}.

A possible way to differentiate between these different scenarios is to study the morphology of galaxies with the position relative to the FMR. 
For instance, \cite{tacchella_eros_2022} analysed the three galaxies at $z\sim8$ from the EROs, finding that the galaxy with the lowest gas-phase metallicity is very compact and consistent with rapid gas accretion. The two other objects with relatively higher gas-phase metallicity show more complex multi-component morphologies on kpc scales, indicating that their recent increase in star-formation rate is driven by mergers or internal gravitational instabilities.
More recently, \cite{langeroodi_fmr_compactness_2023} investigated the correlation between the deviation from the FMR at high redshift and the `compactness' of galaxies, finding more compact sources to be preferentially offset, and interpreting this result as further evidence for enhanced stochasticity in the accretion and star-formation episodes, happening on timescales much shorter than those associated with the ISM enrichment and therefore diluting the metal content of such compact galaxies more than the average scaling relation would predict.


As seen from a different perspective, the apparent inconsistency between the measured oxygen abundance and the FMR predictions reveals the inadequacy of the local formalism when extrapolated to such early epochs and low stellar masses. Within the framework originally described in \cite{curti_massmetallicity_2020} in fact, the SFR dependence is only embedded in the `turnover mass' term (i.e., the value in \mstar\ at which the slope flattens), which is described to scale linearly with the star-formation rate. Therefore, no direct dependence of the MZR slope on the SFR is included in such FMR parametrisation, nor it is, by definition, in any of the two-dimensional projections explored in the literature in the form $\mu_{\alpha}$ = log(\mstar) - $\alpha$log(SFR) versus log(O/H) \citep[e.g.,][]{mannucci_fundamental_2010, andrews_mass-metallicity_2013}, which implicitly assumes no SFR-dependence of the MZR slope. 
In fact, a rotation of the \mstar\ and SFR axis in the 3D space removes the apparent secondary dependence of the MZR on the star-formation rate only if the slopes of the different mass-metallicity relations are invariant for samples with different average star-formation rates.
Then, if galaxies follow the same scaling relations between \mstar, SFR, and O/H over cosmic time (as in the original FMR framework), such a projection is capable to cancel out any apparent evolution of the MZR with redshift.
On the contrary, if such an assumption does not hold anymore, and the low-mass end slope of the MZR is SFR-dependent, 
then a single 2D projection is not capable of capturing the full interplay between gas flows, star-formation an chemical enrichment which manifests in the existence of a relationship between the observational quantities \mstar, SFR, and O/H. 

In \cite{curti_massmetallicity_2020}, such a possibility was briefly discussed, and two different regimes were identified and explored, providing two different parametrisations of the FMR based on the sSFR of galaxies. 
In this scenario, the slope of the MZR is still invariant with SFR, but the relative strength of the SFR-dependence is significantly different in the two regimes, being much more prominent for high sSFR galaxies, which translates into two different $\mu_{\alpha}$ projections, with $\alpha=0.22$ at low sSFR, and $\alpha=0.65$ at high sSFR\footnote{the best-fit $\alpha$ value for the global sample is $\alpha=0.55$}.
However we stress that, being based on SDSS galaxies, these parametrisations are still statistically dominated by galaxies at higher \mstar\, compared to the regime explored in the present work.
For reference, we plot our combined \emph{JWST} sample presented in this work in the O/H versus $\mu_{\alpha}$ projection, for $\alpha=0.65$, in the right-hand panel of Figure~\ref{fig:projections}: such relation, though calibrated on the highest sSFR galaxies in the SDSS, is not yet capable to account for the full redshift evolution in the MZR as seen at $z=3-10$, and most of the objects (as well as their median binned values) clearly fall below the best-fit line (in black).

\section{Summary and Conclusions}
\label{sec:summary}

We have analysed the metallicity properties of a sample of low-mass (log(\mstar/\msun)$\lesssim 8.5$) galaxies at $3<z<10$, exploiting deep NIRSpec spectroscopic observations from the JADES programme.
Detection of multiple emission lines in both PRISM and medium resolution gratings spectra allowed us to derive the gas-phase metallicity in these sources, which we complemented with a sample of sources at log(\mstar/\msun)$\sim 8.5 - 10$ from other programmes and compiled from the literature (CEERS, \citet{nakajima_mzr_ceers_2023}, ERO \citet{curti_smacs_2023}, and GN-z11 \citet{bunker_gnz11_2023}) to provide insights into the cosmic evolution of the metallicity scaling relations in the low-mass regime up to epoch of early galaxy assembly.
Our main findings are summarised as follows:
\begin{itemize}
    \item We find evidence for a mild evolution in the normalisation of the mass-metallicity relation (MZR) at $z>3$, as previously suggested by similar studies \citep{langeroodi_mzr_2022, heintz_fmr_2022, matthee_eiger_2023, nakajima_mzr_ceers_2023,shapley_mzr_2023}.
    Our sample of galaxies spans a wide range in 12+log(O/H) (corresponding to Z$\sim 0.03-0.6$ Z$_{\odot}$), with a median metallicity of 12+log(O/H)=7.77 (Z=0.12 Z$_{\odot}$). 
    Compared to a simple extrapolation in the low \mstar regime of the MZR $z\sim3.3$ from \citep{sanders_mosdef_mzr_2021}, the \emph{JWST} sample show an average offset in log(O/H) of only $0.05$~dex towards lower metallicity (negligible at log(\mstar/\msun)$\lesssim 8$, more prominent at log(\mstar/\msun)$\sim 9$).
    \item An orthogonal distance regression fit performed on the full \emph{JWST} sample suggests a flattening of the slope of low-mass end of the MZR at $z=3-10$ (Figure~\ref{fig:mzr}).
    This is in agreement with previous findings at $z\sim 3$ in the same stellar mass regime as reported by \cite{Li_mzr_dwarfs_z3_2022}, compared to which we measure however a more prominent evolution in the MZR normalisation of $\sim 0.25$~dex.
    The flattening of the MZR slope at low masses could indicate a change in the dominant mechanisms regulating supernovae-driven outflows compared to higher mass galaxies. 
    The slope measured for our combined sample of galaxies at $z=3-10$, $\beta_{\text{3-10}}=0.17\pm0.03$, is in good agreement with predictions from chemical evolution models implementing a `momentum-driven' feedback mode for SNe winds (right panel of Figure~\ref{fig:mzr_simul}). 
    \item Theoretical simulations of the high-redshift Universe generally predict steeper MZR slopes than observed, broadly matching the normalisation of the relation at log(\mstar/\msun)$\sim8-9$ but struggling to reproduce the metallicity at the lowest masses probed by the present work (right panel of Figure~\ref{fig:mzr_simul}).
    On the other hand, analytical chemical evolution models \citep[e.g.][]{peeples_mzr_model_2011} in which the gas fraction is left free to vary according to the same redshift scaling as inferred up to $z\sim3$ \citep{tacconi_phibss_2018} would predict (once fixing all the other parameters) a stronger evolution in the MZR normalisation at low \mstar than observed.
    \item Notably, our galaxy sample is distributed over a similar region (and with a fully consistent inferred slope) of the mass-metallicity plane as low-redshift, low \mstar, metal poor systems with extreme ionisation conditions like `Blueberry' and `Green Pea' galaxies \citep{yang_blueberries_2017, yang_greenpeas_2017} (left panel of Figure~\ref{fig:mzr_simul}).
    These objects have been identified as local analogues of high-redshift galaxies, and we observe that they share similar metallicity properties to those pertaining to the high-z sources analysed in this paper \citep[see also][]{cameron_jades_bpt_2023}.
    \item We find evidence for a deviation of our \emph{JWST} sample from the framework of the fundamental metallicity relation (FMR) as described for local galaxies.
    In particular, galaxies appear significantly metal deficient compared to local galaxies matched in \mstar\ and SFR, and this trend is observed more prominently as redshift increases (Figure~\ref{fig:fmr}).
    This suggests that in these objects metals are either more efficiently removed by SNe-driven outflows, or that the gas is strongly diluted by stochastic accretion of (nearly-)pristine gas, happening on timescales shorter than those associated with the enrichment from subsequent star-formation.
    \item This behaviour also highlights potential inconsistencies of the assumed FMR parametrisation at very high-z, with the strength of the metallicity-SFR dependence at fixed \mstar\ being no longer redshift invariant. 
    In fact, if the low-mass end slope of the MZR depends on the average SFR of galaxies, a simple two-dimensional projection of the FMR in the log(O/H) versus $\mu_{\alpha}$ plane is not capable of capturing the metallicity variations as caused by the interplay of dilution, enrichment, and metal-loaded outflows driven by star-formation (right panel of Figure~\ref{fig:projections}). 
    
\end{itemize}

The observations presented in this paper represent a step to understanding the chemical enrichment processes in place during early galaxy formation. Deep \emph{JWST} observations have already proven to be capable of extending the mass and luminosity regimes probed by previous facilities by orders of magnitudes, extending our investigations up to the highest redshift galaxies ever discovered. Improving the statistical robustness of these samples in the near future will be crucial for forthcoming studies to further reduce the large systematic uncertainties and provide a more comprehensive picture of galaxy formation across different cosmic epochs, a goal fully within the reach of \emph{JWST} capabilities.

\begin{acknowledgements}
We thank the referee for the insightful comments that helped improving this paper.
MC acknowledges support from the ESO Fellowship Programme.
MC, RM, FDE, TJL, JW, JS, LS, and WB acknowledge support by the Science and Technology Facilities Council (STFC), ERC Advanced Grant 695671 "QUENCH".
JW also acknowledges support from the Fondation MERAC.
SC acknowledges support by European Union's HE ERC Starting Grant No. 101040227 - WINGS.
ECL acknowledges support of an STFC Webb Fellowship (ST/W001438/1). 
JC, AJB, AJC, and GCJ acknowledge funding from the “FirstGalaxies” Advanced Grant from the European Research Council (ERC) under the European Union’s Horizon 2020 research and innovation programme (Grant agreement No. 789056). 
H\"{U} gratefully acknowledges support by the Isaac Newton Trust and by the Kavli Foundation through a Newton-Kavli Junior Fellowship.
SA and BRDP acknowledge support from the research project PID2021-127718NB- I00 of the Spanish Ministry of Science and Innovation/State Agency of Research (MICIN/AEI). 
RS acknowledges support from a STFC Ernest Rutherford Fellowship (ST/S004831/1). 
DJE is supported as a Simons Investigator and by JWST/NIRCam contract to the University of Arizona, NAS5- 02015. 
BER, BDJ, EE, MR, and CNAW acknowledge support from the NIRCam Science Team contract to the University of Arizona, NAS5-02015. 
The work of RH is supported by the Johns Hopkins University, Institute for Data Intensive Engineering and Science (IDIES).
The work of CCW is supported by NOIRLab, which is managed by the Association of Universities for Research in Astronomy (AURA) under a cooperative agreement with the National Science Foundation. 
KB acknowledges support from the Australian Research Council Centre of Excellence for All Sky Astrophysics in 3 Dimensions (ASTRO 3D), through project number CE170100013.

MC is grateful to Kimihiko Nakajima for making his measurements of line ratios from the CEERS sample public.
MC also thanks Ivanna Langan and Daniel Ceverino for sharing the data from \textsc{firstlight} simulations.

This work is based on observations made with the NASA/ESA/CSA James Webb Space Telescope. The data were obtained from the Mikulski Archive for Space Telescopes at the Space Telescope Science Institute, which is operated by the Association of Universities for Research in Astronomy, Inc., under NASA contract NAS 5-03127 for JWST. These observations are associated with program \#1180 and \#1210.

\end{acknowledgements}

\section*{Data Availability}
The high-level data products exploited in this paper for the \emph{deep} tier of the JADES-GTO Programme in GOODS-South, i.e., the redshifts, stellar masses, star-formation rates, and metallicities, are reported in Table~\ref{tab:jades_properties}.
Fully reduced spectra and emission line fluxes are available through the MAST database at \url{https://archive.stsci.edu/hlsp/jades}, and are described in \cite{bunker_hst_deep_DR_2023}.

%
%

\bibliographystyle{aa}
\bibliography{biblio}

\begin{thebibliography}{142}
\expandafter\ifx\csname natexlab\endcsname\relax\def\natexlab#1{#1}\fi

\bibitem[{Allende~Prieto {et~al.}(2001)Allende~Prieto, Lambert, \&
  Asplund}]{allende_prieto_forbidden_2001}
Allende~Prieto, C., Lambert, D.~L., \& Asplund, M. 2001, \apjl, 556, L63

\bibitem[{Andrews \& Martini(2013)}]{andrews_mass-metallicity_2013}
Andrews, B.~H. \& Martini, P. 2013, \apj, 765, 140

\bibitem[{{Arellano-C{\'o}rdova} {et~al.}(2022){Arellano-C{\'o}rdova}, {Berg},
  {Chisholm}, {Arrabal Haro}, {Dickinson}, {Finkelstein}, {Leclercq}, {Rogers},
  {Simons}, {Skillman}, {Trump}, \& {Kartaltepe}}]{arellano-corodva_2023}
{Arellano-C{\'o}rdova}, K.~Z., {Berg}, D.~A., {Chisholm}, J., {et~al.} 2022,
  \apjl, 940, L23

\bibitem[{{Asplund} {et~al.}(2009){Asplund}, {Grevesse}, {Sauval}, \&
  {Scott}}]{asplund_solar_2009}
{Asplund}, M., {Grevesse}, N., {Sauval}, A.~J., \& {Scott}, P. 2009, \araa, 47,
  481

\bibitem[{{Baker} \& {Maiolino}(2023)}]{baker_mzr_2023}
{Baker}, W.~M. \& {Maiolino}, R. 2023, \mnras [\eprint[arXiv]{2303.08145}]

\bibitem[{{Baker} {et~al.}(2023{\natexlab{a}}){Baker}, {Maiolino}, {Belfiore},
  {Bluck}, {Curti}, {Wylezalek}, {Bertemes}, {Bothwell}, {Lin}, {Thorp}, \&
  {Pan}}]{baker_sfms_2023}
{Baker}, W.~M., {Maiolino}, R., {Belfiore}, F., {et~al.} 2023{\natexlab{a}},
  \mnras, 518, 4767

\bibitem[{{Baker} {et~al.}(2023{\natexlab{b}}){Baker}, {Maiolino}, {Belfiore},
  {Curti}, {Bluck}, {Lin}, {Ellison}, {Thorp}, \&
  {Pan}}]{baker_resolved_fmr_2023}
{Baker}, W.~M., {Maiolino}, R., {Belfiore}, F., {et~al.} 2023{\natexlab{b}},
  \mnras, 519, 1149

\bibitem[{Baldwin {et~al.}(1981)Baldwin, Phillips, \&
  Terlevich}]{baldwin_classification_1981}
Baldwin, J.~A., Phillips, M.~M., \& Terlevich, R. 1981, \pasp, 93, 5

\bibitem[{Belli {et~al.}(2013)Belli, Jones, Ellis, \&
  Richard}]{belli_testing_2013}
Belli, S., Jones, T., Ellis, R.~S., \& Richard, J. 2013, \apj, 772, 141

\bibitem[{Bian {et~al.}(2018)Bian, Kewley, \&
  Dopita}]{bian_ldquodirectrdquo_2018}
Bian, F., Kewley, L.~J., \& Dopita, M.~A. 2018, \apj, 859, 175

\bibitem[{{Birkmann} {et~al.}(2011){Birkmann}, {B{\"o}ker}, {Ferruit},
  {Giardino}, {Jakobsen}, {de Marchi}, {Sirianni}, {te Plate}, {Savignol},
  {Gnata}, {Wettemann}, {Dorner}, {Cresci}, {Rosales-Ortega}, {Stuhlinger},
  {Cole}, {Tandy}, \& {Brockley-Blatt}}]{birkmann_nirspec_2011}
{Birkmann}, S.~M., {B{\"o}ker}, T., {Ferruit}, P., {et~al.} 2011, in Society of
  Photo-Optical Instrumentation Engineers (SPIE) Conference Series, Vol. 8150,
  Cryogenic Optical Systems and Instruments XIII, ed. J.~B. {Heaney} \& E.~T.
  {Kvamme}, 81500B

\bibitem[{{B{\"o}ker} {et~al.}(2023){B{\"o}ker}, {Beck}, {Birkmann},
  {Giardino}, {Keyes}, {Kumari}, {Muzerolle}, {Rawle}, {Zeidler}, {Abul-Huda},
  {Alves de Oliveira}, {Arribas}, {Bechtold}, {Bhatawdekar}, {Bonaventura},
  {Bunker}, {Cameron}, {Carniani}, {Charlot}, {Curti}, {Espinoza}, {Ferruit},
  {Franx}, {Jakobsen}, {Karakla}, {L{\'o}pez-Caniego}, {L{\"u}tzgendorf},
  {Maiolino}, {Manjavacas}, {Marston}, {Moseley}, {Ogle}, {Perna},
  {Pe{\~n}a-Guerrero}, {Pirzkal}, {Plesha}, {Proffitt}, {Rauscher}, {Rix},
  {Rodr{\'\i}guez del Pino}, {Rustamkulov}, {Sabbi}, {Sing}, {Sirianni}, {te
  Plate}, {{\'U}beda}, {Wahlgren}, {Wislowski}, {Wu}, \&
  {Willott}}]{boeker_nirspec_2023}
{B{\"o}ker}, T., {Beck}, T.~L., {Birkmann}, S.~M., {et~al.} 2023, arXiv
  e-prints, arXiv:2301.13766

\bibitem[{{B{\"o}ker} {et~al.}(2012){B{\"o}ker}, {Birkmann}, {de Marchi},
  {Ferruit}, {Giardino}, {Sirianni}, \& {Beck}}]{boeker_nirspec_2012}
{B{\"o}ker}, T., {Birkmann}, S., {de Marchi}, G., {et~al.} 2012, in Society of
  Photo-Optical Instrumentation Engineers (SPIE) Conference Series, Vol. 8442,
  Space Telescopes and Instrumentation 2012: Optical, Infrared, and Millimeter
  Wave, ed. M.~C. {Clampin}, G.~G. {Fazio}, H.~A. {MacEwen}, \& J.~{Oschmann},
  Jacobus~M., 84423F

\bibitem[{{Bonaventura} {et~al.}(2023){Bonaventura}, {Jakobsen}, {Ferruit},
  {Arribas}, \& {Giardino}}]{bonaventura_empt_2023}
{Bonaventura}, N., {Jakobsen}, P., {Ferruit}, P., {Arribas}, S., \& {Giardino},
  G. 2023, arXiv e-prints, arXiv:2302.10957

\bibitem[{Bruzual \& Charlot(2003)}]{bruzual_stellar_2003}
Bruzual, G. \& Charlot, S. 2003, \mnras, 344, 1000

\bibitem[{{Bunker} {et~al.}(2023{\natexlab{a}}){Bunker}, {Cameron},
  {Curtis-Lake}, {Jakobsen}, {Carniani}, {Curti}, {Witstok}, {Maiolino},
  {D'Eugenio}, {Looser}, {Willott}, {Bonaventura}, {Hainline}, {Uebler},
  {Willmer}, {Saxena}, {Smit}, {Alberts}, {Arribas}, {Baker}, {Baum},
  {Bhatawdekar}, {Bowler}, {Boyett}, {Charlot}, {Chen}, {Chevallard},
  {Circosta}, {DeCoursey}, {de Graaff}, {Egami}, {Eisenstein}, {Endsley},
  {Ferruit}, {Giardino}, {Hausen}, {Helton}, {Hviding}, {Ji}, {Johnson},
  {Jones}, {Kumari}, {Laseter}, {Luetzgendorf}, {Maseda}, {Nelson}, {Parlanti},
  {Perna}, {Rawle}, {Rix}, {Rieke}, {Robertson}, {Rodriguez Del Pino},
  {Sandles}, {Scholtz}, {Sharpe}, {Skarbinski}, {Stark}, {Sun}, {Tacchella},
  {Topping}, {Villanueva}, {Wallace}, {Williams}, \&
  {Woodrum}}]{bunker_hst_deep_DR_2023}
{Bunker}, A.~J., {Cameron}, A.~J., {Curtis-Lake}, E., {et~al.}
  2023{\natexlab{a}}, arXiv e-prints, arXiv:2306.02467

\bibitem[{{Bunker} {et~al.}(2023{\natexlab{b}}){Bunker}, {Saxena}, {Cameron},
  {Willott}, {Curtis-Lake}, {Jakobsen}, {Carniani}, {Smit}, {Maiolino},
  {Witstok}, {Curti}, {D'Eugenio}, {Jones}, {Ferruit}, {Arribas}, {Charlot},
  {Chevallard}, {Giardino}, {de Graaff}, {Looser}, {Luetzgendorf}, {Maseda},
  {Rawle}, {Rix}, {Rodriguez Del Pino}, {Alberts}, {Egami}, {Eisenstein},
  {Endsley}, {Hainline}, {Hausen}, {Johnson}, {Rieke}, {Rieke}, {Robertson},
  {Shivaei}, {Stark}, {Sun}, {Tacchella}, {Tang}, {Williams}, {Willmer},
  {Baker}, {Baum}, {Bhatawdekar}, {Bowler}, {Boyett}, {Chen}, {Circosta},
  {Helton}, {Ji}, {Lyu}, {Nelson}, {Parlanti}, {Perna}, {Sandles}, {Scholtz},
  {Suess}, {Topping}, {Uebler}, {Wallace}, \& {Whitler}}]{bunker_gnz11_2023}
{Bunker}, A.~J., {Saxena}, A., {Cameron}, A.~J., {et~al.} 2023{\natexlab{b}},
  arXiv e-prints, arXiv:2302.07256

\bibitem[{{Cameron} {et~al.}(2023{\natexlab{a}}){Cameron}, {Katz}, {Rey}, \&
  {Saxena}}]{cameron_gnz11_2023}
{Cameron}, A.~J., {Katz}, H., {Rey}, M.~P., \& {Saxena}, A. 2023{\natexlab{a}},
  \mnras, 523, 3516

\bibitem[{{Cameron} {et~al.}(2023{\natexlab{b}}){Cameron}, {Saxena}, {Bunker},
  {D'Eugenio}, {Carniani}, {Maiolino}, {Curtis-Lake}, {Ferruit}, {Jakobsen},
  {Arribas}, {Bonaventura}, {Charlot}, {Chevallard}, {Curti}, {Looser},
  {Maseda}, {Rawle}, {Rodr{\'\i}guez Del Pino}, {Smit}, {{\"U}bler}, {Willott},
  {Witstok}, {Egami}, {Eisenstein}, {Johnson}, {Hainline}, {Rieke},
  {Robertson}, {Stark}, {Tacchella}, {Williams}, {Bhatawdekar}, {Bowler},
  {Boyett}, {Circosta}, {Helton}, {Jones}, {Kumari}, {Ji}, {Nelson},
  {Parlanti}, {Sandles}, {Scholtz}, \& {Sun}}]{cameron_jades_bpt_2023}
{Cameron}, A.~J., {Saxena}, A., {Bunker}, A.~J., {et~al.} 2023{\natexlab{b}},
  arXiv e-prints, arXiv:2302.04298

\bibitem[{Cappellari(2017)}]{cappellari_improving_2017}
Cappellari, M. 2017, \mnras, 466, 798

\bibitem[{{Cappellari}(2022)}]{cappellari_ppxf_2022}
{Cappellari}, M. 2022, MNRAS submitted [\eprint{2208.14974}]

\bibitem[{{Catinella} {et~al.}(2018){Catinella}, {Saintonge}, {Janowiecki},
  {Cortese}, {Dav{\'e}}, {Lemonias}, {Cooper}, {Schiminovich}, {Hummels},
  {Fabello}, {Ger{\'e}b}, {Kilborn}, \& {Wang}}]{catinella_xgass_2018}
{Catinella}, B., {Saintonge}, A., {Janowiecki}, S., {et~al.} 2018, \mnras, 476,
  875

\bibitem[{Chabrier(2003)}]{chabrier_galactic_2003}
Chabrier, G. 2003, \pasp, 115, 763

\bibitem[{{Chevallard} \& {Charlot}(2016)}]{chevallard_beagle_2016}
{Chevallard}, J. \& {Charlot}, S. 2016, \mnras, 462, 1415

\bibitem[{{Choi} {et~al.}(2016){Choi}, {Dotter}, {Conroy}, {Cantiello},
  {Paxton}, \& {Johnson}}]{choi_MIST_2016}
{Choi}, J., {Dotter}, A., {Conroy}, C., {et~al.} 2016, \apj, 823, 102

\bibitem[{{Conroy} {et~al.}(2019){Conroy}, {Naidu}, {Zaritsky}, {Bonaca},
  {Cargile}, {Johnson}, \& {Caldwell}}]{conroy_stellar_halo_2019}
{Conroy}, C., {Naidu}, R.~P., {Zaritsky}, D., {et~al.} 2019, \apj, 887, 237

\bibitem[{Cresci {et~al.}(2018)Cresci, Mannucci, \&
  Curti}]{cresci_fundamental_2018}
Cresci, G., Mannucci, F., \& Curti, M. 2018, arXiv e-prints

\bibitem[{Curti {et~al.}(2017)Curti, Cresci, Mannucci, Marconi, Maiolino, \&
  Esposito}]{curti_new_2017}
Curti, M., Cresci, G., Mannucci, F., {et~al.} 2017, \mnras, 465, 1384

\bibitem[{{Curti} {et~al.}(2023){Curti}, {D'Eugenio}, {Carniani}, {Maiolino},
  {Sandles}, {Witstok}, {Baker}, {Bennett}, {Piotrowska}, {Tacchella},
  {Charlot}, {Nakajima}, {Maheson}, {Mannucci}, {Amiri}, {Arribas}, {Belfiore},
  {Bonaventura}, {Bunker}, {Chevallard}, {Cresci}, {Curtis-Lake},
  {Hayden-Pawson}, {Jones}, {Kumari}, {Laseter}, {Looser}, {Marconi}, {Maseda},
  {Scholtz}, {Smit}, {{\"U}bler}, \& {Wallace}}]{curti_smacs_2023}
{Curti}, M., {D'Eugenio}, F., {Carniani}, S., {et~al.} 2023, \mnras, 518, 425

\bibitem[{Curti {et~al.}(2020{\natexlab{a}})Curti, Maiolino, Cirasuolo,
  Mannucci, Williams, Auger, Mercurio, Hayden-Pawson, Cresci, Marconi,
  Belfiore, Cappellari, Cicone, Cullen, Meneghetti, Ota, Peng, Pettini,
  Swinbank, \& Troncoso}]{curti_klever_2020}
Curti, M., Maiolino, R., Cirasuolo, M., {et~al.} 2020{\natexlab{a}}, Monthly
  Notices of the Royal Astronomical Society, 492, 821, arXiv: 1910.13451

\bibitem[{Curti {et~al.}(2020{\natexlab{b}})Curti, Mannucci, Cresci, \&
  Maiolino}]{curti_massmetallicity_2020}
Curti, M., Mannucci, F., Cresci, G., \& Maiolino, R. 2020{\natexlab{b}},
  Monthly Notices of the Royal Astronomical Society, 491, 944, publisher:
  Oxford Academic

\bibitem[{{Curtis-Lake} {et~al.}(2022){Curtis-Lake}, {Carniani}, {Cameron},
  {Charlot}, {Jakobsen}, {Maiolino}, {Bunker}, {Witstok}, {Smit}, {Chevallard},
  {Willott}, {Ferruit}, {Arribas}, {Bonaventura}, {Curti}, {D'Eugenio},
  {Franx}, {Giardino}, {Looser}, {L{\"u}tzgendorf}, {Maseda}, {Rawle}, {Rix},
  {Rodriguez del Pino}, {{\"U}bler}, {Sirianni}, {Dressler}, {Egami},
  {Eisenstein}, {Endsley}, {Hainline}, {Hausen}, {Johnson}, {Rieke},
  {Robertson}, {Shivaei}, {Stark}, {Tacchella}, {Williams}, {Willmer},
  {Bhatawdekar}, {Bowler}, {Boyett}, {Chen}, {de Graaff}, {Helton}, {Hviding},
  {Jones}, {Kumari}, {Lyu}, {Nelson}, {Perna}, {Sandles}, {Saxena}, {Suess},
  {Sun}, {Topping}, {Wallace}, \& {Whitler}}]{curtis-lake_2023}
{Curtis-Lake}, E., {Carniani}, S., {Cameron}, A., {et~al.} 2022, arXiv
  e-prints, arXiv:2212.04568

\bibitem[{{Dav{\'e}} {et~al.}(2012){Dav{\'e}}, {Finlator}, \&
  {Oppenheimer}}]{dave_mzr_model_2012}
{Dav{\'e}}, R., {Finlator}, K., \& {Oppenheimer}, B.~D. 2012, \mnras, 421, 98

\bibitem[{Davé {et~al.}(2011)Davé, Finlator, \&
  Oppenheimer}]{dave_galaxy_2011}
Davé, R., Finlator, K., \& Oppenheimer, B.~D. 2011, \mnras, 416, 1354

\bibitem[{Davé {et~al.}(2017)Davé, Rafieferantsoa, Thompson, \&
  Hopkins}]{dave_mufasa_2017}
Davé, R., Rafieferantsoa, M.~H., Thompson, R.~J., \& Hopkins, P.~F. 2017,
  \mnras, 467, 115

\bibitem[{Dekel {et~al.}(2009)Dekel, Birnboim, Engel, Freundlich, Goerdt,
  Mumcuoglu, Neistein, Pichon, Teyssier, \& Zinger}]{dekel_cold_2009}
Dekel, A., Birnboim, Y., Engel, G., {et~al.} 2009, \nat, 457, 451

\bibitem[{{Eisenstein} {et~al.}(2023){Eisenstein}, {Willott}, {Alberts},
  {Arribas}, {Bonaventura}, {Bunker}, {Cameron}, {Carniani}, {Charlot},
  {Curtis-Lake}, {D'Eugenio}, {Endsley}, {Ferruit}, {Giardino}, {Hainline},
  {Hausen}, {Jakobsen}, {Johnson}, {Maiolino}, {Rieke}, {Rieke}, {Rix},
  {Robertson}, {Stark}, {Tacchella}, {Williams}, {Willmer}, {Baker}, {Baum},
  {Bhatawdekar}, {Boyett}, {Chen}, {Chevallard}, {Circosta}, {Curti},
  {Danhaive}, {DeCoursey}, {de Graaff}, {Dressler}, {Egami}, {Helton},
  {Hviding}, {Ji}, {Jones}, {Kumari}, {L{\"u}tzgendorf}, {Laseter}, {Looser},
  {Lyu}, {Maseda}, {Nelson}, {Parlanti}, {Perna}, {Pusk{\'a}s}, {Rawle},
  {Rodr{\'\i}guez Del Pino}, {Sandles}, {Saxena}, {Scholtz}, {Sharpe},
  {Shivaei}, {Silcock}, {Simmonds}, {Skarbinski}, {Smit}, {Stone}, {Suess},
  {Sun}, {Tang}, {Topping}, {{\"U}bler}, {Villanueva}, {Wallace}, {Whitler},
  {Witstok}, \& {Woodrum}}]{Eisenstein_JADES_2023}
{Eisenstein}, D.~J., {Willott}, C., {Alberts}, S., {et~al.} 2023, arXiv
  e-prints, arXiv:2306.02465

\bibitem[{Ellison {et~al.}(2008)Ellison, Patton, Simard, \&
  McConnachie}]{ellison_clues_2008}
Ellison, S.~L., Patton, D.~R., Simard, L., \& McConnachie, A.~W. 2008, \apjl,
  672, L107

\bibitem[{Erb {et~al.}(2006)Erb, Shapley, Pettini, Steidel, Reddy, \&
  Adelberger}]{erb_mass-metallicity_2006}
Erb, D.~K., Shapley, A.~E., Pettini, M., {et~al.} 2006, \apj, 644, 813

\bibitem[{{Ferruit} {et~al.}(2022){Ferruit}, {Jakobsen}, {Giardino}, {Rawle},
  {Alves de Oliveira}, {Arribas}, {Beck}, {Birkmann}, {B{\"o}ker}, {Bunker},
  {Charlot}, {de Marchi}, {Franx}, {Henry}, {Karakla}, {Kassin}, {Kumari},
  {L{\'o}pez-Caniego}, {L{\"u}tzgendorf}, {Maiolino}, {Manjavacas}, {Marston},
  {Moseley}, {Muzerolle}, {Pirzkal}, {Rauscher}, {Rix}, {Sabbi}, {Sirianni},
  {te Plate}, {Valenti}, {Willott}, \& {Zeidler}}]{ferruit_nirspec_2022}
{Ferruit}, P., {Jakobsen}, P., {Giardino}, G., {et~al.} 2022, \aap, 661, A81

\bibitem[{{Finkelstein} {et~al.}(2022){Finkelstein}, {Bagley}, {Ferguson},
  {Wilkins}, {Kartaltepe}, {Papovich}, {Yung}, {Arrabal Haro}, {Behroozi},
  {Dickinson}, {Kocevski}, {Koekemoer}, {Larson}, {Le Bail}, {Morales},
  {Perez-Gonzalez}, {Burgarella}, {Dave}, {Hirschmann}, {Somerville}, {Wuyts},
  {Bromm}, {Casey}, {Fontana}, {Fujimoto}, {Gardner}, {Giavalisco}, {Grazian},
  {Grogin}, {Hathi}, {Hutchison}, {Jha}, {Jogee}, {Kewley}, {Kirkpatrick},
  {Long}, {Lotz}, {Pentericci}, {Pierel}, {Pirzkal}, {Ravindranath}, {Ryan},
  {Trump}, {Yang}, {Bhatawdekar}, {Bisigello}, {Buat}, {Calabro}, {Castellano},
  {Cleri}, {Cooper}, {Croton}, {Daddi}, {Dekel}, {Elbaz}, {Franco}, {Gawiser},
  {Holwerda}, {Huertas-Company}, {Jaskot}, {Leung}, {Lucas}, {Mobasher},
  {Pandya}, {Tacchella}, {Weiner}, \& {Zavala}}]{finkelstein_ceers_paper1_2022}
{Finkelstein}, S.~L., {Bagley}, M.~B., {Ferguson}, H.~C., {et~al.} 2022, arXiv
  e-prints, arXiv:2211.05792

\bibitem[{{Finlator} \& {Dav{\'e}}(2008)}]{finlator_dave_2008}
{Finlator}, K. \& {Dav{\'e}}, R. 2008, \mnras, 385, 2181

\bibitem[{{Forbes} {et~al.}(2014){Forbes}, {Krumholz}, {Burkert}, \&
  {Dekel}}]{forbes_fmr_models_2014}
{Forbes}, J.~C., {Krumholz}, M.~R., {Burkert}, A., \& {Dekel}, A. 2014, \mnras,
  443, 168

\bibitem[{{Foreman-Mackey} {et~al.}(2013){Foreman-Mackey}, {Hogg}, {Lang}, \&
  {Goodman}}]{emcee_2013}
{Foreman-Mackey}, D., {Hogg}, D.~W., {Lang}, D., \& {Goodman}, J. 2013, \pasp,
  125, 306

\bibitem[{Fraternali \& Binney(2008)}]{fraternali_accretion_2008}
Fraternali, F. \& Binney, J.~J. 2008, \mnras, 386, 935

\bibitem[{{Giardino} {et~al.}(2019){Giardino}, {Birkmann}, {Robberto},
  {Ferruit}, {Rauscher}, {Sirianni}, {Alves de Oliveira}, {Boeker},
  {Luetzgendorf}, {te Plate}, {Puga}, \& {Rawle}}]{giardino_nirspec_CRs_2019}
{Giardino}, G., {Birkmann}, S., {Robberto}, M., {et~al.} 2019, \pasp, 131,
  094503

\bibitem[{{Giavalisco} {et~al.}(2004){Giavalisco}, {Ferguson}, {Koekemoer},
  {Dickinson}, {Alexander}, {Bauer}, {Bergeron}, {Biagetti}, {Brandt},
  {Casertano}, {Cesarsky}, {Chatzichristou}, {Conselice}, {Cristiani}, {Da
  Costa}, {Dahlen}, {de Mello}, {Eisenhardt}, {Erben}, {Fall}, {Fassnacht},
  {Fosbury}, {Fruchter}, {Gardner}, {Grogin}, {Hook}, {Hornschemeier}, {Idzi},
  {Jogee}, {Kretchmer}, {Laidler}, {Lee}, {Livio}, {Lucas}, {Madau},
  {Mobasher}, {Moustakas}, {Nonino}, {Padovani}, {Papovich}, {Park},
  {Ravindranath}, {Renzini}, {Richardson}, {Riess}, {Rosati}, {Schirmer},
  {Schreier}, {Somerville}, {Spinrad}, {Stern}, {Stiavelli}, {Strolger},
  {Urry}, {Vandame}, {Williams}, \& {Wolf}}]{giavalisco_goodss_2004}
{Giavalisco}, M., {Ferguson}, H.~C., {Koekemoer}, A.~M., {et~al.} 2004, \apjl,
  600, L93

\bibitem[{{Gordon} {et~al.}(2003){Gordon}, {Clayton}, {Misselt}, {Landolt}, \&
  {Wolff}}]{gordon_LMC_attenuation_2003}
{Gordon}, K.~D., {Clayton}, G.~C., {Misselt}, K.~A., {Landolt}, A.~U., \&
  {Wolff}, M.~J. 2003, \apj, 594, 279

\bibitem[{Guo {et~al.}(2016)Guo, Koo, Lu, Forbes, Rafelski, Trump, Amorín,
  Barro, Davé, Faber, Hathi, Yesuf, Cooper, Dekel, Guhathakurta, Kirby,
  Koekemoer, Pérez-González, Lin, Newman, Primack, Rosario, Willmer, \&
  Yan}]{guo_stellar_2016}
Guo, Y., Koo, D.~C., Lu, Y., {et~al.} 2016, \apj, 822, 103

\bibitem[{{Hayden-Pawson} {et~al.}(2022){Hayden-Pawson}, {Curti}, {Maiolino},
  {Cirasuolo}, {Belfiore}, {Cappellari}, {Concas}, {Cresci}, {Cullen},
  {Kobayashi}, {Mannucci}, {Marconi}, {Meneghetti}, {Mercurio}, {Peng},
  {Swinbank}, \& {Vincenzo}}]{hayden-pawson_NO_2022}
{Hayden-Pawson}, C., {Curti}, M., {Maiolino}, R., {et~al.} 2022, \mnras, 512,
  2867

\bibitem[{{Heintz} {et~al.}(2022){Heintz}, {Brammer}, {Gim{\'e}nez-Arteaga},
  {Strait}, {Lagos}, {Vijayan}, {Matthee}, {Watson}, {Mason}, {Hutter}, {Toft},
  {Fynbo}, \& {Oesch}}]{heintz_fmr_2022}
{Heintz}, K.~E., {Brammer}, G.~B., {Gim{\'e}nez-Arteaga}, C., {et~al.} 2022,
  arXiv e-prints, arXiv:2212.02890

\bibitem[{{Henry} {et~al.}(2013{\natexlab{a}}){Henry}, {Martin}, {Finlator}, \&
  {Dressler}}]{henry_low_mass_mzr_2013}
{Henry}, A., {Martin}, C.~L., {Finlator}, K., \& {Dressler}, A.
  2013{\natexlab{a}}, \apj, 769, 148

\bibitem[{{Henry} {et~al.}(2013{\natexlab{b}}){Henry}, {Scarlata},
  {Dom{\'\i}nguez}, {Malkan}, {Martin}, {Siana}, {Atek}, {Bedregal}, {Colbert},
  {Rafelski}, {Ross}, {Teplitz}, {Bunker}, {Dressler}, {Hathi}, {Masters},
  {McCarthy}, \& {Straughn}}]{henry_mzr_2013}
{Henry}, A., {Scarlata}, C., {Dom{\'\i}nguez}, A., {et~al.} 2013{\natexlab{b}},
  \apjl, 776, L27

\bibitem[{Hirschauer {et~al.}(2018)Hirschauer, Salzer, Janowiecki, \&
  Wegner}]{hirschauer_metal_2018}
Hirschauer, A.~S., Salzer, J.~J., Janowiecki, S., \& Wegner, G.~A. 2018, \aj,
  155, 82

\bibitem[{{Jakobsen} {et~al.}(2022){Jakobsen}, {Ferruit}, {Alves de Oliveira},
  {Arribas}, {Bagnasco}, {Barho}, {Beck}, {Birkmann}, {B{\"o}ker}, {Bunker},
  {Charlot}, {de Jong}, {de Marchi}, {Ehrenwinkler}, {Falcolini}, {Fels},
  {Franx}, {Franz}, {Funke}, {Giardino}, {Gnata}, {Holota}, {Honnen}, {Jensen},
  {Jentsch}, {Johnson}, {Jollet}, {Karl}, {Kling}, {K{\"o}hler}, {Kolm},
  {Kumari}, {Lander}, {Lemke}, {L{\'o}pez-Caniego}, {L{\"u}tzgendorf},
  {Maiolino}, {Manjavacas}, {Marston}, {Maschmann}, {Maurer}, {Messerschmidt},
  {Moseley}, {Mosner}, {Mott}, {Muzerolle}, {Pirzkal}, {Pittet}, {Plitzke},
  {Posselt}, {Rapp}, {Rauscher}, {Rawle}, {Rix}, {R{\"o}del}, {Rumler},
  {Sabbi}, {Salvignol}, {Schmid}, {Sirianni}, {Smith}, {Strada}, {te Plate},
  {Valenti}, {Wettemann}, {Wiehe}, {Wiesmayer}, {Willott}, {Wright}, {Zeidler},
  \& {Zincke}}]{jakobsen_nirspec_2022}
{Jakobsen}, P., {Ferruit}, P., {Alves de Oliveira}, C., {et~al.} 2022, \aap,
  661, A80

\bibitem[{Johnson {et~al.}(2021)Johnson, Leja, Conroy, \&
  Speagle}]{Johnson_prospector_2021}
Johnson, B.~D., Leja, J., Conroy, C., \& Speagle, J.~S. 2021, The Astrophysical
  Journal Supplement Series, 254, 22

\bibitem[{Kauffmann {et~al.}(2003)Kauffmann, Heckman, White, Charlot, Tremonti,
  Peng, Seibert, Brinkmann, Nichol, SubbaRao, \&
  York}]{kauffmann_dependence_2003}
Kauffmann, G., Heckman, T.~M., White, S. D.~M., {et~al.} 2003, \mnras, 341, 54

\bibitem[{Kennicutt \& Evans(2012)}]{kennicutt_star_2012}
Kennicutt, R.~C. \& Evans, N.~J. 2012, \araa, 50, 531

\bibitem[{Kewley {et~al.}(2001)Kewley, Dopita, Sutherland, Heisler, \&
  Trevena}]{kewley_theoretical_2001}
Kewley, L.~J., Dopita, M.~A., Sutherland, R.~S., Heisler, C.~A., \& Trevena, J.
  2001, \apj, 556, 121

\bibitem[{Kewley {et~al.}(2013)Kewley, Maier, Yabe, Ohta, Akiyama, Dopita, \&
  Yuan}]{kewley_cosmic_2013}
Kewley, L.~J., Maier, C., Yabe, K., {et~al.} 2013, \apjl, 774, L10

\bibitem[{{Kobayashi} \& {Taylor}(2023)}]{kobayashi_taylor_chemodyn_2023}
{Kobayashi}, C. \& {Taylor}, P. 2023, arXiv e-prints, arXiv:2302.07255

\bibitem[{{Langan} {et~al.}(2020){Langan}, {Ceverino}, \&
  {Finlator}}]{langan_theo_mzr_2020}
{Langan}, I., {Ceverino}, D., \& {Finlator}, K. 2020, \mnras, 494, 1988

\bibitem[{{Langeroodi} \& {Hjorth}(2023)}]{langeroodi_fmr_compactness_2023}
{Langeroodi}, D. \& {Hjorth}, J. 2023, arXiv e-prints, arXiv:2307.06336

\bibitem[{{Langeroodi} {et~al.}(2022){Langeroodi}, {Hjorth}, {Chen}, {Kelly},
  {Williams}, {Lin}, {Scarlata}, {Zitrin}, {Broadhurst}, {Diego}, {Huang},
  {Filippenko}, {Foley}, {Jha}, {Koekemoer}, {Oguri}, {Perez-Fournon},
  {Pierel}, {Poidevin}, \& {Strolger}}]{langeroodi_mzr_2022}
{Langeroodi}, D., {Hjorth}, J., {Chen}, W., {et~al.} 2022, arXiv e-prints,
  arXiv:2212.02491

\bibitem[{{Laseter} {et~al.}(2023){Laseter}, {Maseda}, {Curti}, {Maiolino},
  {D'Eugenio}, {Cameron}, {Looser}, {Arribas}, {Baker}, {Bhatawdekar},
  {Boyett}, {Bunker}, {Carniani}, {Charlot}, {Chevallard}, {Curtis-lake},
  {Egami}, {Eisenstein}, {Hainline}, {Hausen}, {Ji}, {Kumari}, {Perna},
  {Rawle}, {Rix}, {Robertson}, {Rodr{\'\i}guez Del Pino}, {Sandles}, {Scholtz},
  {Smit}, {Tacchella}, {{\"U}bler}, {Williams}, {Willott}, \&
  {Witstok}}]{laseter_auroral_jades_2023}
{Laseter}, I.~H., {Maseda}, M.~V., {Curti}, M., {et~al.} 2023, arXiv e-prints,
  arXiv:2306.03120

\bibitem[{Lee {et~al.}(2006)Lee, Skillman, Cannon, Jackson, Gehrz, Polomski, \&
  Woodward}]{lee_extending_2006}
Lee, H., Skillman, E.~D., Cannon, J.~M., {et~al.} 2006, \apj, 647, 970

\bibitem[{Lequeux {et~al.}(1979)Lequeux, Peimbert, Rayo, Serrano, \&
  Torres-Peimbert}]{lequeux_chemical_1979}
Lequeux, J., Peimbert, M., Rayo, J.~F., Serrano, A., \& Torres-Peimbert, S.
  1979, \aap, 80, 155

\bibitem[{{Li} {et~al.}(2022){Li}, {Cai}, {Bian}, {Lin}, {Li}, {Wu}, {Sun},
  {Zhang}, {Zou}, {Fan}, {Egami}, {Charlot}, {Bruzual}, \&
  {Chevallard}}]{Li_mzr_dwarfs_z3_2022}
{Li}, M., {Cai}, Z., {Bian}, F., {et~al.} 2022, arXiv e-prints,
  arXiv:2211.01382

\bibitem[{Lilly {et~al.}(2013)Lilly, Carollo, Pipino, Renzini, \&
  Peng}]{lilly_gas_2013}
Lilly, S.~J., Carollo, C.~M., Pipino, A., Renzini, A., \& Peng, Y. 2013, \apj,
  772, 119

\bibitem[{{Looser} {et~al.}(2023){Looser}, {D'Eugenio}, {Maiolino}, {Witstok},
  {Sandles}, {Curtis-Lake}, {Chevallard}, {Tacchella}, {Johnson}, {Baker},
  {Suess}, {Carniani}, {Ferruit}, {Arribas}, {Bonaventura}, {Bunker},
  {Cameron}, {Charlot}, {Curti}, {de Graaff}, {Maseda}, {Rawle}, {Rix},
  {Rodriguez Del Pino}, {Smit}, {{\"U}bler}, {Willott}, {Alberts}, {Egami},
  {Eisenstein}, {Endsley}, {Hausen}, {Rieke}, {Robertson}, {Shivaei},
  {Williams}, {Boyett}, {Chen}, {Ji}, {Jones}, {Kumari}, {Nelson}, {Perna},
  {Saxena}, \& {Scholtz}}]{looser_gs73_2023}
{Looser}, T.~J., {D'Eugenio}, F., {Maiolino}, R., {et~al.} 2023, arXiv
  e-prints, arXiv:2302.14155

\bibitem[{Ma {et~al.}(2016)Ma, Hopkins, Faucher-Giguère, Zolman, Muratov,
  Kereš, \& Quataert}]{ma_origin_2016}
Ma, X., Hopkins, P.~F., Faucher-Giguère, C.-A., {et~al.} 2016, \mnras, 456,
  2140

\bibitem[{{Maier} {et~al.}(2014){Maier}, {Lilly}, {Ziegler}, {Contini},
  {P{\'e}rez Montero}, {Peng}, \& {Balestra}}]{maier_fmr_z2_2014}
{Maier}, C., {Lilly}, S.~J., {Ziegler}, B.~L., {et~al.} 2014, \apj, 792, 3

\bibitem[{Maiolino \& Mannucci(2019)}]{maiolino_re_2019}
Maiolino, R. \& Mannucci, F. 2019, \aapr, 27, 3

\bibitem[{Maiolino {et~al.}(2008)Maiolino, Nagao, Grazian, Cocchia, Marconi,
  Mannucci, Cimatti, Pipino, Ballero, Calura, Chiappini, Fontana, Granato,
  Matteucci, Pastorini, Pentericci, Risaliti, Salvati, \&
  Silva}]{maiolino_amaze_2008}
Maiolino, R., Nagao, T., Grazian, A., {et~al.} 2008, \aap, 488, 463

\bibitem[{{Maiolino} {et~al.}(2023{\natexlab{a}}){Maiolino}, {Scholtz},
  {Curtis-Lake}, {Carniani}, {Baker}, {de Graaff}, {Tacchella}, {{\"U}bler},
  {D'Eugenio}, {Witstok}, {Curti}, {Arribas}, {Bunker}, {Charlot},
  {Chevallard}, {Eisenstein}, {Egami}, {Ji}, {Jones}, {Lyu}, {Rawle},
  {Robertson}, {Rujopakarn}, {Perna}, {Sun}, {Venturi}, {Williams}, \&
  {Willott}}]{maiolino_jades_agn_2023}
{Maiolino}, R., {Scholtz}, J., {Curtis-Lake}, E., {et~al.} 2023{\natexlab{a}},
  arXiv e-prints, arXiv:2308.01230

\bibitem[{{Maiolino} {et~al.}(2023{\natexlab{b}}){Maiolino}, {Scholtz},
  {Witstok}, {Carniani}, {D'Eugenio}, {de Graaff}, {Uebler}, {Tacchella},
  {Curtis-Lake}, {Arribas}, {Bunker}, {Charlot}, {Chevallard}, {Curti},
  {Looser}, {Maseda}, {Rawle}, {Rodriguez Del Pino}, {Willott}, {Egami},
  {Eisenstein}, {Hainline}, {Robertson}, {Williams}, {Willmer}, {Baker},
  {Boyett}, {DeCoursey}, {Fabian}, {Helton}, {Ji}, {Jones}, {Kumari},
  {Laporte}, {Nelson}, {Perna}, {Sandles}, {Shivaei}, \&
  {Sun}}]{maiolino_gnz11_2023}
{Maiolino}, R., {Scholtz}, J., {Witstok}, J., {et~al.} 2023{\natexlab{b}},
  arXiv e-prints, arXiv:2305.12492

\bibitem[{Mannucci {et~al.}(2010)Mannucci, Cresci, Maiolino, Marconi, \&
  Gnerucci}]{mannucci_fundamental_2010}
Mannucci, F., Cresci, G., Maiolino, R., Marconi, A., \& Gnerucci, A. 2010,
  \mnras, 408, 2115

\bibitem[{Mannucci {et~al.}(2009)Mannucci, Cresci, Maiolino, Marconi,
  Pastorini, Pozzetti, Gnerucci, Risaliti, Schneider, Lehnert, \&
  Salvati}]{mannucci_lsd_2009}
Mannucci, F., Cresci, G., Maiolino, R., {et~al.} 2009, \mnras, 398, 1915

\bibitem[{{Mascia} {et~al.}(2023){Mascia}, {Pentericci}, {Calabro'}, {Treu},
  {Santini}, {Yang}, {Napolitano}, {Roberts-Borsani}, {Bergamini}, {Grillo},
  {Rosati}, {Vulcani}, {Castellano}, {Boyett}, {Fontana}, {Glazebrook},
  {Henry}, {Mason}, {Merlin}, {Morishita}, {Nanayakkara}, {Paris}, {Roy},
  {Williams}, {Wang}, {Brammer}, {Bradac}, {Chen}, {Kelly}, {Koekemoer},
  {Trenti}, \& {Windhorst}}]{mascia_reionisation_2023}
{Mascia}, S., {Pentericci}, L., {Calabro'}, A., {et~al.} 2023, arXiv e-prints,
  arXiv:2301.02816

\bibitem[{{Matthee} {et~al.}(2022){Matthee}, {Mackenzie}, {Simcoe}, {Kashino},
  {Lilly}, {Bordoloi}, \& {Eilers}}]{matthee_eiger_2023}
{Matthee}, J., {Mackenzie}, R., {Simcoe}, R.~A., {et~al.} 2022, arXiv e-prints,
  arXiv:2211.08255

\bibitem[{Nakajima \& Ouchi(2014)}]{nakajima_ionization_2014}
Nakajima, K. \& Ouchi, M. 2014, \mnras, 442, 900

\bibitem[{{Nakajima} {et~al.}(2023){Nakajima}, {Ouchi}, {Isobe}, {Harikane},
  {Zhang}, {Ono}, {Umeda}, \& {Oguri}}]{nakajima_mzr_ceers_2023}
{Nakajima}, K., {Ouchi}, M., {Isobe}, Y., {et~al.} 2023, arXiv e-prints,
  arXiv:2301.12825

\bibitem[{{Nakajima} {et~al.}(2022){Nakajima}, {Ouchi}, {Xu}, {Rauch},
  {Harikane}, {Nishigaki}, {Isobe}, {Kusakabe}, {Nagao}, {Ono}, {Onodera},
  {Sugahara}, {Kim}, {Komiyama}, {Lee}, \& {Zahedy}}]{nakajima_empress_2022}
{Nakajima}, K., {Ouchi}, M., {Xu}, Y., {et~al.} 2022, arXiv e-prints,
  arXiv:2206.02824

\bibitem[{Noeske {et~al.}(2007)Noeske, Faber, Weiner, Koo, Primack, Dekel,
  Papovich, Conselice, Le~Floc'h, Rieke, Coil, Lotz, Somerville, \&
  Bundy}]{noeske_star_2007}
Noeske, K.~G., Faber, S.~M., Weiner, B.~J., {et~al.} 2007, \apjl, 660, L47

\bibitem[{{Oesch} {et~al.}(2016){Oesch}, {Brammer}, {van Dokkum},
  {Illingworth}, {Bouwens}, {Labb{\'e}}, {Franx}, {Momcheva}, {Ashby}, {Fazio},
  {Gonzalez}, {Holden}, {Magee}, {Skelton}, {Smit}, {Spitler}, {Trenti}, \&
  {Willner}}]{oesch_gnz11_2016}
{Oesch}, P.~A., {Brammer}, G., {van Dokkum}, P.~G., {et~al.} 2016, \apj, 819,
  129

\bibitem[{Osterbrock \& Ferland(2006)}]{osterbrock_astrophysics_2006}
Osterbrock, D.~E. \& Ferland, G.~J. 2006, Astrophysics of {Gaseous} {Nebulae}
  and {Active} {Galactic} {Nuclei}, 2nd edn. (University Science Books)

\bibitem[{{Pallottini} {et~al.}(2022){Pallottini}, {Ferrara}, {Gallerani},
  {Behrens}, {Kohandel}, {Carniani}, {Vallini}, {Salvadori}, {Gelli},
  {Sommovigo}, {D'Odorico}, {Di Mascia}, \& {Pizzati}}]{pallottini_serra_2022}
{Pallottini}, A., {Ferrara}, A., {Gallerani}, S., {et~al.} 2022, \mnras, 513,
  5621

\bibitem[{{Peeples} \& {Shankar}(2011)}]{peeples_mzr_model_2011}
{Peeples}, M.~S. \& {Shankar}, F. 2011, \mnras, 417, 2962

\bibitem[{{P{\'e}roux} {et~al.}(2020){P{\'e}roux}, {Nelson}, {van de Voort},
  {Pillepich}, {Marinacci}, {Vogelsberger}, \& {Hernquist}}]{peroux_2020}
{P{\'e}roux}, C., {Nelson}, D., {van de Voort}, F., {et~al.} 2020, \mnras, 499,
  2462

\bibitem[{Pettini \& Pagel(2004)}]{pettini_oiiinii_2004}
Pettini, M. \& Pagel, B. E.~J. 2004, \mnras, 348, L59

\bibitem[{Pilyugin {et~al.}(2009)Pilyugin, Mattsson, Vílchez, \&
  Cedrés}]{pilyugin_electron_2009}
Pilyugin, L.~S., Mattsson, L., Vílchez, J.~M., \& Cedrés, B. 2009, \mnras,
  398, 485

\bibitem[{{Planck Collaboration} {et~al.}(2020){Planck Collaboration},
  {Aghanim}, {Akrami}, {Ashdown}, {Aumont}, {Baccigalupi}, {Ballardini},
  {Banday}, {Barreiro}, {Bartolo}, {Basak}, {Battye}, {Benabed}, {Bernard},
  {Bersanelli}, {Bielewicz}, {Bock}, {Bond}, {Borrill}, {Bouchet}, {Boulanger},
  {Bucher}, {Burigana}, {Butler}, {Calabrese}, {Cardoso}, {Carron},
  {Challinor}, {Chiang}, {Chluba}, {Colombo}, {Combet}, {Contreras}, {Crill},
  {Cuttaia}, {de Bernardis}, {de Zotti}, {Delabrouille}, {Delouis}, {Di
  Valentino}, {Diego}, {Dor{\'e}}, {Douspis}, {Ducout}, {Dupac}, {Dusini},
  {Efstathiou}, {Elsner}, {En{\ss}lin}, {Eriksen}, {Fantaye}, {Farhang},
  {Fergusson}, {Fernandez-Cobos}, {Finelli}, {Forastieri}, {Frailis},
  {Fraisse}, {Franceschi}, {Frolov}, {Galeotta}, {Galli}, {Ganga},
  {G{\'e}nova-Santos}, {Gerbino}, {Ghosh}, {Gonz{\'a}lez-Nuevo}, {G{\'o}rski},
  {Gratton}, {Gruppuso}, {Gudmundsson}, {Hamann}, {Handley}, {Hansen},
  {Herranz}, {Hildebrandt}, {Hivon}, {Huang}, {Jaffe}, {Jones}, {Karakci},
  {Keih{\"a}nen}, {Keskitalo}, {Kiiveri}, {Kim}, {Kisner}, {Knox},
  {Krachmalnicoff}, {Kunz}, {Kurki-Suonio}, {Lagache}, {Lamarre}, {Lasenby},
  {Lattanzi}, {Lawrence}, {Le Jeune}, {Lemos}, {Lesgourgues}, {Levrier},
  {Lewis}, {Liguori}, {Lilje}, {Lilley}, {Lindholm}, {L{\'o}pez-Caniego},
  {Lubin}, {Ma}, {Mac{\'\i}as-P{\'e}rez}, {Maggio}, {Maino}, {Mandolesi},
  {Mangilli}, {Marcos-Caballero}, {Maris}, {Martin}, {Martinelli},
  {Mart{\'\i}nez-Gonz{\'a}lez}, {Matarrese}, {Mauri}, {McEwen}, {Meinhold},
  {Melchiorri}, {Mennella}, {Migliaccio}, {Millea}, {Mitra},
  {Miville-Desch{\^e}nes}, {Molinari}, {Montier}, {Morgante}, {Moss}, {Natoli},
  {N{\o}rgaard-Nielsen}, {Pagano}, {Paoletti}, {Partridge}, {Patanchon},
  {Peiris}, {Perrotta}, {Pettorino}, {Piacentini}, {Polastri}, {Polenta},
  {Puget}, {Rachen}, {Reinecke}, {Remazeilles}, {Renzi}, {Rocha}, {Rosset},
  {Roudier}, {Rubi{\~n}o-Mart{\'\i}n}, {Ruiz-Granados}, {Salvati}, {Sandri},
  {Savelainen}, {Scott}, {Shellard}, {Sirignano}, {Sirri}, {Spencer},
  {Sunyaev}, {Suur-Uski}, {Tauber}, {Tavagnacco}, {Tenti}, {Toffolatti},
  {Tomasi}, {Trombetti}, {Valenziano}, {Valiviita}, {Van Tent}, {Vibert},
  {Vielva}, {Villa}, {Vittorio}, {Wandelt}, {Wehus}, {White}, {White},
  {Zacchei}, \& {Zonca}}]{planck_2020}
{Planck Collaboration}, {Aghanim}, N., {Akrami}, Y., {et~al.} 2020, \aap, 641,
  A6

\bibitem[{{Pontoppidan} {et~al.}(2022){Pontoppidan}, {Blome}, {Braun}, {Brown},
  {Carruthers}, {Coe}, {DePasquale}, {Espinoza}, {Garcia Marin}, {Gordon},
  {Henry}, {Hustak}, {James}, {Koekemoer}, {LaMassa}, {Law}, {Lockwood},
  {Moro-Martin}, {Mullally}, {Pagan}, {Player}, {Proffitt}, {Pulliam},
  {Ramsay}, {Ravindranath}, {Reid}, {Robberto}, {Sabbi}, \&
  {Ubeda}}]{pontoppidan_ERO_2022}
{Pontoppidan}, K., {Blome}, C., {Braun}, H., {et~al.} 2022, arXiv e-prints,
  arXiv:2207.13067

\bibitem[{{Popesso} {et~al.}(2023){Popesso}, {Concas}, {Cresci}, {Belli},
  {Rodighiero}, {Inami}, {Dickinson}, {Ilbert}, {Pannella}, \&
  {Elbaz}}]{popesso_SFMS_2023}
{Popesso}, P., {Concas}, A., {Cresci}, G., {et~al.} 2023, \mnras, 519, 1526

\bibitem[{{Reddy} {et~al.}(2022){Reddy}, {Topping}, {Shapley}, {Steidel},
  {Sanders}, {Du}, {Coil}, {Mobasher}, {Price}, \& {Shivaei}}]{reddy_lyA_2022}
{Reddy}, N.~A., {Topping}, M.~W., {Shapley}, A.~E., {et~al.} 2022, \apj, 926,
  31

\bibitem[{Renzini \& Peng(2015)}]{renzini_objective_2015}
Renzini, A. \& Peng, Y.-j. 2015, \apjl, 801, L29

\bibitem[{{Rhoads} {et~al.}(2023){Rhoads}, {Wold}, {Harish}, {Kim}, {Pharo},
  {Malhotra}, {Gabrielpillai}, {Jiang}, \& {Yang}}]{rhoads_2023}
{Rhoads}, J.~E., {Wold}, I. G.~B., {Harish}, S., {et~al.} 2023, \apjl, 942, L14

\bibitem[{{Rieke} \& {the JADES Collaboration}(2023)}]{rieke_jades_DR_2023}
{Rieke}, M. \& {the JADES Collaboration}. 2023, arXiv e-prints,
  arXiv:2306.02466

\bibitem[{{Robertson} {et~al.}(2022){Robertson}, {Tacchella}, {Johnson},
  {Hainline}, {Whitler}, {Eisenstein}, {Endsley}, {Rieke}, {Stark}, {Alberts},
  {Dressler}, {Egami}, {Hausen}, {Rieke}, {Shivaei}, {Williams}, {Willmer},
  {Arribas}, {Bonaventura}, {Bunker}, {Cameron}, {Carniani}, {Charlot},
  {Chevallard}, {Curti}, {Curtis-Lake}, {D'Eugenio}, {Jakobsen}, {Looser},
  {L{\"u}tzgendorf}, {Maiolino}, {Maseda}, {Rawle}, {Rix}, {Smit}, {{\"U}bler},
  {Willott}, {Witstok}, {Baum}, {Bhatawdekar}, {Boyett}, {Chen}, {de Graaff},
  {Florian}, {Helton}, {Hviding}, {Ji}, {Kumari}, {Lyu}, {Nelson}, {Sandles},
  {Saxena}, {Suess}, {Sun}, {Topping}, \& {Wallace}}]{robertson_jades_2022}
{Robertson}, B.~E., {Tacchella}, S., {Johnson}, B.~D., {et~al.} 2022, arXiv
  e-prints, arXiv:2212.04480

\bibitem[{{Saintonge} {et~al.}(2016){Saintonge}, {Catinella}, {Cortese},
  {Genzel}, {Giovanelli}, {Haynes}, {Janowiecki}, {Kramer}, {Lutz},
  {Schiminovich}, {Tacconi}, {Wuyts}, \& {Accurso}}]{saintonge_2016}
{Saintonge}, A., {Catinella}, B., {Cortese}, L., {et~al.} 2016, \mnras, 462,
  1749

\bibitem[{{Saintonge} {et~al.}(2017){Saintonge}, {Catinella}, {Tacconi},
  {Kauffmann}, {Genzel}, {Cortese}, {Dav{\'e}}, {Fletcher},
  {Graci{\'a}-Carpio}, {Kramer}, {Heckman}, {Janowiecki}, {Lutz}, {Rosario},
  {Schiminovich}, {Schuster}, {Wang}, {Wuyts}, {Borthakur}, {Lamperti}, \&
  {Roberts-Borsani}}]{saintonge_xcoldgass_2017}
{Saintonge}, A., {Catinella}, B., {Tacconi}, L.~J., {et~al.} 2017, \apjs, 233,
  22

\bibitem[{Salim {et~al.}(2015)Salim, Lee, Davé, \&
  Dickinson}]{salim_mass-metallicity-star_2015}
Salim, S., Lee, J.~C., Davé, R., \& Dickinson, M. 2015, \apj, 808, 25

\bibitem[{{Sanders} {et~al.}(2021){Sanders}, {Shapley}, {Jones}, {Reddy},
  {Kriek}, {Siana}, {Coil}, {Mobasher}, {Shivaei}, {Dav{\'e}}, {Azadi},
  {Price}, {Leung}, {Freeman}, {Fetherolf}, {de Groot}, {Zick}, \&
  {Barro}}]{sanders_mosdef_mzr_2021}
{Sanders}, R.~L., {Shapley}, A.~E., {Jones}, T., {et~al.} 2021, \apj, 914, 19

\bibitem[{Sanders {et~al.}(2018)Sanders, Shapley, Kriek, Freeman, Reddy, Siana,
  Coil, Mobasher, Davé, Shivaei, Azadi, Price, Leung, Fetherholf, de~Groot,
  Zick, Fornasini, \& Barro}]{sanders_mosdef_2018}
Sanders, R.~L., Shapley, A.~E., Kriek, M., {et~al.} 2018, \apj, 858, 99

\bibitem[{{Sanders} {et~al.}(2023{\natexlab{a}}){Sanders}, {Shapley},
  {Topping}, {Reddy}, \& {Brammer}}]{sanders_calibrations_2023}
{Sanders}, R.~L., {Shapley}, A.~E., {Topping}, M.~W., {Reddy}, N.~A., \&
  {Brammer}, G.~B. 2023{\natexlab{a}}, arXiv e-prints, arXiv:2303.08149

\bibitem[{{Sanders} {et~al.}(2023{\natexlab{b}}){Sanders}, {Shapley},
  {Topping}, {Reddy}, \& {Brammer}}]{sanders_ceers_2023}
{Sanders}, R.~L., {Shapley}, A.~E., {Topping}, M.~W., {Reddy}, N.~A., \&
  {Brammer}, G.~B. 2023{\natexlab{b}}, arXiv e-prints, arXiv:2301.06696

\bibitem[{{Schaerer} {et~al.}(2022){Schaerer}, {Marques-Chaves}, {Barrufet},
  {Oesch}, {Izotov}, {Naidu}, {Guseva}, \& {Brammer}}]{Schaerer_ero_2022}
{Schaerer}, D., {Marques-Chaves}, R., {Barrufet}, L., {et~al.} 2022, \aap, 665,
  L4

\bibitem[{{Scoville} {et~al.}(2017){Scoville}, {Lee}, {Vanden Bout},
  {Diaz-Santos}, {Sanders}, {Darvish}, {Bongiorno}, {Casey}, {Murchikova},
  {Koda}, {Capak}, {Vlahakis}, {Ilbert}, {Sheth}, {Morokuma-Matsui}, {Ivison},
  {Aussel}, {Laigle}, {McCracken}, {Armus}, {Pope}, {Toft}, \&
  {Masters}}]{scoville_2017}
{Scoville}, N., {Lee}, N., {Vanden Bout}, P., {et~al.} 2017, \apj, 837, 150

\bibitem[{Shapley {et~al.}(2005)Shapley, Coil, Ma, \&
  Bundy}]{shapley_chemical_2005}
Shapley, A.~E., Coil, A.~L., Ma, C.-P., \& Bundy, K. 2005, \apj, 635, 1006

\bibitem[{{Shapley} {et~al.}(2023{\natexlab{a}}){Shapley}, {Reddy}, {Sanders},
  {Topping}, \& {Brammer}}]{shapley_mzr_2023}
{Shapley}, A.~E., {Reddy}, N.~A., {Sanders}, R.~L., {Topping}, M.~W., \&
  {Brammer}, G.~B. 2023{\natexlab{a}}, arXiv e-prints, arXiv:2303.00410

\bibitem[{{Shapley} {et~al.}(2023{\natexlab{b}}){Shapley}, {Sanders}, {Reddy},
  {Topping}, \& {Brammer}}]{shapley_balmer_2023}
{Shapley}, A.~E., {Sanders}, R.~L., {Reddy}, N.~A., {Topping}, M.~W., \&
  {Brammer}, G.~B. 2023{\natexlab{b}}, arXiv e-prints, arXiv:2301.03241

\bibitem[{{Somerville} \& {Dav{\'e}}(2015)}]{somerville_dave_2015}
{Somerville}, R.~S. \& {Dav{\'e}}, R. 2015, \araa, 53, 51

\bibitem[{Speagle {et~al.}(2014)Speagle, Steinhardt, Capak, \&
  Silverman}]{speagle_highly_2014}
Speagle, J.~S., Steinhardt, C.~L., Capak, P.~L., \& Silverman, J.~D. 2014,
  \apjs, 214, 15

\bibitem[{Steidel {et~al.}(2014)Steidel, Rudie, Strom, Pettini, Reddy, Shapley,
  Trainor, Erb, Turner, Konidaris, Kulas, Mace, Matthews, \&
  McLean}]{steidel_strong_2014}
Steidel, C.~C., Rudie, G.~C., Strom, A.~L., {et~al.} 2014, \apj, 795, 165

\bibitem[{Strom {et~al.}(2017)Strom, Steidel, Rudie, Trainor, Pettini, \&
  Reddy}]{strom_nebular_2017}
Strom, A.~L., Steidel, C.~C., Rudie, G.~C., {et~al.} 2017, \apj, 836, 164

\bibitem[{{Tacchella} {et~al.}(2023){Tacchella}, {Eisenstein}, {Hainline},
  {Johnson}, {Baker}, {Helton}, {Robertson}, {Suess}, {Chen}, {Nelson},
  {Pusk{\'a}s}, {Sun}, {Alberts}, {Egami}, {Hausen}, {Rieke}, {Rieke},
  {Shivaei}, {Williams}, {Willmer}, {Bunker}, {Cameron}, {Carniani}, {Charlot},
  {Curti}, {Curtis-Lake}, {Looser}, {Maiolino}, {Maseda}, {Rawle}, {Rix},
  {Smit}, {{\"U}bler}, {Willott}, {Witstok}, {Baum}, {Bhatawdekar}, {Boyett},
  {Danhaive}, {de Graaff}, {Endsley}, {Ji}, {Lyu}, {Sandles}, {Saxena},
  {Scholtz}, {Topping}, \& {Whitler}}]{tacchella_gnz11_2023}
{Tacchella}, S., {Eisenstein}, D.~J., {Hainline}, K., {et~al.} 2023, arXiv
  e-prints, arXiv:2302.07234

\bibitem[{{Tacchella} {et~al.}(2022){Tacchella}, {Johnson}, {Robertson},
  {Carniani}, {D'Eugenio}, {Kumar}, {Maiolino}, {Nelson}, {Suess}, {{\"U}bler},
  {Williams}, {Adebusola}, {Alberts}, {Arribas}, {Bhatawdekar}, {Bonaventura},
  {Bowler}, {Bunker}, {Cameron}, {Curti}, {Egami}, {Eisenstein}, {Frye},
  {Hainline}, {Helton}, {Ji}, {Looser}, {Lyu}, {Perna}, {Rawle}, {Rieke},
  {Rieke}, {Saxena}, {Sandles}, {Shivaei}, {Simmonds}, {Sun}, {Willmer},
  {Willott}, \& {Witstok}}]{tacchella_eros_2022}
{Tacchella}, S., {Johnson}, B.~D., {Robertson}, B.~E., {et~al.} 2022, arXiv
  e-prints, arXiv:2208.03281

\bibitem[{Tacconi {et~al.}(2018)Tacconi, Genzel, Saintonge, Combes,
  García-Burillo, Neri, Bolatto, Contini, Förster~Schreiber, Lilly, Lutz,
  Wuyts, Accurso, Boissier, Boone, Bouché, Bournaud, Burkert, Carollo, Cooper,
  Cox, Feruglio, Freundlich, Herrera-Camus, Juneau, Lippa, Naab, Renzini,
  Salome, Sternberg, Tadaki, Übler, Walter, Weiner, \&
  Weiss}]{tacconi_phibss_2018}
Tacconi, L.~J., Genzel, R., Saintonge, A., {et~al.} 2018, \apj, 853, 179

\bibitem[{{Tacconi} {et~al.}(2020){Tacconi}, {Genzel}, \&
  {Sternberg}}]{tacconi_review_2020}
{Tacconi}, L.~J., {Genzel}, R., \& {Sternberg}, A. 2020, \araa, 58, 157

\bibitem[{{Taylor} {et~al.}(2022){Taylor}, {Barger}, \& {Cowie}}]{taylor_2022}
{Taylor}, A.~J., {Barger}, A.~J., \& {Cowie}, L.~L. 2022, \apjl, 939, L3

\bibitem[{Telford {et~al.}(2016)Telford, Dalcanton, Skillman, \&
  Conroy}]{telford_exploring_2016}
Telford, O.~G., Dalcanton, J.~J., Skillman, E.~D., \& Conroy, C. 2016, \apj,
  827, 35

\bibitem[{{Topping} {et~al.}(2021){Topping}, {Shapley}, {Sanders}, {Kriek},
  {Reddy}, {Coil}, {Mobasher}, {Siana}, {Freeman}, {Shivaei}, {Azadi}, {Price},
  {Leung}, {Fetherolf}, {de Groot}, {Zick}, {Fornasini}, {Barro}, \&
  {Runco}}]{topping_mzr_2021}
{Topping}, M.~W., {Shapley}, A.~E., {Sanders}, R.~L., {et~al.} 2021, \mnras,
  506, 1237

\bibitem[{{Torrey} {et~al.}(2019){Torrey}, {Vogelsberger}, {Marinacci},
  {Pakmor}, {Springel}, {Nelson}, {Naiman}, {Pillepich}, {Genel}, {Weinberger},
  \& {Hernquist}}]{torrey_mzr_TNG_2019}
{Torrey}, P., {Vogelsberger}, M., {Marinacci}, F., {et~al.} 2019, \mnras, 484,
  5587

\bibitem[{Tremonti {et~al.}(2004)Tremonti, Heckman, Kauffmann, Brinchmann,
  Charlot, White, Seibert, Peng, Schlegel, Uomoto, Fukugita, \&
  Brinkmann}]{tremonti_origin_2004}
Tremonti, C.~A., Heckman, T.~M., Kauffmann, G., {et~al.} 2004, \apj, 613, 898

\bibitem[{{Treu} {et~al.}(2022){Treu}, {Roberts-Borsani}, {Bradac}, {Brammer},
  {Fontana}, {Henry}, {Mason}, {Morishita}, {Pentericci}, {Wang}, {Acebron},
  {Bagley}, {Bergamini}, {Belfiori}, {Bonchi}, {Boyett}, {Boutsia},
  {Calabr{\'o}}, {Caminha}, {Castellano}, {Dressler}, {Glazebrook}, {Grillo},
  {Jacobs}, {Jones}, {Kelly}, {Leethochawalit}, {Malkan}, {Marchesini},
  {Mascia}, {Mercurio}, {Merlin}, {Nanayakkara}, {Nonino}, {Paris},
  {Poggianti}, {Rosati}, {Santini}, {Scarlata}, {Shipley}, {Strait}, {Trenti},
  {Tubthong}, {Vanzella}, {Vulcani}, \& {Yang}}]{treu_glass_survey_2022}
{Treu}, T., {Roberts-Borsani}, G., {Bradac}, M., {et~al.} 2022, \apj, 935, 110

\bibitem[{{Trump} {et~al.}(2022){Trump}, {Arrabal Haro}, {Simons}, {Backhaus},
  {Amor{\'\i}n}, {Dickinson}, {Fern{\'a}ndez}, {Papovich}, {Nicholls},
  {Kewley}, {Brunker}, {Salzer}, {Wilkins}, {Almaini}, {Bagley}, {Berg},
  {Bhatawdekar}, {Bisigello}, {Buat}, {Burgarella}, {Calabr{\`o}}, {Casey},
  {Ciesla}, {Cleri}, {Cole}, {Cooper}, {Cooray}, {Costantin}, {Ferguson},
  {Finkelstein}, {Fujimoto}, {Gardner}, {Gawiser}, {Giavalisco}, {Grazian},
  {Grogin}, {Hathi}, {Hirschmann}, {Holwerda}, {Huertas-Company}, {Hutchison},
  {Jogee}, {Juneau}, {Jung}, {Kartaltepe}, {Kirkpatrick}, {Koekemoer}, {Lotz},
  {Lucas}, {Magnelli}, {Matharu}, {P{\'e}rez-Gonz{\'a}lez}, {Pirzkal},
  {Rafelski}, {Rose}, {Seill{\'e}}, {Somerville}, {Straughn}, {Tacchella},
  {Vanderhoof}, {Weiner}, {Wuyts}, {Yung}, \& {Zavala}}]{trump_2022}
{Trump}, J.~R., {Arrabal Haro}, P., {Simons}, R.~C., {et~al.} 2022, arXiv
  e-prints, arXiv:2207.12388

\bibitem[{{{\"U}bler} {et~al.}(2023){{\"U}bler}, {Maiolino}, {Curtis-Lake},
  {P{\'e}rez-Gonz{\'a}lez}, {Curti}, {Perna}, {Arribas}, {Charlot}, {Marshall},
  {D'Eugenio}, {Scholtz}, {Bunker}, {Carniani}, {Ferruit}, {Jakobsen}, {Rix},
  {Rodr{\'\i}guez Del Pino}, {Willott}, {B{\"o}ker}, {Cresci}, {Jones},
  {Kumari}, \& {Rawle}}]{Ubler_AGN_z5_2023}
{{\"U}bler}, H., {Maiolino}, R., {Curtis-Lake}, E., {et~al.} 2023, arXiv
  e-prints, arXiv:2302.06647

\bibitem[{{Ucci} {et~al.}(2023){Ucci}, {Dayal}, {Hutter}, {Kobayashi},
  {Gottl{\"o}ber}, {Yepes}, {Hunt}, {Legrand}, \&
  {Tortora}}]{ucci_astraeus_2023}
{Ucci}, G., {Dayal}, P., {Hutter}, A., {et~al.} 2023, \mnras, 518, 3557

\bibitem[{{Vidal-Garc{\'\i}a} {et~al.}(2017){Vidal-Garc{\'\i}a}, {Charlot},
  {Bruzual}, \& {Hubeny}}]{vidal_garcia_2017}
{Vidal-Garc{\'\i}a}, A., {Charlot}, S., {Bruzual}, G., \& {Hubeny}, I. 2017,
  \mnras, 470, 3532

\bibitem[{Vincenzo {et~al.}(2016)Vincenzo, Matteucci, Belfiore, \&
  Maiolino}]{vincenzo_modern_2016}
Vincenzo, F., Matteucci, F., Belfiore, F., \& Maiolino, R. 2016, \mnras, 455,
  4183

\bibitem[{{Williams} {et~al.}(2022){Williams}, {Kelly}, {Chen}, {Brammer},
  {Zitrin}, {Treu}, {Scarlata}, {Koekemoer}, {Oguri}, {Lin}, {Diego}, {Nonino},
  {Hjorth}, {Langeroodi}, {Broadhurst}, {Rogers}, {Perez-Fournon}, {Foley},
  {Jha}, {Filippenko}, {Strolger}, {Pierel}, {Poidevin}, \&
  {Yang}}]{williams_z9_2022}
{Williams}, H., {Kelly}, P.~L., {Chen}, W., {et~al.} 2022, arXiv e-prints,
  arXiv:2210.15699

\bibitem[{{Witstok} {et~al.}(2021){Witstok}, {Smit}, {Maiolino}, {Curti},
  {Laporte}, {Massey}, {Richard}, \& {Swinbank}}]{witstok_lensed_z5_2021}
{Witstok}, J., {Smit}, R., {Maiolino}, R., {et~al.} 2021, \mnras, 508, 1686

\bibitem[{Wuyts {et~al.}(2014)Wuyts, Kurk, Förster~Schreiber, Genzel,
  Wisnioski, Bandara, Wuyts, Beifiori, Bender, Brammer, Burkert, Buschkamp,
  Carollo, Chan, Davies, Eisenhauer, Fossati, Kulkarni, Lang, Lilly, Lutz,
  Mancini, Mendel, Momcheva, Naab, Nelson, Renzini, Rosario, Saglia, Seitz,
  Sharples, Sternberg, Tacchella, Tacconi, van Dokkum, \&
  Wilman}]{wuyts_consistent_2014}
Wuyts, E., Kurk, J., Förster~Schreiber, N.~M., {et~al.} 2014, \apjl, 789, L40

\bibitem[{{Wuyts} {et~al.}(2012){Wuyts}, {Rigby}, {Sharon}, \&
  {Gladders}}]{wuyts_low_mass_mzr_2012}
{Wuyts}, E., {Rigby}, J.~R., {Sharon}, K., \& {Gladders}, M.~D. 2012, \apj,
  755, 73

\bibitem[{{Yabe} {et~al.}(2014){Yabe}, {Ohta}, {Iwamuro}, {Akiyama}, {Tamura},
  {Yuma}, {Kimura}, {Takato}, {Moritani}, {Sumiyoshi}, {Maihara}, {Silverman},
  {Dalton}, {Lewis}, {Bonfield}, {Lee}, {Curtis-Lake}, {Macaulay}, \&
  {Clarke}}]{yabe_mzr_z1_2014}
{Yabe}, K., {Ohta}, K., {Iwamuro}, F., {et~al.} 2014, \mnras, 437, 3647

\bibitem[{{Yang} {et~al.}(2017{\natexlab{a}}){Yang}, {Malhotra}, {Gronke},
  {Rhoads}, {Leitherer}, {Wofford}, {Jiang}, {Dijkstra}, {Tilvi}, \&
  {Wang}}]{yang_greenpeas_2017}
{Yang}, H., {Malhotra}, S., {Gronke}, M., {et~al.} 2017{\natexlab{a}}, \apj,
  844, 171

\bibitem[{{Yang} {et~al.}(2017{\natexlab{b}}){Yang}, {Malhotra}, {Rhoads}, \&
  {Wang}}]{yang_blueberries_2017}
{Yang}, H., {Malhotra}, S., {Rhoads}, J.~E., \& {Wang}, J. 2017{\natexlab{b}},
  \apj, 847, 38

\bibitem[{Yates {et~al.}(2012)Yates, Kauffmann, \& Guo}]{yates_relation_2012}
Yates, R.~M., Kauffmann, G., \& Guo, Q. 2012, \mnras, 422, 215

\bibitem[{Yates {et~al.}(2019)Yates, Schady, Chen, Schweyer, \&
  Wiseman}]{yates_present-day_2019}
Yates, R.~M., Schady, P., Chen, T.-W., Schweyer, T., \& Wiseman, P. 2019, arXiv
  e-prints

\bibitem[{Zahid {et~al.}(2012)Zahid, Bresolin, Kewley, Coil, \&
  Davé}]{zahid_metallicities_2012}
Zahid, H.~J., Bresolin, F., Kewley, L.~J., Coil, A.~L., \& Davé, R. 2012,
  \apj, 750, 120

\bibitem[{Zahid {et~al.}(2014)Zahid, Dima, Kudritzki, Kewley, Geller, Hwang,
  Silverman, \& Kashino}]{zahid_universal_2014}
Zahid, H.~J., Dima, G.~I., Kudritzki, R.-P., {et~al.} 2014, \apj, 791, 130

\bibitem[{Zahid {et~al.}(2011)Zahid, Kewley, \&
  Bresolin}]{zahid_mass-metallicity_2011}
Zahid, H.~J., Kewley, L.~J., \& Bresolin, F. 2011, \apj, 730, 137

\end{thebibliography}

\appendix

\section{Metallicity scaling relations under different assumptions on sample selection and derivation of galaxy properties}
\label{sec:appendix_A}

In the main body of the paper, we discuss the mass-metallicity (MZR) and the fundamental metallicity (FMR) relations for a combined sample of galaxies observed with the \emph{JWST} from different observational programmes. 
Here, we investigate whether any of the main results presented in the paper change if we adopt different choices than our set of fiducial assumptions, particularly regarding sample selection, SFR measurement, and FMR parametrisation.

\begin{figure*}[h!]
\begin{center}
    
\includegraphics[width=0.48\textwidth]{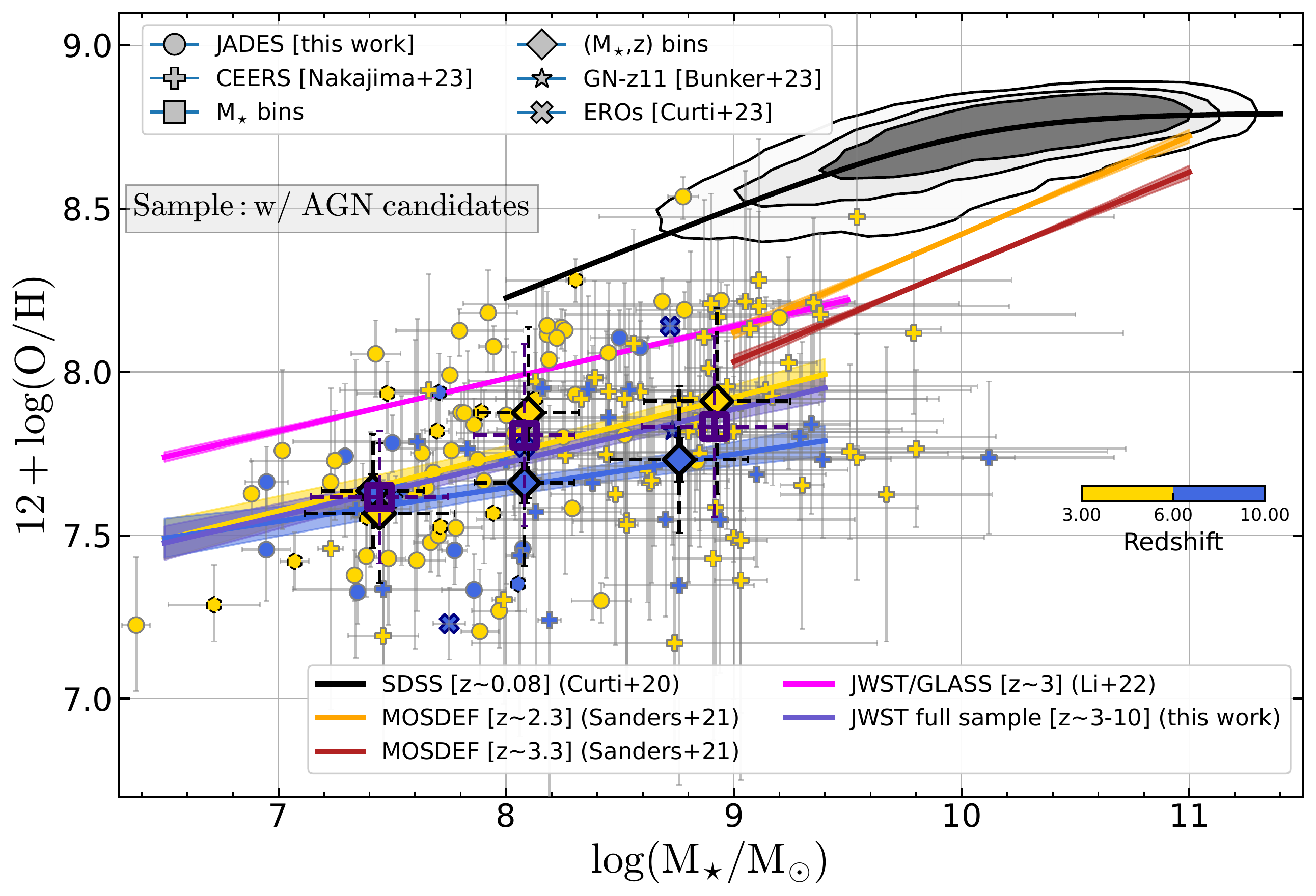}
\hspace{0.05cm}
\includegraphics[width=0.48\textwidth]{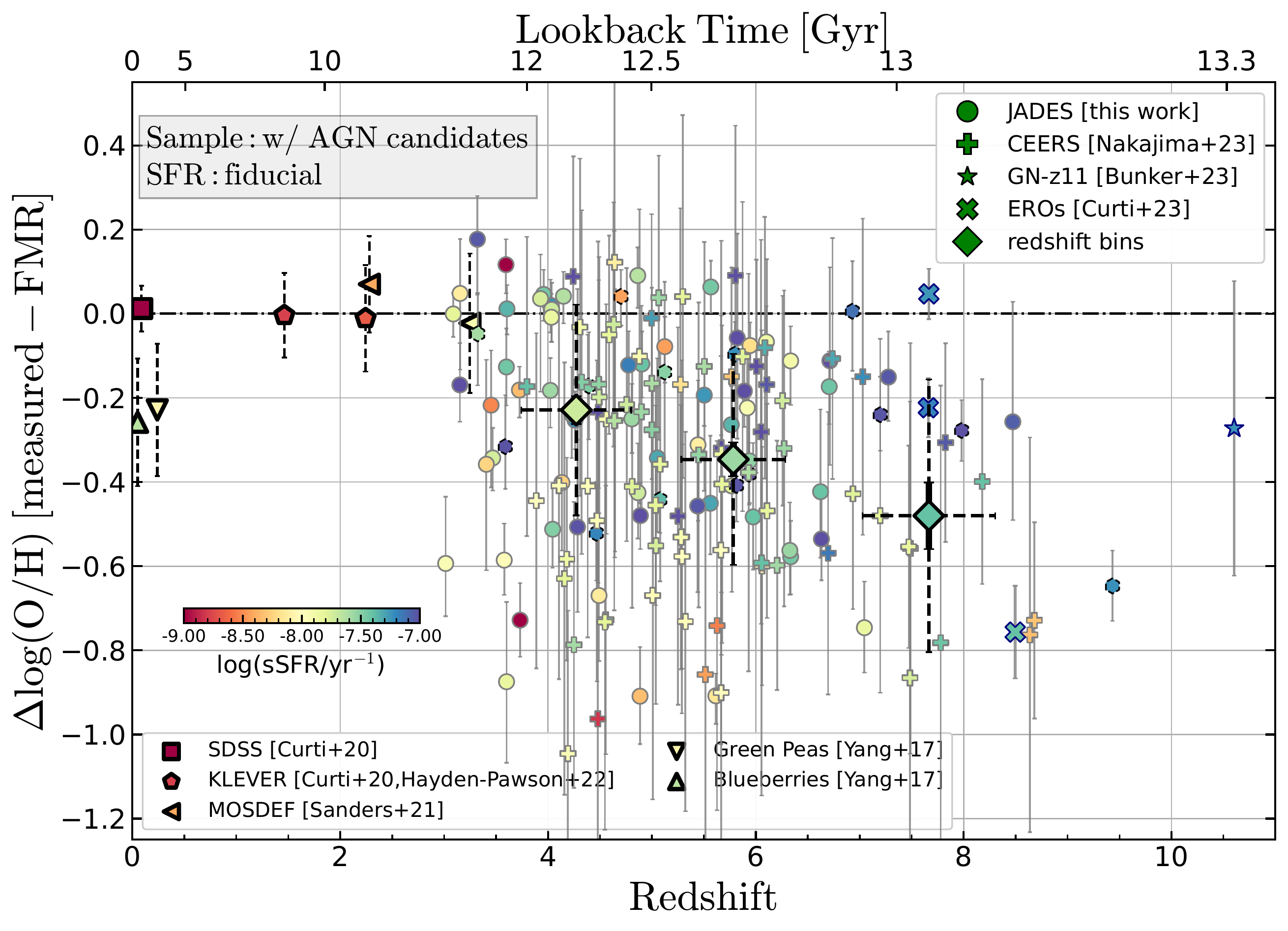}

\end{center}    
\caption{Similar to Figure~\ref{fig:mzr} (left-hand panel) and Figure~\ref{fig:fmr} (right-hand panel), but including the sample of $14$ (narrow-line) AGN candidates in \emph{JADES-deep} (Scholtz et al., in preparation), here marked as dashed hexagons. 
No clear change in the median trends of the metallicity scaling relations is found, and the main results reported in the paper are unaffected.
}
\label{fig:including_agns}
\end{figure*}

In Figure~\ref{fig:including_agns} we show the MZR (left-hand panel) and deviations from the FMR (right-hand panel) for our sample of galaxies which now includes $14$ objects from JADES identified as AGN candidates. For these galaxies, we here assume a mild contribution of AGN ionisation to the emission line spectrum, so that the standard metallicity calibrations adopted for ``normal'' star-forming galaxies can be applied. Stellar masses and SFRs are also derived assuming negligible AGN contribution to the continuum level. No significant change to the observed evolution in the metallicity scaling relations is found by including these objects. 

\begin{figure*}
\begin{center}   
\includegraphics[width=0.48\textwidth]{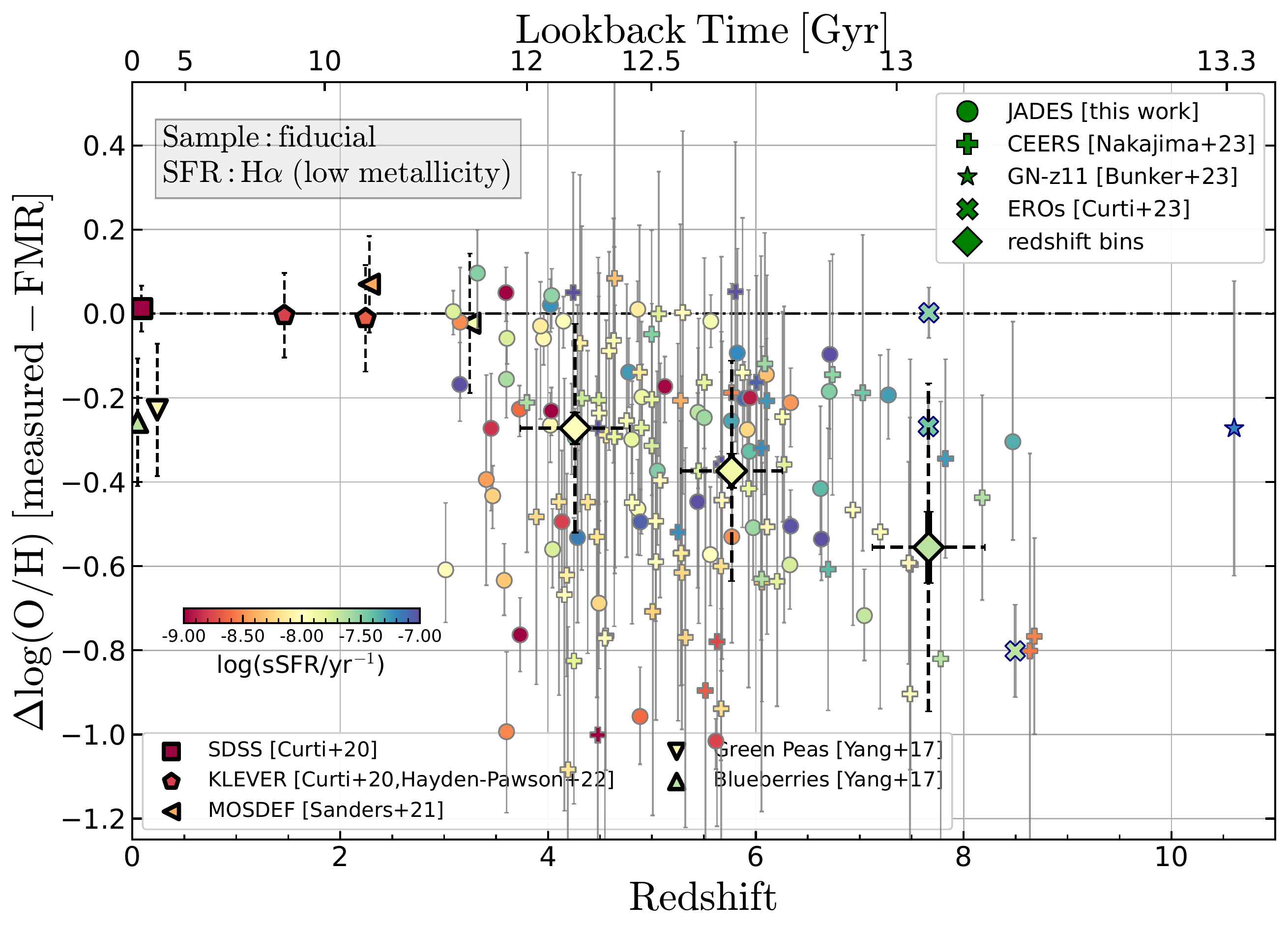}
\includegraphics[width=0.48\textwidth]{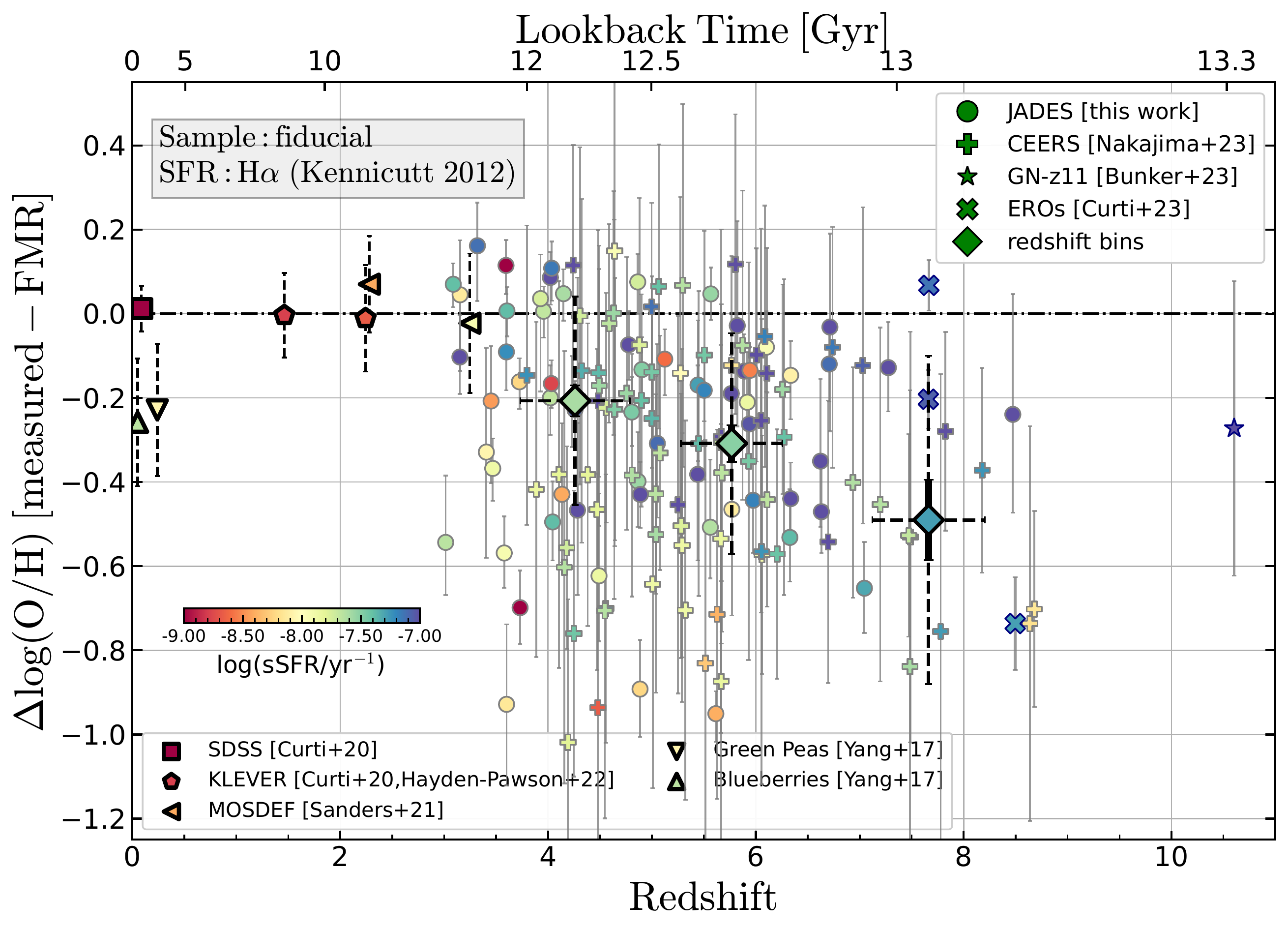}
\end{center}  
\caption{Same as Figure~\ref{fig:fmr}, but with SFR inferred from the (attenuation corrected) \Ha\ flux (or \Hb, at $z\gtrsim7$).
The calibration for low metallicity galaxies described in \citealt{reddy_lyA_2022, shapley_balmer_2023} is  adopted for the entire \emph{JWST} sample in the left-hand panel, while in the right-hand panel the SFR is derived following the calibration from \citealt{kennicutt_star_2012} for solar metallicity. }
\label{fig:fmr_sfr_HA}
\end{figure*}

In Figure~\ref{fig:fmr_sfr_HA} instead we report, similarly to Figure~\ref{fig:fmr}, the deviations from the predictions of the local FMR, but this time assuming different estimates for the SFR of our entire \emph{JWST} sample. In particular, we adopt the calibration for metal poor galaxies described in \cite{reddy_lyA_2022, shapley_balmer_2023} in the left-hand panel, whereas the calibration from \cite{kennicutt_star_2012}, and suited for solar metallicities, is assumed in the right-hand panel.
As already discussed in Section~\ref{sec:mass_sfr} of the main text, once estimated on JADES galaxies the former provides on average lower (s)SFRs by $0.14$~dex, whereas the latter delivers higher (s)SFR by $0.23$~dex, compared to our fiducial values based on \textsc{beagle} fitting to both PRISM spectra and photometry.
This means that the predicted metallicities from the FMR (hence $\Delta$log(O/H)) are, on average, higher (lower) than our fiducial case when the low-metallicity (solar metallicity) calibration to convert the \Ha\ flux to SFR is adopted.
Nonetheless, the median trends and the observed evolution in the FMR at high-z are robust against the choice of the SFR calibration.

\begin{figure*}
\begin{center}   
\includegraphics[width=0.33\textwidth]{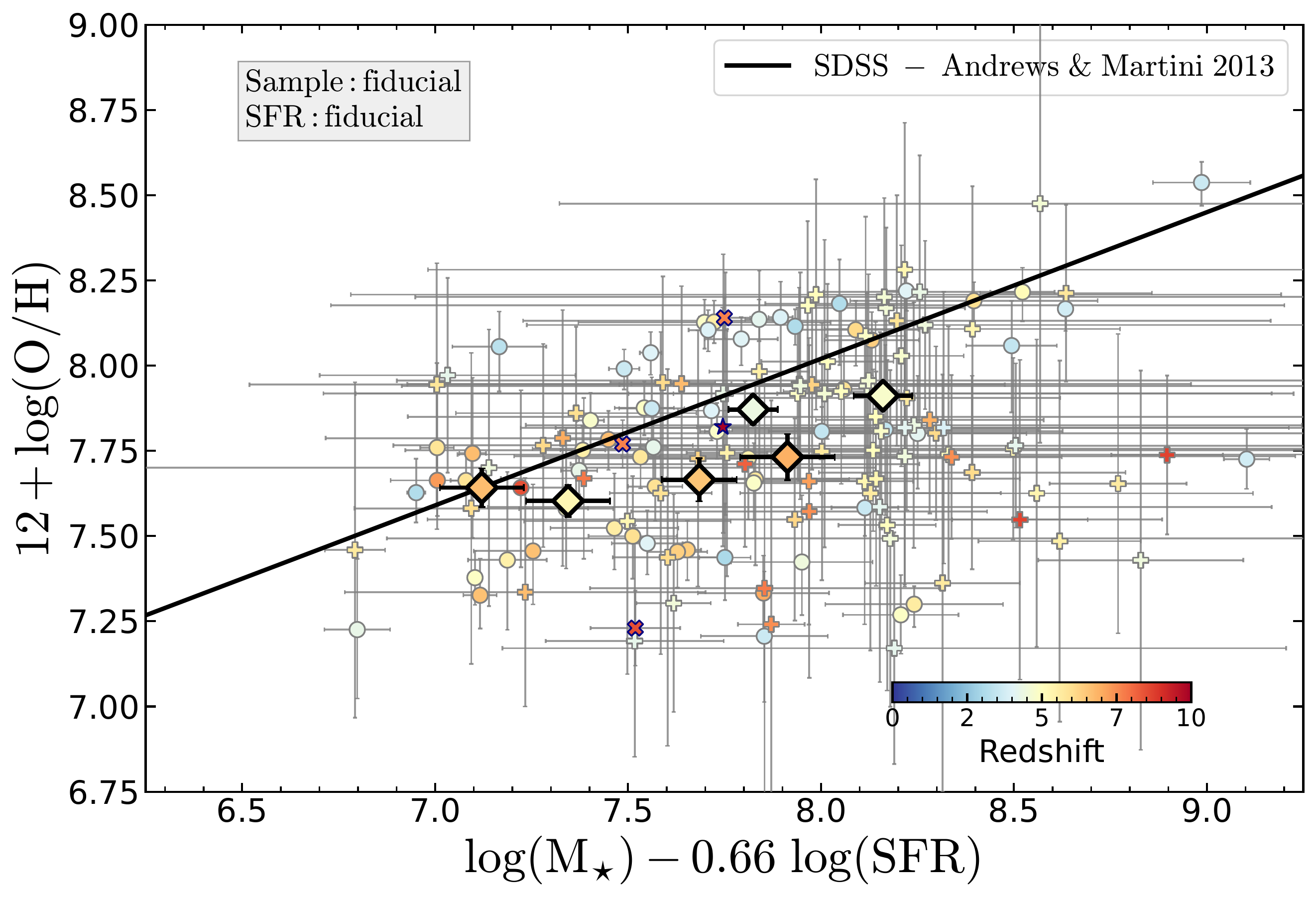}
\includegraphics[width=0.33\textwidth]{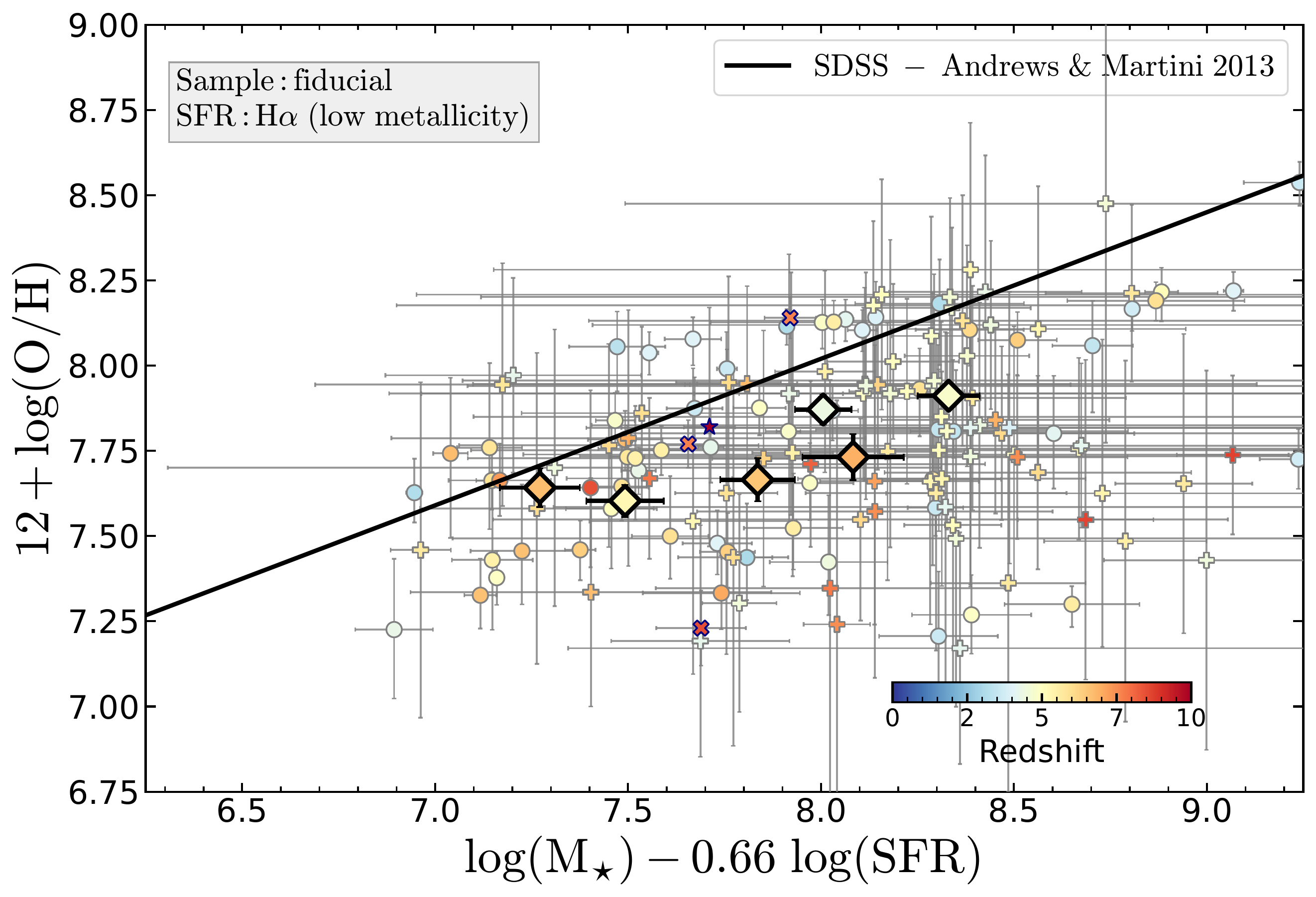}
\includegraphics[width=0.33\textwidth]{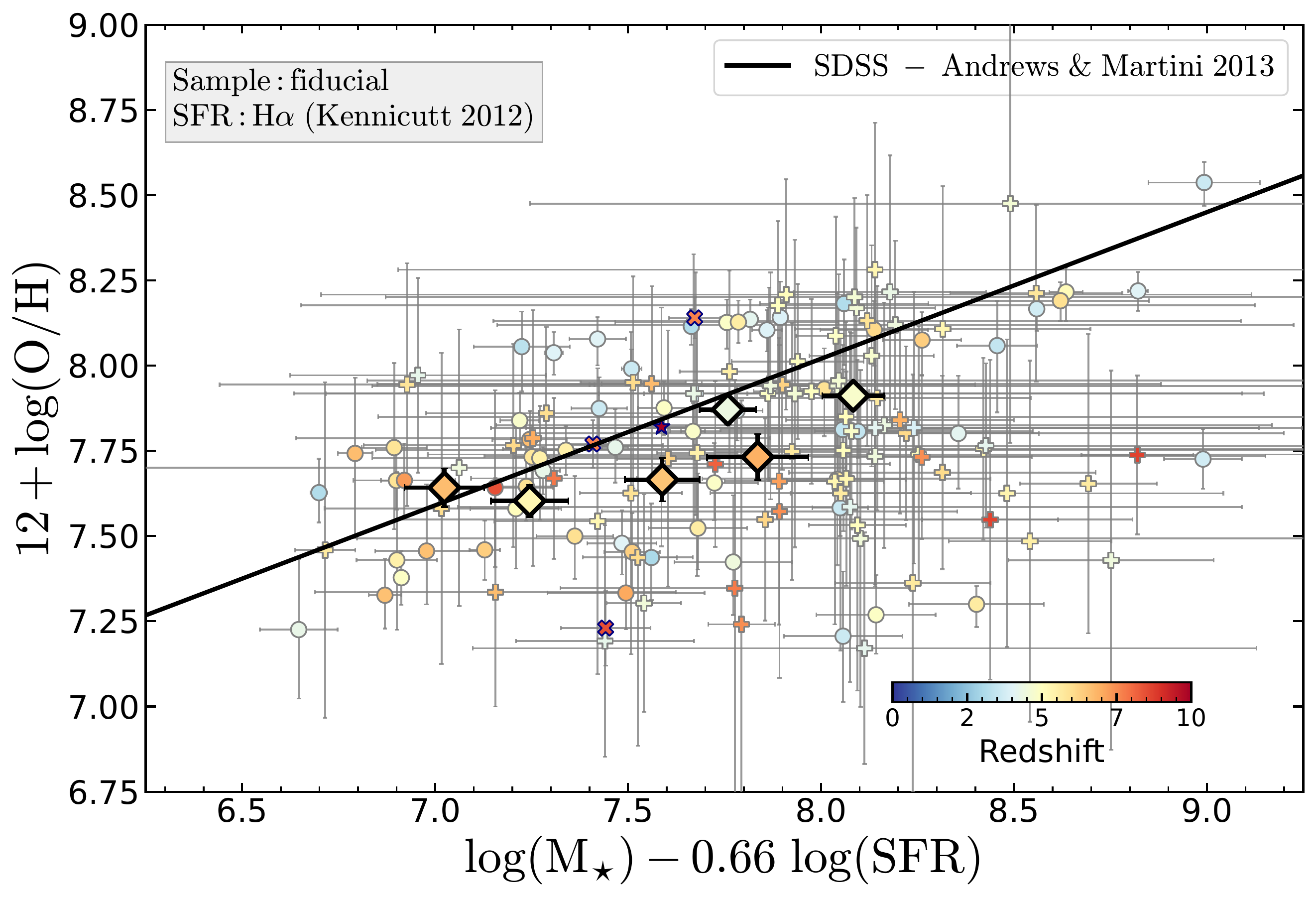}
\end{center}  
\caption{Same as Figure~\ref{fig:projections}, but adopting the formalism from \citealt{andrews_mass-metallicity_2013}, based on SDSS stacked spectra in bins of \mstar and SFR. 
The three panels show, from left to right, the measured versus FMR-predicted metallicities based on three different SFR estimates, i.e., the fiducial value adopted throughout the paper (and based on \textsc{beagle} fitting to both R100 spectra and photometry), the SFR from the \Ha\ flux and the calibration for low-metallicity galaxies from \citealt{reddy_lyA_2022, shapley_balmer_2023}, and the \Ha-based SFR based on the calibration from \citealt{kennicutt_star_2012}, respectively. In the latter case, the SFRs are higher on average by $\sim 0.23$~dex than our fiducial values, and by $\sim 0.37$~dex compared to the calibration for metal poor galaxies, decreasing $\mu_{\alpha}$ and improving the agreement with the predictions of the \citealt{andrews_mass-metallicity_2013} parametrisation of the FMR.}
\label{fig:fmr_AM13}
\end{figure*}

Finally, in the three panels of Figure~\ref{fig:fmr_AM13} we compare (from left to right) the observed metallicity for our \emph{JWST} sample to the predictions of the FMR adopting the parametrisation of \cite{andrews_mass-metallicity_2013} (based on stacked spectra of SDSS galaxies in bins of \mstar and SFR), and in particular considering the projection on the log(O/H) versus $\mu_{\alpha}$ plane, under different estimates for the SFR of galaxies, i.e., fiducial, \Ha-based with low-metallicity calibration, \Ha-based with solar metallicity calibration.
In all cases, the relation from \cite{andrews_mass-metallicity_2013} fails to match the bulk of individual galaxies and the median values in bins of \mstar and redshift, despite embedding a stronger secondary dependence on SFR ($\alpha$=0.66) into $\mu_{\alpha}$, and a steeper slope, compared to different realisations of the FMR \citep[][but see the discussion in Section~\ref{sec:fmr_evol}]{mannucci_fundamental_2010, curti_massmetallicity_2020}.
However, we note that the case with \Ha-based SFR and the calibration from \cite{kennicutt_star_2012}, which significantly reduces the number of ionizing photons per unit SFR compared to both our fiducial and the \Ha-based, low-metallicity calibration scenario, provides the best agreement with the FMR predictions of \cite{andrews_mass-metallicity_2013}, as driven by the higher estimates of the SFR which reduce $\mu_{\alpha}$ and move the points towards the left on the x-axis of the diagram.

\section{Summary of properties for JADES galaxies}

\begin{sidewaystable*}
    \centering    
\caption{Summary of derived properties for JADES galaxies analysed in this paper.}
\begin{tabular}{ *{11}{c|} }
\hline
NIRSpec ID & JADES ID & Redshift & log M$_{\star}$ & SFR$^{a}$ & SFR$^{b}$ [\Ha] & 12 + log(O/H) & C & Metallicity & NC & AGN \\
 & JADES-GS & & [M$_{\odot}$] & [M$_{\odot}$ yr$^{-1}$] & [M$_{\odot}$ yr$^{-1}$] & & ($\star$)  & diagnostics & ($\dagger$) & ($\ddagger$) \\
\hline
2923 & +53.15407-27.82094 & 3.015 & 7.39$\substack{+0.05 \\ -0.05}$ & 0.28$\substack{+0.02 \\ -0.02}$ & 0.23$\pm0.02$ & 7.44$\substack{+0.16 \\ -0.13}$ & P & R3 , R2 , \^R , O32 & 1 &  \\
21150 & +53.16995-27.76840 & 3.088 & 8.18$\substack{+0.01 \\ -0.01}$ & 2.38$\substack{+0.09 \\ -0.08}$ & 2.57$\pm0.04$ & 8.12$\substack{+0.05 \\ -0.05}$ & M & R3 , R2 , \^R , O32 , Ne3O2 & 0 &  \\
18322 & +53.13171-27.77412 & 3.151 & 6.88$\substack{+0.02 \\ -0.02}$ & 0.79$\substack{+0.02 \\ -0.02}$ & 0.80$\pm0.02$ & 7.63$\substack{+0.10 \\ -0.09}$ & M & R3 , R2 , \^R , O32 , Ne3O2 & 0 &  \\
10040 & +53.12304-27.79564 & 3.154 & 7.92$\substack{+0.13 \\ -0.13}$ & 0.64$\substack{+0.17 \\ -0.11}$ & 0.26$\pm0.08$ & 8.18$\substack{+0.13 \\ -0.18}$ & M & R3 , R2 , \^R , O32 & 1 &  \\
19431 & +53.16292-27.77141 & 3.320 & 7.43$\substack{+0.11 \\ -0.11}$ & 2.49$\substack{+0.11 \\ -0.11}$ & 0.86$\pm0.05$ & 8.06$\substack{+0.10 \\ -0.13}$ & M & R3 , R2 , \^R , O32 & 1 &  \\
10013597 & +53.12325-27.76939 & 3.323 & 7.48$\substack{+0.13 \\ -0.13}$ & 0.77$\substack{+0.20 \\ -0.12}$ & 0.28$\pm0.01$ & 7.94$\substack{+0.10 \\ -0.12}$ & P & R3 , R2 , \^R , O32 & 0 & $\ddagger$ \\
3322 & +53.12574-27.81921 & 3.406 & 8.04$\substack{+0.25 \\ -0.25}$ & 0.63$\substack{+0.48 \\ -0.15}$ & 0.39$\pm0.09$ & 7.81$\substack{+0.25 \\ -0.25}$ & P & R3 & 1 &  \\
7629 & +53.16983-27.80344 & 3.452 & 8.45$\substack{+0.09 \\ -0.09}$ & 0.86$\substack{+0.10 \\ -0.07}$ & 0.41$\pm0.02$ & 8.06$\substack{+0.13 \\ -0.20}$ & P & R3 , R2 , \^R , O32 & 1 &  \\
3184 & +53.15010-27.81971 & 3.468 & 8.52$\substack{+0.11 \\ -0.11}$ & 6.11$\substack{+0.66 \\ -0.61}$ & 1.85$\pm0.03$ & 7.81$\substack{+0.07 \\ -0.08}$ & P & R3 , R2 , \^R , O32 , Ne3O2 & 0 &  \\
7809 & +53.12972-27.80274 & 3.578 & 8.29$\substack{+0.05 \\ -0.05}$ & 1.85$\substack{+0.31 \\ -0.23}$ & 0.98$\pm0.09$ & 7.58$\substack{+0.09 \\ -0.08}$ & P & R3 , R2 , \^R , O32 , Ne3O2 & 0 &  \\
10035295 & +53.11434-27.81549 & 3.589 & 7.39$\substack{+0.07 \\ -0.07}$ & 2.34$\substack{+0.06 \\ -0.05}$ & 1.94$\pm0.03$ & 7.56$\substack{+0.12 \\ -0.10}$ & M & R3 , R2 , \^R , O32 , Ne3O2 & 0 & $\ddagger$ \\
6460 & +53.12808-27.80739 & 3.595 & 8.78$\substack{+0.07 \\ -0.07}$ & 0.48$\substack{+0.11 \\ -0.09}$ & 0.20$\pm0.05$ & 8.54$\substack{+0.06 \\ -0.07}$ & P & R3 , R2 , \^R , O32 & 0 &  \\
10009506 & +53.15301-27.80607 & 3.599 & 7.82$\substack{+0.07 \\ -0.07}$ & 2.43$\substack{+0.14 \\ -0.13}$ & 1.65$\pm0.02$ & 7.87$\substack{+0.09 \\ -0.09}$ & M & R3 , R2 , \^R , O32 , Ne3O2 & 0 &  \\
4282 & +53.12051-27.81523 & 3.601 & 7.88$\substack{+0.08 \\ -0.08}$ & 1.12$\substack{+0.36 \\ -0.28}$ & 0.23$\pm0.06$ & 7.21$\substack{+0.19 \\ -0.19}$ & P & R3 & 1 &  \\
19519 & +53.16609-27.77126 & 3.604 & 7.75$\substack{+0.02 \\ -0.02}$ & 2.50$\substack{+0.14 \\ -0.16}$ & 0.99$\pm0.01$ & 7.99$\substack{+0.06 \\ -0.06}$ & P & R3 , R2 , \^R , O32 , Ne3O2 & 0 &  \\
18970 & +53.15650-27.77227 & 3.725 & 9.20$\substack{+0.00 \\ -0.00}$ & 7.22$\substack{+0.15 \\ -0.14}$ & 3.95$\pm0.08$ & 8.17$\substack{+0.06 \\ -0.07}$ & M & R3 , R2 , \^R , O32 , Ne3O2 & 0 &  \\
18028 & +53.12869-27.77482 & 3.731 & 8.84$\substack{+0.04 \\ -0.04}$ & 0.40$\substack{+0.03 \\ -0.02}$ & 0.25$\pm0.05$ & 7.73$\substack{+0.09 \\ -0.09}$ & P & R3 , R2 , \^R , O32 , Ne3O2 & 0 &  \\
10016186 & +53.11695-27.80283 & 3.928 & 8.18$\substack{+0.01 \\ -0.01}$ & 2.70$\substack{+0.15 \\ -0.21}$ & 1.14$\pm0.02$ & 8.14$\substack{+0.11 \\ -0.22}$ & M & R3 , R2 , \^R , O32 & 0 &  \\
10015193 & +53.10889-27.79840 & 3.957 & 8.22$\substack{+0.04 \\ -0.04}$ & 6.03$\substack{+0.31 \\ -0.35}$ & 1.50$\pm0.02$ & 8.10$\substack{+0.06 \\ -0.06}$ & P & R3 , R2 , \^R , O32 , Ne3O2 & 0 &  \\
10013578 & +53.11875-27.77811 & 4.022 & 8.00$\substack{+0.05 \\ -0.05}$ & 2.71$\substack{+0.50 \\ -0.23}$ & 0.91$\pm0.06$ & 7.87$\substack{+0.08 \\ -0.09}$ & P & R3 , R2 , \^R , O32 , Ne3O2 & 0 &  \\
4270 & +53.16871-27.81516 & 4.023 & 8.19$\substack{+0.02 \\ -0.02}$ & 9.00$\substack{+0.25 \\ -0.26}$ & 9.12$\pm0.11$ & 8.04$\substack{+0.06 \\ -0.06}$ & M & R3 , R2 , \^R , O32 , Ne3O2 & 0 &  \\
10015344 & +53.11438-27.77158 & 4.032 & 8.94$\substack{+0.02 \\ -0.02}$ & 12.45$\substack{+1.31 \\ -1.47}$ & 0.65$\pm0.01$ & 8.22$\substack{+0.06 \\ -0.06}$ & P & R3 , R2 , \^R , O32 , Ne3O2 & 0 &  \\
10013545 & +53.12413-27.79914 & 4.035 & 7.95$\substack{+0.06 \\ -0.06}$ & 1.70$\substack{+0.18 \\ -0.17}$ & 2.63$\pm0.08$ & 8.08$\substack{+0.06 \\ -0.08}$ & M & R3 , R2 , \^R , O32 , Ne3O2 & 0 &  \\
7507 & +53.15548-27.80388 & 4.044 & 7.67$\substack{+0.07 \\ -0.07}$ & 1.51$\substack{+0.10 \\ -0.08}$ & 0.80$\pm0.04$ & 7.48$\substack{+0.10 \\ -0.09}$ & P & R3 , R2 , \^R , O32 & 1 &  \\
17777 & +53.16743-27.77548 & 4.134 & 8.25$\substack{+0.19 \\ -0.19}$ & 1.00$\substack{+0.45 \\ -0.18}$ & 0.29$\pm0.02$ & 7.80$\substack{+0.17 \\ -0.16}$ & M & R3 , R2 , \^R , O32 , Ne3O2 & 1 &  \\
7762 & +53.11333-27.80299 & 4.149 & 8.24$\substack{+0.02 \\ -0.02}$ & 4.10$\substack{+0.29 \\ -0.28}$ & 1.87$\pm0.03$ & 8.14$\substack{+0.06 \\ -0.06}$ & M & R3 , R2 , \^R , O32 , Ne3O2 & 0 &  \\
7892 & +53.15003-27.80250 & 4.229 & 7.76$\substack{+0.03 \\ -0.03}$ & 1.96$\substack{+0.05 \\ -0.05}$ & 1.17$\pm0.02$ & 7.76$\substack{+0.11 \\ -0.10}$ & M & R3 , R2 , \^R , O32 , Ne3O2 & 0 &  \\
6519 & +53.15832-27.80724 & 4.260 & 7.68$\substack{+0.04 \\ -0.04}$ & 2.92$\substack{+0.08 \\ -0.06}$ & 1.71$\pm0.03$ & 7.69$\substack{+0.07 \\ -0.07}$ & P & R3 , R2 , \^R , O32 , Ne3O2 & 0 &  \\
10001916 & +53.13228-27.79811 & 4.283 & 6.38$\substack{+0.06 \\ -0.06}$ & 0.23$\substack{+0.02 \\ -0.02}$ & 0.16$\pm0.02$ & 7.23$\substack{+0.21 \\ -0.20}$ & M & R3 & 1 &  \\
8073 & +53.15119-27.80195 & 4.392 & 7.70$\substack{+0.05 \\ -0.05}$ & 1.68$\substack{+0.24 \\ -0.17}$ & 0.50$\pm0.04$ & 7.82$\substack{+0.13 \\ -0.13}$ & P & R3 , R2 , \^R , O32 & 1 & $\ddagger$ \\
10000626 & +53.14700-27.81303 & 4.464 & 6.72$\substack{+0.20 \\ -0.20}$ & 0.34$\substack{+0.02 \\ -0.02}$ & 0.29$\pm0.02$ & 7.29$\substack{+0.12 \\ -0.11}$ & M & R3 & 1 & $\ddagger$ \\
7304 & +53.16083-27.80455 & 4.490 & 7.61$\substack{+0.13 \\ -0.13}$ & 0.30$\substack{+0.07 \\ -0.05}$ & 0.24$\pm0.02$ & 7.42$\substack{+0.20 \\ -0.16}$ & P & R3 & 1 &  \\
17072 & +53.17022-27.77739 & 4.702 & 8.30$\substack{+0.04 \\ -0.04}$ & 0.75$\substack{+0.08 \\ -0.06}$ & 0.53$\pm0.06$ & 8.28$\substack{+0.06 \\ -0.07}$ & P & R3 , R2 , \^R , O32 & 1 & $\ddagger$ \\
\hline
\end{tabular}
    \label{tab:jades_properties}
\end{sidewaystable*}

\addtocounter{table}{-1}

\begin{sidewaystable*}
    \centering    
\caption{(cont.)}
\begin{tabular}{ *{11}{c|} }
\hline
NIRSpec ID & JADES ID & Redshift & log M$_{\star}$ & SFR & SFR [\Ha] & 12 + log(O/H) & C & Metallicity & NC & AGN \\
 & JADES-GS & & [M$_{\odot}$] & [M$_{\odot}$ yr$^{-1}$] & [M$_{\odot}$ yr$^{-1}$] & & ($\star$)  & diagnostics & ($\dagger$) & ($\ddagger$) \\
\hline
18090 & +53.16718-27.77462 & 4.775 & 7.86$\substack{+0.03 \\ -0.03}$ & 4.91$\substack{+0.10 \\ -0.10}$ & 3.93$\pm0.29$ & 7.84$\substack{+0.08 \\ -0.08}$ & M & R3 , R2 , \^R , O32 , Ne3O2 & 0 &  \\
7938 & +53.16268-27.80237 & 4.806 & 8.07$\substack{+0.03 \\ -0.03}$ & 3.22$\substack{+0.07 \\ -0.07}$ & 1.69$\pm0.03$ & 7.81$\substack{+0.08 \\ -0.08}$ & M & R3 , R2 , \^R , O32 , Ne3O2 & 0 &  \\
5457 & +53.11667-27.81093 & 4.863 & 7.79$\substack{+0.28 \\ -0.28}$ & 1.40$\substack{+0.26 \\ -0.20}$ & 0.48$\pm0.02$ & 8.13$\substack{+0.07 \\ -0.08}$ & P & R3 , R2 , \^R , O32 , Ne3O2 & 1 &  \\
4009 & +53.15705-27.81629 & 4.866 & 8.04$\substack{+0.10 \\ -0.10}$ & 2.11$\substack{+0.41 \\ -0.37}$ & 1.27$\pm0.07$ & 7.66$\substack{+0.12 \\ -0.11}$ & P & R3 , R2 , \^R , O32 , Ne3O2 & 1 &  \\
17260 & +53.12689-27.77689 & 4.885 & 7.97$\substack{+0.12 \\ -0.12}$ & 0.44$\substack{+0.05 \\ -0.04}$ & 0.23$\pm0.03$ & 7.27$\substack{+0.12 \\ -0.11}$ & P & R3 & 1 &  \\
10005217 & +53.17351-27.77187 & 4.888 & 7.33$\substack{+0.01 \\ -0.01}$ & 2.24$\substack{+0.05 \\ -0.04}$ & 1.84$\pm0.01$ & 7.38$\substack{+0.08 \\ -0.08}$ & P & R3 , R2 , \^R , O32 , Ne3O2 & 0 &  \\
8113 & +53.12300-27.80176 & 4.903 & 7.80$\substack{+0.05 \\ -0.05}$ & 2.48$\substack{+0.25 \\ -0.16}$ & 0.87$\pm0.04$ & 7.88$\substack{+0.08 \\ -0.08}$ & P & R3 , R2 , \^R , O32 , Ne3O2 & 0 &  \\
5759 & +53.14946-27.80979 & 5.052 & 7.40$\substack{+0.10 \\ -0.10}$ & 1.25$\substack{+0.08 \\ -0.09}$ & 0.83$\pm0.04$ & 7.58$\substack{+0.24 \\ -0.18}$ & M & R3 , R2 , \^R , O32 & 1 &  \\
10015338 & +53.11535-27.77289 & 5.076 & 7.94$\substack{+0.04 \\ -0.04}$ & 3.64$\substack{+0.15 \\ -0.19}$ & 1.89$\pm0.05$ & 7.57$\substack{+0.08 \\ -0.08}$ & P & R3 , R2 , \^R , O32 , Ne3O2 & 0 & $\ddagger$ \\
9452 & +53.11583-27.79755 & 5.122 & 8.13$\substack{+0.17 \\ -0.17}$ & 4.17$\substack{+1.17 \\ -0.97}$ & 1.19$\pm0.11$ & 7.92$\substack{+0.21 \\ -0.22}$ & P & R3 & 0 & $\ddagger$ \\
4902 & +53.11852-27.81297 & 5.123 & 8.69$\substack{+0.03 \\ -0.03}$ & 1.77$\substack{+0.41 \\ -0.24}$ & 0.50$\pm0.02$ & 8.22$\substack{+0.07 \\ -0.09}$ & P & R3 , R2 , \^R , O32 & 1 &  \\
9743 & +53.12300-27.79661 & 5.440 & 7.48$\substack{+0.08 \\ -0.08}$ & 2.80$\substack{+0.28 \\ -0.23}$ & 3.21$\pm0.32$ & 7.43$\substack{+0.26 \\ -0.20}$ & M & R3 & 0 &  \\
9343 & +53.12874-27.79787 & 5.443 & 7.25$\substack{+0.37 \\ -0.37}$ & 0.14$\substack{+0.10 \\ -0.06}$ & 0.39$\pm0.09$ & 7.73$\substack{+0.15 \\ -0.18}$ & P & R2 , Ne3O2 & 0 &  \\
10016374 & +53.11572-27.77496 & 5.503 & 7.63$\substack{+0.05 \\ -0.05}$ & 2.38$\substack{+0.09 \\ -0.07}$ & 1.17$\pm0.03$ & 7.75$\substack{+0.07 \\ -0.07}$ & P & R3 , R2 , \^R , O32 , Ne3O2 & 0 &  \\
6246 & +53.12972-27.80818 & 5.562 & 7.78$\substack{+0.11 \\ -0.11}$ & 3.00$\substack{+0.55 \\ -0.53}$ & 0.59$\pm0.03$ & 7.52$\substack{+0.16 \\ -0.12}$ & M & R3 , R2 , \^R , O32 & 1 &  \\
16745 & +53.13002-27.77839 & 5.567 & 8.26$\substack{+0.03 \\ -0.03}$ & 6.45$\substack{+0.26 \\ -0.27}$ & 2.19$\pm0.04$ & 8.13$\substack{+0.06 \\ -0.06}$ & P & R3 , R2 , \^R , O32 , Ne3O2 & 0 &  \\
6384 & +53.13059-27.80771 & 5.615 & 8.42$\substack{+0.13 \\ -0.13}$ & 1.84$\substack{+0.77 \\ -0.53}$ & 0.44$\pm0.07$ & 7.30$\substack{+0.05 \\ -0.07}$ & P & R3 & 1 &  \\
4404 & +53.11537-27.81477 & 5.764 & 7.87$\substack{+0.04 \\ -0.04}$ & 3.30$\substack{+0.19 \\ -0.13}$ & 3.72$\pm0.12$ & 7.73$\substack{+0.09 \\ -0.09}$ & M & R3 , R2 , \^R , O32 , Ne3O2 & 0 &  \\
3968 & +53.14505-27.81643 & 5.769 & 7.90$\substack{+0.16 \\ -0.16}$ & 1.29$\substack{+0.55 \\ -0.24}$ & 0.26$\pm0.02$ & 7.67$\substack{+0.28 \\ -0.25}$ & P & R3 & 1 &  \\
22251 & +53.15407-27.76607 & 5.798 & 7.89$\substack{+0.03 \\ -0.03}$ & 4.68$\substack{+0.12 \\ -0.13}$ & 4.75$\pm0.11$ & 7.88$\substack{+0.08 \\ -0.07}$ & M & R3 , R2 , \^R , O32 , Ne3O2 & 0 & $\ddagger$ \\
10056849 & +53.11351-27.77284 & 5.814 & 7.07$\substack{+0.06 \\ -0.06}$ & 1.12$\substack{+0.05 \\ -0.04}$ & 1.98$\pm0.17$ & 7.42$\substack{+0.09 \\ -0.09}$ & P & R3 , R2 , \^R , O32 , Ne3O2 & 0 & $\ddagger$ \\
10005113 & +53.16730-27.80287 & 5.821 & 7.02$\substack{+0.03 \\ -0.03}$ & 1.04$\substack{+0.05 \\ -0.05}$ & 0.65$\pm0.12$ & 7.76$\substack{+0.25 \\ -0.24}$ & M & R3 & 1 &  \\
19606 & +53.17655-27.77111 & 5.889 & 7.23$\substack{+0.03 \\ -0.03}$ & 1.68$\substack{+0.04 \\ -0.04}$ & 1.33$\pm0.04$ & 7.66$\substack{+0.09 \\ -0.08}$ & P & R3 , R2 , \^R , O32 , Ne3O2 & 1 &  \\
10013620 & +53.12259-27.76057 & 5.918 & 8.30$\substack{+0.05 \\ -0.05}$ & 2.35$\substack{+0.20 \\ -0.17}$ & 1.18$\pm0.06$ & 7.93$\substack{+0.12 \\ -0.14}$ & M & R3 , R2 , \^R , O32 , Ne3O2 & 0 &  \\
9422 & +53.12175-27.79763 & 5.936 & 7.71$\substack{+0.00 \\ -0.00}$ & 5.35$\substack{+0.04 \\ -0.03}$ & 20.38$\pm0.78$ & 7.53$\substack{+0.07 \\ -0.07}$ & M & R3 , R2 , \^R , O32 , Ne3O2 & 0 & $\ddagger$ \\
6002 & +53.11041-27.80892 & 5.937 & 7.65$\substack{+0.05 \\ -0.05}$ & 1.32$\substack{+0.09 \\ -0.08}$ & 1.79$\pm0.08$ & 7.65$\substack{+0.12 \\ -0.10}$ & M & R3 , R2 , \^R , O32 , Ne3O2 & 1 &  \\
10013618 & +53.11911-27.76080 & 5.944 & 8.78$\substack{+0.21 \\ -0.21}$ & 3.85$\substack{+0.75 \\ -2.12}$ & 0.74$\pm0.04$ & 8.19$\substack{+0.05 \\ -0.06}$ & M & R3 , R2 , \^R , O32 , Ne3O2 & 0 &  \\
19342 & +53.16062-27.77161 & 5.974 & 7.70$\substack{+0.09 \\ -0.09}$ & 1.95$\substack{+0.19 \\ -0.17}$ & 1.39$\pm0.06$ & 7.50$\substack{+0.18 \\ -0.12}$ & M & R3 & 1 &  \\
17566 & +53.15613-27.77584 & 6.102 & 8.50$\substack{+0.13 \\ -0.13}$ & 4.15$\substack{+0.59 \\ -0.44}$ & 1.48$\pm0.07$ & 8.11$\substack{+0.09 \\ -0.11}$ & M & R3 , R2 , \^R , O32 & 0 &  \\
18976 & +53.16660-27.77240 & 6.327 & 7.77$\substack{+0.06 \\ -0.06}$ & 1.66$\substack{+0.13 \\ -0.12}$ & 1.06$\pm0.06$ & 7.45$\substack{+0.11 \\ -0.11}$ & M & R3 , R2 , \^R , O32 & 1 &  \\
18846 & +53.13492-27.77271 & 6.335 & 8.07$\substack{+0.03 \\ -0.03}$ & 4.32$\substack{+0.09 \\ -0.08}$ & 11.39$\pm0.40$ & 7.46$\substack{+0.09 \\ -0.09}$ & P & R3 , R2 , \^R , O32 , Ne3O2 & 0 &  \\
18179 & +53.17582-27.77446 & 6.335 & 8.59$\substack{+0.08 \\ -0.08}$ & 4.93$\substack{+0.68 \\ -0.64}$ & 1.32$\pm0.08$ & 8.07$\substack{+0.08 \\ -0.10}$ & P & R3 , R2 , \^R , O32 & 1 &  \\
\hline
\end{tabular}
\end{sidewaystable*}

\addtocounter{table}{-1}

\begin{sidewaystable*}
    \centering    
\caption{(cont.)}
\begin{tabular}{ *{11}{c|} }
\hline
NIRSpec ID & JADES ID & Redshift & log M$_{\star}$ & SFR & SFR [\Ha] & 12 + log(O/H) & C & Metallicity & NC & AGN \\
 & JADES-GS & & [M$_{\odot}$] & [M$_{\odot}$ yr$^{-1}$] & [M$_{\odot}$ yr$^{-1}$] & & ($\star$)  & diagnostics & ($\dagger$) & ($\ddagger$) \\
 \hline
10005447 & +53.16288-27.76928 & 6.623 & 6.95$\substack{+0.11 \\ -0.11}$ & 0.34$\substack{+0.06 \\ -0.05}$ & 0.38$\pm0.04$ & 7.46$\substack{+0.20 \\ -0.16}$ & P & R3 & 1 &  \\
16625 & +53.16904-27.77884 & 6.631 & 7.35$\substack{+0.03 \\ -0.03}$ & 2.24$\substack{+0.09 \\ -0.07}$ & 2.23$\pm0.07$ & 7.33$\substack{+0.11 \\ -0.10}$ & M & R3 & 0 &  \\
3334 & +53.15138-27.81917 & 6.706 & 7.50$\substack{+0.08 \\ -0.08}$ & 1.19$\substack{+0.12 \\ -0.10}$ & 1.03$\pm0.08$ & 7.78$\substack{+0.08 \\ -0.09}$ & P & R3 , R2 , \^R , O32 , Ne3O2 & 1 &  \\
4297 & +53.15579-27.81520 & 6.713 & 7.29$\substack{+0.03 \\ -0.03}$ & 1.98$\substack{+0.07 \\ -0.05}$ & 2.42$\pm0.16$ & 7.74$\substack{+0.22 \\ -0.25}$ & M & R3 & 0 &  \\
10013609 & +53.11730-27.76408 & 6.929 & 7.71$\substack{+0.06 \\ -0.06}$ & 3.85$\substack{+0.19 \\ -0.18}$ & 5.39$\pm0.31$ & 7.94$\substack{+0.12 \\ -0.14}$ & M & R3 , R2 , \^R , O32 , Ne3O2 & 0 & $\ddagger$ \\
20961 & +53.13423-27.76891 & 7.045 & 7.86$\substack{+0.14 \\ -0.14}$ & 1.03$\substack{+0.12 \\ -0.09}$ & 1.50$\pm0.33$ & 7.33$\substack{+0.11 \\ -0.11}$ & M & R3 , R2 , \^R , O32 & 1 &  \\
10013905 & +53.11833-27.76901 & 7.197 & 7.37$\substack{+0.05 \\ -0.05}$ & 2.20$\substack{+0.14 \\ -0.12}$ & 4.98$\pm0.67$ & 7.63$\substack{+0.09 \\ -0.09}$ & P & R3 , R2 , \^R , O32 , Ne3O2 & 0 & $\ddagger$ \\
10013682 & +53.16746-27.77201 & 7.275 & 6.95$\substack{+0.09 \\ -0.09}$ & 0.82$\substack{+0.08 \\ -0.07}$ & 0.47$\pm0.06$ & 7.66$\substack{+0.11 \\ -0.10}$ & P & R3 , R2 , \^R , O32 , Ne3O2 & 1 &  \\
21842 & +53.15682-27.76716 & 7.981 & 7.51$\substack{+0.04 \\ -0.04}$ & 3.21$\substack{+0.25 \\ -0.25}$ & 1.72$\pm0.08$ & 7.61$\substack{+0.07 \\ -0.07}$ & P & R3 , R2 , \^R , O32 , Ne3O2 & 1 & $\ddagger$ \\
8013 & +53.16446-27.80218 & 8.473 & 7.54$\substack{+0.08 \\ -0.08}$ & 3.03$\substack{+0.32 \\ -0.26}$ & 1.61$\pm0.37$ & 7.64$\substack{+0.29 \\ -0.23}$ & M & R3 & 1 &  \\
10058975 & +53.11243-27.77461 & 9.433 & 8.05$\substack{+0.03 \\ -0.03}$ & 6.62$\substack{+0.20 \\ -0.20}$ & 13.79$\pm1.36$ & 7.35$\substack{+0.08 \\ -0.08}$ & M & R3 , R2 , \^R , O32 , Ne3O2 & 0 & $\ddagger$ \\
\hline
\end{tabular}
    \\
    \vspace{0.2cm}
Notes: 
$^{a}$ derived via \textsc{beagle} fitting to PRISM spectra and NIRCAM photometry (where available, see `NC' column).\
$^{b}$ derived from \Ha (\Hb) flux adopting the calibration for low-metallicity galaxies from \cite{reddy_lyA_2022}.\
$\star$ Spectral configuration (P=prism; M=medium resolution gratings) considered for deriving the metallicity. See Section~\ref{sec:metallicity} for more details.\ 
$\dagger$ NIRCAM photometry: for galaxies with flag=1 the photometry has been included in the \mstar and SFR derivation with \textsc{beagle}; objects with flag=0 have \mstar and SFR measured only from fitting to the PRISM spectra. \ 
$\ddagger$ Galaxies identified as candidate (narrow-lines) AGN based on the criteria described in Scholtz et al. (in preparation). These objects are removed from the fiducial analysis presented in the main body of the paper due to the unknown AGN contribution and the potential biases introduced in the \mstar, SFR, and metallicity derivation, nonetheless their main properties (estimated self-consistently as for the other sources) are included here (and shown in Section~\ref{sec:appendix_A}).
\end{sidewaystable*}

\end{document}